\documentclass[a4paper,11pt]{article}
\pdfoutput=1 % if your are submitting a pdflatex (i.e. if you have
\usepackage{jheppub} % for details on the use of the package, please
\usepackage{amsmath}
\allowdisplaybreaks[4]        
\usepackage{amssymb}
\usepackage{euscript}     
\usepackage{color}         
\usepackage{tensor}        
\usepackage{amsthm}
\usepackage{tcolorbox}
\usepackage[header,title,page,titletoc]{appendix} 
\usepackage[T1]{fontenc} % if needed
\usepackage[numbers]{natbib}  
\usepackage{tasks}
\usepackage{pifont}

\usepackage{tikz}
\usetikzlibrary{angles,quotes}
\usetikzlibrary{arrows.meta}

\renewcommand{\(}{\left(}
\renewcommand{\)}{\right)}
\renewcommand{\[}{\left[}
\renewcommand{\]}{\right]}
 
\def\tr{{\text{Tr}}}
\newcommand{\Gn}{G_\mt{N}}

\newcommand{\eg}{{\it e.g.,}\ }
\newcommand{\ie}{{\it i.e.,}\ }
\newcommand{\viz}{{\it viz.}\ }

\newcommand{\mt}[1]{\textrm{\tiny #1}}

\newcommand{\bra}[1]{\left< #1 \right|}
\newcommand{\ket}[1]{\left| #1 \right>}

\newcommand{\hr}{\hat{\rho}}
\newcommand{\hs}{\hat{\sigma}}

\newcommand{\hK}{\hat{K}}

\newcommand{\mE}{\mathcal{E}}

\newcommand{\s}{\mathbf{s}}
\newcommand{\R}{\mathbf{R}}

\newcommand{\mC}{\mathcal{C}}
\newcommand{\mP}{\mathcal{P}}

\newcommand{\mO}{\mathcal{O}}

\newcommand{\mH}{\mathcal{H}}

\newcommand{\mA}{{\scriptscriptstyle\mathcal{A}}}
\newcommand{\mB}{{\scriptscriptstyle\mathcal{B}}}
\newcommand{\bmA}{\mathcal{A}}
\newcommand{\bmB}{\mathcal{B}}

\newcommand{\xmark}{\text{\ding{55}}}
\newcommand{\cmark}{\text{\ding{51}}}

\def\comp{\ensuremath\mathop{\scalebox{.6}{$\circ$}}}

\title{\boldmath Purification Complexity without Purifications}

\author[a,b]{Shan-Ming Ruan}

\affiliation[a]{Perimeter Institute for Theoretical Physics, 31 Caroline Street North, Waterloo, Ontario N2L 2Y5, Canada}

\affiliation[b]{Department of Physics and Astronomy,
	University of Waterloo, Waterloo, ON N2L 3G1, Canada}

\emailAdd{sruan@perimeterinstitute.ca}

\abstract{ 
We generalize the Fubini-Study method for pure-state complexity to generic quantum states by taking Bures metric or quantum Fisher information metric (QFIM) on the space of density matrices as the complexity measure. Due to Uhlmann's theorem, we show that the mixed-state complexity exactly equals the purification complexity measured by the Fubini-Study metric for purified states but without explicitly applying any purification. We also find the purification complexity is non-increasing under any trace-preserving quantum operations. We also study the mixed Gaussian states as an example to explicitly illustrate our conclusions for purification complexity.
}


\begin{document} 
	
	\maketitle
	\flushbottom

	\section {Introduction and Motivations}
   
   Quantum information concepts and perspectives have inspired surprising new insights into the understanding of the gravitational holography, \eg \cite{Ryu:2006bv,Rangamani:2016dms,Blanco:2013joa,Faulkner:2013ica,Almheiri:2014lwa,Swingle:2009bg,VanRaamsdonk:2010pw,Harlow:2018fse}. One fascinating concept that has recently drawn increasing attention is quantum circuit complexity \cite{Susskind:2018pmk,qft1} which is defined as the minimal number of simple operations required in order to prepare the target state $\ket{\Phi_\mt{T}}$ starting from a given reference state $\ket{\Psi_{\mt{R}}}$ using a set of elementary gates. There exist several proposals for the holographic dual of circuit complexity of a boundary state such as the complexity=volume (CV) conjecture \cite{Susskind:2014rva,Stanford:2014jda} and the complexity=action (CA) conjecture \cite{Brown:2015bva}. Due to the absence of a well-posed definition for the boundary dual of holographic complexity, some progresses have been made toward defining the computational complexity of states in quantum field theory in recent years, \eg Nielsen's geometric method \cite{Nielsen,Nielsen02,qft1}, Fubini-Study method \cite{qft2} and path-integral complexity proposal \cite{Caputa:2017urj,Czech:2017ryf}. See also \cite{coherent,Hackl:2018ptj,Khan:2018rzm,Bhattacharyya:2018bbv,Chapman:2018hou,Ali:2018fcz,Caputa:2017yrh,Bhattacharyya:2018wym,Camargo:2018eof,Camargo:2019isp,Caputa:2018xuf,Erdmenger:2020sup,Flory:2020eot} and references therein for more recent developments on various proposals. In light of the definitions for the complexity between two pure states, it is natural to generalize it to the case of mixed states. Several proposals have been made to define mixed-state complexity in \cite{Agon:2018zso}. 
   
   More explicitly, we would like to explore the mixed-state complexity between arbitrary quantum states, \viz
   \begin{equation}
   \mC\( \hs_{\mt{R}}, \hr_{\mt{T}}\) : \hs_{\mt{R}} \longrightarrow \hr_{\mt{T}}\,,
   \end{equation}
   in this paper. Different from the complexity of pure states, \ie $\mC\(\ket{\Phi_\mt{R}}, \ket{\Psi_\mt{T}}\)$ for which the unitary operations are sufficient to construct the transformation from a reference state to a target state, we need to introduce the non-unitary operation if the target state is a mixed state in the Hilbert space $\mH_{\mA}$, \eg the quantum states associated with a subregion in QFT. The non-unitary operations call for the ancillae. In order to respect on unitary evolution, we can consider the complexity for purified states with the help of an auxiliary system $\mH_{\mA^c}$. More generally, we can also start from purified reference state if it is also not pure. 
   In light of the non-uniqueness of the purification, a natural definition of mixed-state complexity between the reference $\hs_\mt{R}$ and the target state $\hr_\mt{T}$ is called {\it purification complexity}  $\mP$ that is defined to be 
   	\begin{equation}\label{purification_complexity}
   	\begin{split}
   	&\mP\( \hs_{\mt{R}}, \hr_{\mt{T}}\) \equiv   \min\limits_{\Phi} \min\limits_{\Psi} \, \mC \(  \ket{\Phi_{\mt{R}}}, \ket{\Psi_{\mt{T}}}\) \,, \\
   	\text{with}\qquad & \tr\(\ket{\Phi_{\mt{R}}} \bra{\Phi_{\mt{R}}}\)= \hs_{\mt{R}}\,, \quad  \tr\(\ket{\Psi_{\mt{T}}} \bra{\Psi_{\mt{T}}}\)= \hr_{\mt{T}} \,,
   	\end{split}
   	\end{equation} 
   where the minimization is performed over all possible purifications $\ket{\Phi_{\mt{R}}}, \ket{\Psi_{\mt{T}}}$ of $\hs_{\mt{R}},\hr_{\mt{T}}$, respectively, and $\mC$ denotes a specific pure-state complexity we are interested in. Taking Nielsen's geometric method, the purification complexity with various cost functions has been explored in \cite{purification} by focusing on Gaussian states.
   
   However, the purification complexity is based on the triple minimizations. First of all, we need to minimize all paths to find the optimal circuit for a given purified reference state and target state. Secondly, we also have to search for the optimal purifications twice by minimizing the complexity for all free parameters due to the freedom in purification. With these tips from the purification complexity based on Nielsen's geometric method, it is natural to explore the similar purification complexity by taking account of the Fubini-Study metric as the complexity measure for purified states. As shown in \cite{purification}, finding the optimal purification for mixed states in QFT is a challenging task even for Gaussian states due to the huge number of free parameters in purification. 
   	
\begin{figure}[htbp]
   	\centering\includegraphics[width=5.0in]{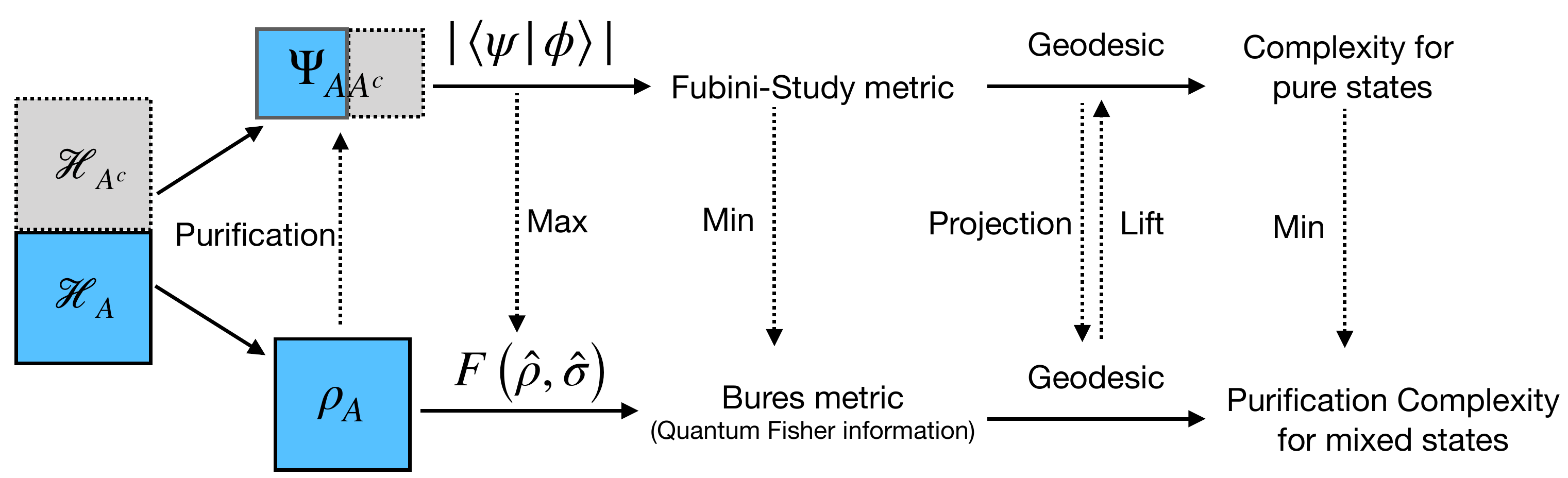}
   	\caption{The connections between pure-state complexity from the Fubini-Study metric and purification complexity derived from the Bures metric (or quantum Fisher information metric). The Hilbert space consisting of quantum states $\hr_{\mA}$ we are interested in is denoted by $\mH_{\mA}$ and the extended Hilbert space with introducing auxiliary system is represented by $\mH_{\mA}\otimes \mH_{\mA^c}$.}\label{fig:purification_complexity_Bures}
   \end{figure}

   In view of the difficulties in the minimization for purification complexity, we would like to generalize the Fubini-Study metric method for pure-state complexity to arbitrary quantum states $\hr_{\mA}$ by defining the geodesic distance in the space of density matrix equipped with a special metric as the complexity measure for mixed states. Different from pure states where the Fubini-Study metric serves as one unique definition \cite{bengtsson2017geometry}, there are too many similar definitions of finite distance and also corresponding local metrics for mixed states \cite{bengtsson2017geometry,chruscinski2012geometric,NielsenChuang}. We propose to consider the dubbed \textcolor{blue}{Bures metric} or \textcolor{blue}{quantum Fisher information metric} (QFIM) as the complexity measure for generic quantum states. Thanks to Uhlmann's fidelity theorem \cite{uhlmann1976transition,NielsenChuang}, we find that the complexity from the quantum Fisher information metric can be exactly explained as the purification complexity $\mP$ with the Fubini-Study metric acting as the complexity measure of purified states. As a result, our proposal avoids the explicit process for purification and also minimization. The connections are summarized in the figure \ref{fig:purification_complexity_Bures}.
   
    \section{Definition: Towards Defining Complexity for Mixed States}
    \subsection{Uhlmann's fidelity and Quantum Fisher Information Metric}
    In \cite{qft2} the circuit complexity connecting a pure reference state and a pure target state is proposed to be the energy or length of a geodesic measured by the Fubini-Study metric on the space of pure states. Before discussing its generalization to mixed states, let's first simply review several concepts associated with the Fubini-Study metric. Stating from a family of pure states $|\Psi(\lambda)\rangle$ with parameters $\lambda^\mu$, one can derive the quantum information metric (fidelity susceptibility) $g_{\mu\nu}$, \eg \cite{NielsenChuang,fidelity_review} by
   \begin{equation}\label{fidel2}
   F(\lambda,\lambda+d\lambda)=1-\frac12 g_{\mu\nu}
   \,d\lambda^\mu\,d\lambda^\nu +\mathcal{O}(d\lambda^3) \,,
   \end{equation}
   where $F(\lambda,\lambda')$ is the quantum fidelity defined as the inner product between two states, \ie 
   \begin{equation}\label{fidel}
   F(\lambda,\lambda')  = |\langle\Psi(\lambda)|\Psi(\lambda')\rangle|\,.
   \end{equation}
   For pure states, it is easy to show that the quantum information metric is equivalent to the Fubini-Study metric 
   \begin{equation}\label{FSmetric}
   g_{\mu\nu}^{\mt{FS}}=\frac{1}{2}\( \langle\partial_\mu \Psi|\partial_\nu\Psi\rangle +  \langle\partial_\nu \Psi|\partial_\mu\Psi\rangle  \) - \langle\partial_\mu \Psi|\Psi\rangle  \langle\Psi|\partial_\nu\Psi\rangle\,.
   \end{equation}
   In the sense of distance, Fubini-Study metric can be considered as the infinitesimal version of the finite distance $(1-F(\lambda, \lambda'))$ between arbitrary two pure states $\ket{\Psi(\lambda)}, \ket{\Psi(\lambda')}$.
   Following \cite{qft2, coherent},  we can define the complexity of pure states as the length of the geodesic $\lambda^\mu(s)$  connecting a reference state $\ket{\Phi_\mt{R}}$ and a target state $\ket{\Psi_\mt{R}}$, \ie 
   \begin{equation}\label{Def_FScomplexity}
   \begin{split}
   \mathcal{C}_\mt{FS} \( \ket{\Phi_\mt{R}},  \ket{\Psi_\mt{R}}\)&= \int_0^1\!\! d\s\   \sqrt{ 2g_{\mu\nu}^{\mt{FS}}\(\lambda \)\, \dot{\lambda}^\mu\, \dot{\lambda}^\nu}\,,\\
   \end{split}
   \end{equation}
   where the boundary conditions are determined by the reference state and target state, $\dot{\lambda}{}^\mu(s) = \frac{d\lambda^\mu(s)}{d\s}$ denotes the tangent vector to the trajectory and we artificially add a factor $2$ to make this definition consistent with the results from $F_2$ norm in Nielson's geometric method \cite{coherent} for Gaussian states. Similar to the $\kappa=2$ cost function used in the Nielsen's geometric method \cite{qft1,purification,coherent}, we can also use the Fubini-Study metric to define the energy of the geodesic as the complexity of pure states by
   \begin{equation}
   \mathcal{C}_\mt{FS}^{\kappa=2} =2 \int_0^1\!\! d\s\   g_{\mu\nu}^{\mt{FS}}\, \dot{\lambda}^\mu\, \dot{\lambda}^\nu = \(\mathcal{C}_\mt{FS} \)^2\,,
   \end{equation}
   in order to match the divergence of holographic complexity for vacuum state \cite{qft1,qft2}. Note the second equality $\mathcal{C}_\mt{FS}^{\kappa=2}=\(\mathcal{C}_\mt{FS} \)^2$ always holds  since we have assumed that the integral is done with the on-shell solution and then the integrand is a constant with respect to affine parameter $\s$ along the geodesic. Due to the Riemannian structure, it is easy to find that the Euler-Lagrangian equation leads to the conclusion that $\frac{d}{d \s} \(g_{\mu\nu}^{\mt{FS}}\, \dot{\lambda}^\mu\, \dot{\lambda}^\nu\) =0$. Equipped with the complexity measure $\mC_{\mt{FS}}$ for pure states, we can also define the specific purification complexity as  
   \begin{equation}\label{purification_FS01}
   \mP_{\mt{FS}}\( \hs_{\mA},\hat{\rho}_\mA \)=  \min\limits_{\Phi}  \min\limits_{\Psi} \, \mC_{\mt{FS}} \(  \ket{\Phi_{\mA\mA^c}}, \ket{\Psi_{\mA\mA^c}}\) \,,
   \end{equation}
   where $\ket{\Phi_{\mA\mA^c}}, \ket{\Psi_{\mA\mA^c}}$ denote the purifications of the two density operators $\hs_{\mA}, \hr_{\mA}$, respectively.
   
   Inspired by the connections between quantum fidelity and circuit complexity proposal \cite{qft2}, we would like extend the Fubini-Study method to more generic quantum states but avoiding the challenges in purification complexity due to the minimization over all purifications. Obviously, the key question is how to define an analog of the Fubini-Study metric for mixed states. We ask for the help of the quantum fidelity \footnote{In some literatures, the quantum fidelity may be defined as $F(\hat{\rho}, \hat{\sigma})^2$. } between two general quantum states. In this paper, we focus on the fidelity of two quantum states $\hr$ and $\hs$ defined by \cite{NielsenChuang} 
    \footnote{For a positive semi-definite operator, its square root uniquely exits and is also positive semi-definite.}
   \begin{equation}\label{fidelity_def}
   F(\hat{\rho}, \hat{\sigma})\equiv  \tr\(\sqrt{\sqrt{\hat{\rho}}\hat{\sigma}\sqrt{\hat{\rho}}}\)= || \sqrt{\hr}\,\sqrt{\hs}||_1\,, 
   \end{equation}
   which is simply reduced to the overlap $|\langle \phi | \psi \rangle |$ for pure states. If at least one of the two states is pure, the quantum fidelity $F$ reduces to the overlap between two density matrices 
   \begin{equation}
    F(\hat{\rho}, \hat{\sigma}) =\sqrt{ \tr\( \hat{\rho}\,\hat{\sigma}\) } = \sqrt{ \bra{\psi} \hat{\rho} \ket{\psi}} \,, \qquad   \hat{\sigma} = \ket{\psi} \bra{\psi} \,.
   \end{equation}
   This quantum fidelity $F(\hat{\rho}, \hat{\sigma})$ can be naturally interpreted as a generalization of the transition probability for pure states. For later use, we also list some interesting and nice properties of the quantum fidelity as follows:
   \begin{tasks}(1)
   	\task  $0 \le  F(\hat{\rho}, \hat{\sigma})  \le 1$\,;
   	\task   $F(\hat{\rho}, \hat{\sigma})=1  \Longleftrightarrow  \hr=\hs$; \,\, $F(\hat{\rho}, \hat{\sigma})=0  \Longleftrightarrow  \hr \perp \hs$\,;
   	\task  Symmetric : $ F(\hat{\rho}, \hat{\sigma}) = F(\hs,\hr) $\,;
   	\task  Concavity  :  $ F\(\hs, \sum_i  p_i \hr_i \) \ge \sum_i p_i  F\(\hs,\hr_i\) $ for all $0\le p_i\le 1$ such that $\sum_i p_i =1$\,;
   	\task Strong Concavity:  $F\(\sum_{i} p_i\hs_i, \sum_i q_i\hr_i \) \ge \sum_i\sqrt{p_i q_i} F\(\hs_i,\hr_i \)$ for all $0\le p_i, q_i \le 1$ such that $\sum_i p_i =1 =\sum_i q_i$\,;
   	\task  Multiplicativity : $F\( \hr_1 \otimes \hr_2, \hs_1 \otimes \hs_2 \) =F\( \hr_1 , \hs_1 \)F\( \hr_2, \hs_2 \)$\,;
   	\task Unitary invariance : $ F(\hat{\rho}, \hat{\sigma}) = F(U\hat{\rho}U^\dagger, U\hat{\sigma}U^\dagger) $.
   \end{tasks}
   The proofs for those properties can be found in textbooks on quantum information, \eg \cite{NielsenChuang,watrous2018theory,wilde2013quantum} or original references \eg \cite{jozsa1994fidelity,PhysRevLett.76.2818,Nielsen:1996uh}.
   There are also some other definitions for the quantum fidelity or distance between two density matrices. However, we prefer the definition in \eqref{fidelity_def} because there is an important theorem called {\it Uhlmann's theorem} which states that \\
    \begin{tcolorbox}[title={Uhlmann's theorem }]
   	For any possible purification $\ket{\psi}$ and $\ket{\phi} $ in system $\bmA\bmA^c$ with respect to $\hat{\rho}$ and $\hat{\sigma}$ in system $\bmA$, respectively \footnote{So it means that we have the constrains $\tr_{\mA^c} \ket{\psi}\bra{\psi}= \hat{\rho}$ and $\tr_{\mA^c} \ket{\phi}\bra{\phi}= \hat{\sigma}$.}, the quantum fidelity satisfies 
   	\begin{equation}\label{Uhlmann_theorem}
   	\begin{split}
   	F(\hat{\rho}, \hat{\sigma})&\equiv \tr\(\sqrt{\sqrt{\hat{\rho}}\hat{\sigma}\sqrt{\hat{\rho}}}\)= \max\limits_{\ket{\psi},\ket{\phi}}  | \langle{\phi} \ket{\psi}| \ge | \langle{\phi} \ket{\psi}| \,,\\
   	\end{split}
   	\end{equation}
   	where the maximization is over all purifications of $\hr, \hs$ and the last equality can always be saturated by some appropriate purifications (called parallel purification).   
   \end{tcolorbox}
   Uhlmann's theorem plays an important role in connecting the complexity from quantum Fisher information metric to the purification complexity $\mP$. 
   Another crucial property for the quantum fidelity is associated with the lowest bound of fidelity and its meaning in distinguishing states. Let $\{E_a\}$ with $\sum_a E_a= \mathbb{I}$ be an arbitrary generalized measurement, \ie {\it positive operator-valued measure} (POVM), the quantum fidelity between two density operators satisfies
   \begin{equation}\label{lowest}
   F(\hat{\rho}, \hat{\sigma}) = \min\limits_{\{E_a\}}  \sum_a \sqrt{\tr(\hat{\rho}E_a)} \sqrt{\tr(\hat{\sigma}E_a)} \,,
   \end{equation}	
   where the minimization is performed with respect to all sets of positive operators $\{E_a\}$ and we can call the POVM saturating the bound as the optimal POVM.  
  Considering two distributions $P_1(a)= \tr(\hat{\rho}E_a)$ and $P_2(a)= \tr(\hat{\sigma}E_a)$, it is clear that the definition of quantum fidelity in  \eqref{fidelity_def} is the analogue of the statistical overlap and is actually the minimal overlap between these two probability distributions. In view of the importance of this inequality, let's sketch the proof to convince the readers who are not familiar with that. Starting from any POVM and unitary operator $U$, one can find \cite{PhysRevLett.76.2818}
  \begin{equation}
  \begin{aligned} \sum_{a} \sqrt{\tr\left(\hr E_{a}\right)} \sqrt{\tr\left(\hs E_{a}\right)} &=\sum_{a} \sqrt{\tr\left(U \sqrt{\hr} E_{a} \sqrt{\hr} U^{\dagger}\right)} \sqrt{\tr\left( \sqrt{\hs} E_{a} \sqrt{\hs}\right)} 
  \\ & 
  \geq \sum_{a}\left|\tr\left(U \sqrt{\hr}  \sqrt{E_{a}} \sqrt{E_{a}} \sqrt{\hs} \right)\right| =\left|\tr\left(U \sqrt{\hr}\sqrt{\hs} \right)\right| \end{aligned} \,,
  \end{equation}
  where we only need the cyclic property of the trace and Schwarz inequality (\ie the Schatten $2$-norm is sub-multiplicative.). In consideration of the fact that the maximization over all unitary operator $U$, namely
  \begin{equation}
   \max\limits_{U} \, \left| \tr\(U O\) \right| = \tr\(\sqrt{O^\dagger O} \)\,,
  \end{equation}
  is saturated if and only if $U O = e^{i \phi}\sqrt{O^\dagger O}$, we finally arrive at the conclusion for the quantum fidelity, \ie \eqref{lowest} by applying that maximization to operator $\sqrt{\hr}\sqrt{\hs}$. Furthermore, when one can also find that the optimal POVM is the special positive semi-definite operator  with spectral decomposition 
  \begin{equation}
  \hat{E} \equiv \sum_a \lambda_a E_a=  \sum_a \lambda_a \ket{a}\bra{a}= \(\hs\)^{-\frac{1}{2}} \sqrt{\sqrt{\hs} \hr \sqrt{\hs}}  \(\hs\)^{-\frac{1}{2}} \,,
  \end{equation}
  which is nothing but the geometric mean of $\hat{\rho}$ and $\hs^{-1}$.
  Interpreting the quantum fidelity \eqref{fidelity_def} as the minimization of statical overlap, one can prove some other interesting properties, \eg the non-broadcasting of non-commuting mixed states \cite{PhysRevLett.76.2818}. Here we stress its another application that the quantum fidelity $F\(\hs,\hr\) $ is non-decreasing under any quantum operations.  Similar to the purifications of mixed states, we can introduce the bipartite system $\mathcal{H}_{\mA}\otimes \mathcal{H}_{\mB} $ and have the corresponding density matrices in the two subsystems such that 
  \begin{equation}
  \hat{\rho}_{\mA\mB} \in \mathcal{H}_{\mA}\otimes \mathcal{H}_{\mB} \,, \quad \tr_{\mB}\( \hat{\rho}_{\mA\mB} \) = \hr_{\mA}\,, \qquad \tr_{\mA}\( \hat{\rho}_{\mA\mB} \) = \hr_{\mB}\,.
  \end{equation}
  The minimization in \eqref{lowest} implies we have the monotonicity of quantum fidelity \footnote{Obviously, this monotonicity is a consequence of Uhlmann's theorem.  Since the optimal purification of $\hat{\rho}_{\mA\mB}$ and  $\hat{\sigma}_{\mA\mB}$ are surely the purification of $\hat{\rho}_{\mA}$ and  $\hat{\sigma}_{\mA}$, respectively, they may not be the optimal ones with respect to $\hat{\rho}_{\mA}$ and $\hat{\sigma}_{\mA}$.}
  \begin{equation}\label{nondecreasing}
     F\(\hat{\rho}_{\mA\mB}, \hat{\sigma}_{\mA\mB}\) \le F\(\hat{\rho}_{\mA}, \hat{\sigma}_{\mA}\) \,,
  \end{equation}
  which means that any partial trace can not reduce Uhlmann's fidelity and also indicates that the density operators in a subsystem are less distinguishable than those in a larger system. More generally, we can also explain this property in the way associated with quantum operation. As it is known \cite{NielsenChuang}, the quantum operation (quantum channel) $\mathcal{E}$ defined by {\it completely positive trace-preserving} (CPTP) map can be explained in different ways (see appendix \ref{sec:app_operation} for more details). For example, we can realize quantum operations $\mathcal{E}\( \hr_{\mA}\)$ on density operators $\hr_{\mA}$ by the unitary transformations acting on the extended Hilbert space $\mathcal{H}_{\mA}\otimes \mathcal{H}_{\mA^c} $ with some ancillae (or environment), \ie 
  \begin{equation}
  \mathcal{E}\( \hr_{\mA}\)= \tr_{\mA^c} \( U_{\mA\mA^c}\(  \hat{\rho}_{\mA}\otimes \hr_{\mA^c}\)U^\dagger_{\mA\mA^c}  \)\,,
  \end{equation}
  where the $\hr_{\mA^c}$ is the initial state for the ancillae and $\tr_{\mA^c}$ refers to tracing out the ancilla part. On the other hand, we can also rewrite the equivalent quantum operations in the operator sum representation by
  \begin{equation}
  \mathcal{E}\( \hr_{\mA}\)= \sum_{a} \hat{M}_a \hat{\rho}_{\mA} \hat{M}_a^\dagger\,,\quad  \sum_{a}\hat{M}_a^\dagger \hat{M}_a  =\mathbb{I} \,,
  \end{equation} 
  where $\mathbb{I}$ denotes the identity matrix.
  So finally, in the sense of quantum operations, one can understand the non-decreasing of quantum fidelity as 
  \begin{equation}
   F\(\mathcal{E}\( \hr_{\mA}\), \mathcal{E}\( \hs_{\mA}\)\)\ge  F\(\hat{\rho}_{\mA}\otimes \hr_{\mA^c}, \hat{\sigma}_{\mA}\otimes \hs_{\mA^c}\) = F\(\hat{\rho}_{\mA}, \hat{\sigma}_{\mA}\)F\(\hat{\rho}_{\mA^c}, \hat{\sigma}_{\mA^c}\)\,,
  \end{equation}
  where we use the non-decrease of the quantum fidelity under partial trace and its unitary invariance in the first inequality and its multiplicativity to derive the second equality. Taking the ancilla part for the two density operators as the same, we can arrive at a monotonic form 
	\begin{equation}\label{fidelity_nondecreasing}
	F\(\mathcal{E}\( \hr_{\mA}\), \mathcal{E}\( \hs_{\mA}\)\) \ge F\(\hat{\rho}_{\mA}, \hat{\sigma}_{\mA}\)\ge F\(\hat{\rho}_{\mA\mB}, \hat{\sigma}_{\mA\mB}\)\,,
	\end{equation}     
	indicating the quantum operation can not decrease the fidelity. Physically, the above inequality also implies physical process can not increase the distinguishability between quantum states. The first inequality holds for any trace-preserving quantum operation (quantum channel) defined by 
$\mathcal{E} : \hr \rightarrow \mathcal{E}\( \hr\)$ and can be understood as the quantum analog of the classical information-processing inequality.

  After introducing the quantum fidelity between density matrices, we move on to our new proposal for the circuit complexity between two generic quantum states. 
  Similar to the pure-state complexity based on the Fubini-Study metric, we can parametrize the space of quantum states by density operators $\hr\( \lambda^\mu\)$ with independent parameters $\lambda^\mu$.  Then our proposal to the complexity from any reference state $\hs_{\mt{R}}\( \lambda^\mu_0\)$ to any target state $\hr_{\mt{T}}\( \lambda^\mu_1\)$ is the following 
   \begin{equation}\label{def_IFcomplexity}
   \begin{split}
   \mathcal{C}_\mt{IM} \( \hs_{\mt{R}}, \hr_{\mt{T}} \)&= \int_0^1\!\! d\s\,   \sqrt{ 2g^{\mt{IM}}_{\mu\nu}\, \dot{\lambda}^\mu\, \dot{\lambda}^\nu}\,,\qquad  \dot{\lambda}^\mu= \frac{d\lambda^\mu(\s) }{d\s }\,,\\
   \mC_{\mt{IM}}^{\kappa=2}\( \hs_{\mt{R}}, \hr_{\mt{T}} \)&= 2\int_0^1\!\! d\s\,   g^{\mt{IM}}_{\mu\nu}\, \dot{\lambda}^\mu\, \dot{\lambda}^\nu= \(  \mathcal{C}_\mt{IM}\)^2\,,
   \end{split}
   \end{equation}
   where the integral is taken along the optimal circuit \ie the geodesic  $\gamma_{\mt{IM}}$ measured by the fidelity susceptibility $g^\mt{IM} _{\mu\nu}$ with reference state and target state as the endpoints. If the geodesics are not unique, we should choose the one minimizing the distance between the reference state and the target state. 
   The quantum fidelity susceptibility $g_{\mu\nu}^\mt{IM}$ can be derived from the expansion of Uhlmann's fidelity between two nearby quantum states, \ie 
   \begin{equation}\label{def_IM}
   g_{\mu\nu}^{\mt{IM}}\( \lambda \) d \lambda^\mu d\lambda^\nu=_2 2\(1 - 	F\(\hat{\rho}(\lambda), \hat{\rho} (\lambda+\delta \lambda\) \) =_2 1-F(\hat{\rho}, \hat{\rho}+\delta \hr)^2  \\\,,
   \end{equation}
   where the equality is taken at the second order of $\delta \lambda$ and the quantum fidelity $F$ for mixed states is defined to be \eqref{fidelity_def}. We have used subscript "IM" for this metric because it equals the {\it quantum Fisher information metric} (QFIM). See appendix \ref{sec:app_metric} for more explicit forms of QFIM. Technically, it is also convenient to derive the quantum Fisher information metric by 
   \begin{equation}\label{IM_derivative}
   g_{\mu\nu}^{\mt{IM}}= -\lim_{ \lambda' \rightarrow \lambda } \frac{\partial^2 F(\lambda,\lambda') }{\partial{\lambda^\mu}\partial{\lambda^\nu}}  \,.
   \end{equation}
   For mixed states, the quantum Fisher information metric or quantum fidelity susceptibility is also known as Bures metric \cite{bures1969extension,bengtsson2017geometry,jozsa1994fidelity,chruscinski2012geometric,Liu:2019xfr} \footnote{In literatures in quantum information field, quantum Fisher information metric and Bures metric are different by a factor 4. Because we need to normalize the metric in order to measure complexity, we ignore this factor and do not distinguish the two metrics in this paper.} which is derived from the finite Bures distance defined by $1-F\(\hat{\rho}, \hat{\sigma} \)$. We leave the similar deviation from the viewpoint of Bures distance in section \ref{sec:Bures}. 
   
   To close this subsection, we should stress that the choice $\hat{\rho}(\lambda^\mu)$ is not arbitrary and in principle, it is determined by the set of gates on the whole system. From the viewpoint of the quantum circuit with ancillae (\eg figure \ref{fig:circuit_fidelity}), the whole Hilbert space is defined by $\hr(\lambda^\mu)= \tr_{\mA^c} \( \ket{\Psi_{\mA\mA^c}} \bra{\Psi_{\mA\mA^c}} \)$ where the pure states are constrained by the set of gates, \ie all possible unitary operations $U_{\mA\mA^c}$ from $\ket{\Psi_{\mA\mA^c}}= U_{\mA\mA^c} \ket{\Phi_{\mt{R}}}$. The last thing we want to point out is the different meanings of "optimal" states. Uhlmann's fidelity provides a criterion for the optimal purification with respect to any two states. However, the circuit complexity is based on the optimal path in the space of states, \ie geodesic $\gamma_{\mt{IM}}$. The quantum fidelity only quantifies the local measure while the geodesic length indicates a global optimization for a given reference state and a target state. 
 
  \subsection{Purification Complexity without Purifications} \label{sec:purification_complexity}
  \subsubsection{It is Purification Complexity}
\begin{figure}[htbp]
   	\centering\includegraphics[width=6.0in]{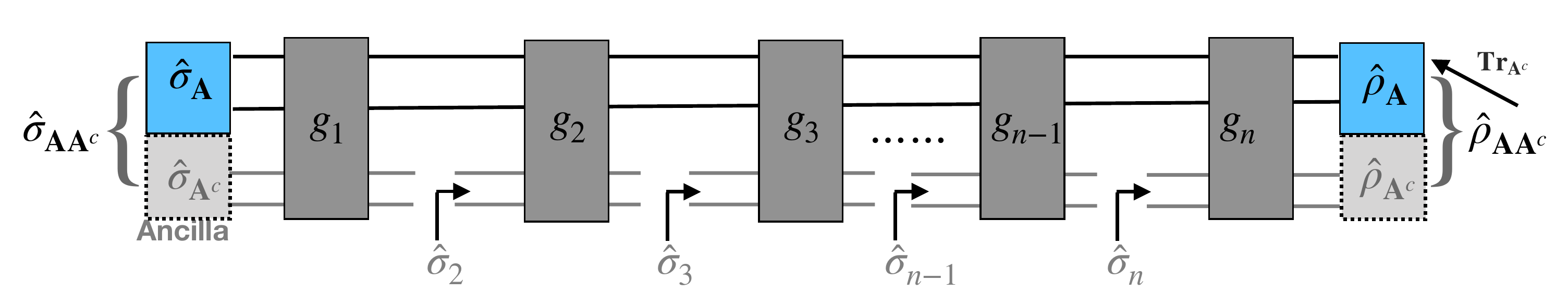}
   	\caption{A lift of evolution $\hr_{\mA}\(\s\)$ to the extended Hilbert space: A general circuit connecting purified state $\hs_{\mA\mA^c}$ to $\hr_{\mA\mA^c}$ with different ancillae after every step because we do not count the cost of introducing ancilla and tracing out the auxiliary system. More importantly, it is based on the fact that any trace-preserving quantum operation is equivalently described by the unitary evolution with ancilla.}\label{fig:circuit_fidelity}
   \end{figure}

\begin{figure}[htbp]
	\centering\includegraphics[width=5.0in]{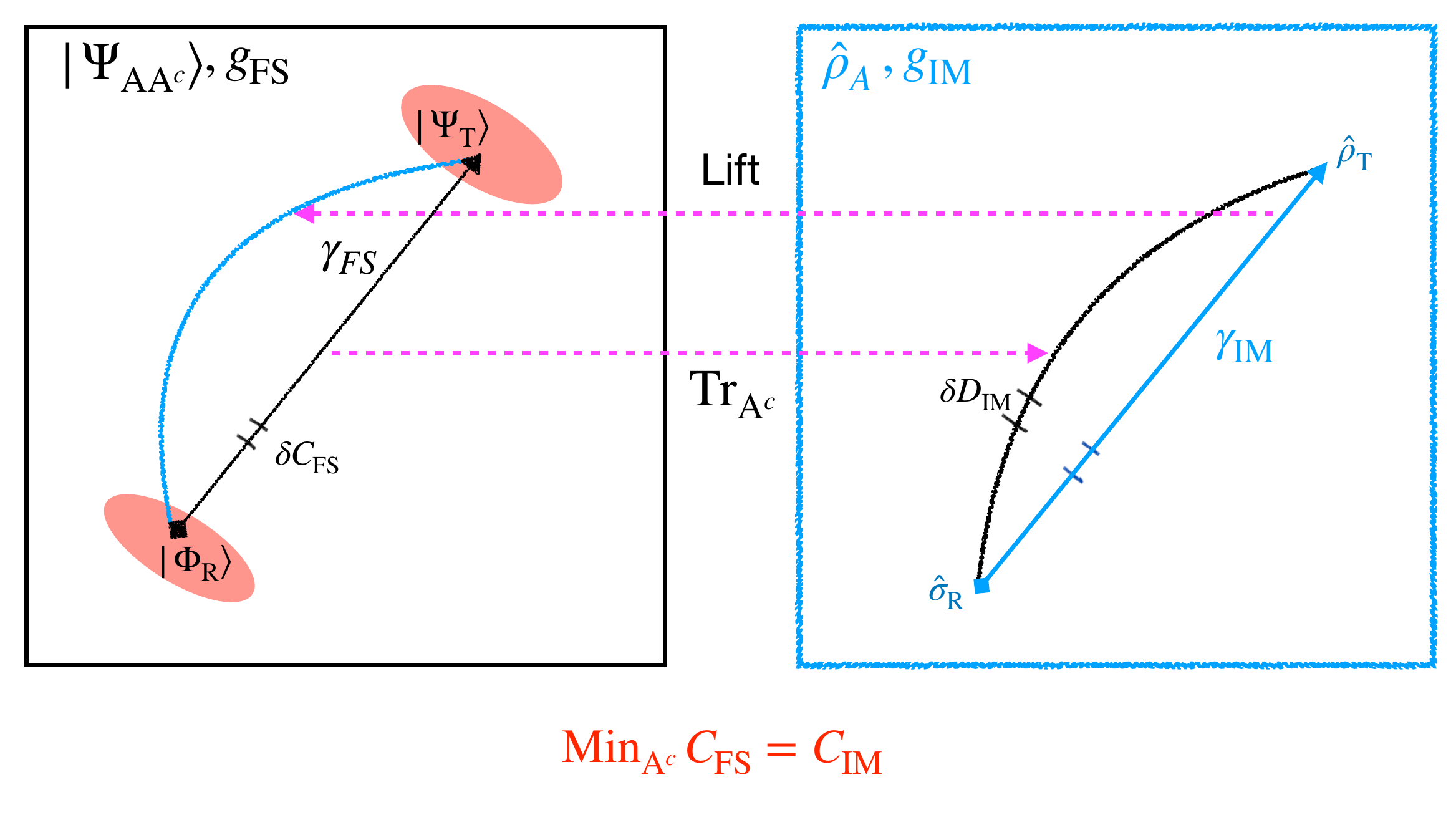}
	\caption{Left side is the Hilbert space $\mathcal{H}_{\mA\mA^c}$ of purified state $\ket{\Psi_{\mA\mA^c}}$, which is equipped with the Fubini-study metric $g_{\mt{FS}}$ as the complexity measure. The black line is referred to as the geodesic $\gamma_{\mt{FS}}$ in this space. The right side represents the Hilbert space $\mH_{\mA}$ for density matrices $\hat{\rho}_{\mA}$ with the quantum Fisher information metric $g_{\mt{IM}}$ defined in \eqref{def_IM} as the complexity measure. The corresponding geodesic $\gamma_{\mt{IM}}$ is indicated by the blue line. By tracing out the ancillary part $\bmA^c$, we can find the projection-map from $\gamma_{\mt{FS}}$ to a path in the space of $\hat{\rho}_{\mA}$ which is shown as the black curve. According to Uhlmann's theorem, we can also construct a lift-map from $\mathcal{H}_{\mA}$ to $\mathcal{H}_{\mA\mA^c}$ with the fidelity bound is always saturating.} \label{fig:map_inequality}
\end{figure}

   In the last subsection, we have seen that Uhlmann's theorem \eqref{Uhlmann_theorem} naturally relates the quantum fidelity between two mixed states to the fidelity from their "optimal" purifications.
   It may remind you of the idea about purification complexity \cite{purification,Agon:2018zso} by introducing ancillae in the quantum circuit and defining the minimal complexity of optimal purifications as the complexity for respective mixed states. Here, we would like to show that the complexity derived from the quantum Fisher information metric is actually the purification complexity where the pure-state complexity is measured by the Fubini-Study metric. Generally, we can take arbitrary mixed states $\hs_{\mA},\hr_{\mA}$ in system $\bmA$ as our reference state and target state, respectively. 
   First of all, let's think that we have found a specific purification  $\ket{\Psi_{\mA\mA^c}}$ by introducing an ancillary system $\bmA^c$. Considering the Fubini-Study metric as the complexity measure for pure states, we can search for the optimal purification and define the corresponding purification complexity as 
   \begin{equation}\label{purification_FS}
  \mP_{\mt{FS}}\( \hs_{\mA},\hat{\rho}_\mA \)=  \min\limits_{\Phi}  \min\limits_{\Psi} \, \mC_{\mt{FS}} \(  \ket{\Phi_{\mA\mA^c}}, \ket{\Psi_{\mA\mA^c}}\) = \min\limits_{\Phi}  \min\limits_{\Psi} \,  \int_0^1\!\! d\s\  \sqrt{ 2g_{\mu\nu}^{\mt{FS}}\, \dot{\lambda}^\mu\dot{\lambda}^\nu}\,, 
   \end{equation}
   where $g_{\mu\nu}^{\mt{FS}}$ is the Fubini-Study metric defined in \eqref{FSmetric} and the minimization is employed over all purifications for the target state $\hr_{\mA}$ and reference state $\hs_{\mA}$. We can assume the optimal purification from this point of view as $\ket{\tilde{\Psi}_{\mA\mA^c}}$. Let's just focus on an arbitrary infinitesimal step in the optimal quantum circuit, \ie the geodesic on the space of $\ket{\Psi_{\mA\mA^c}}$. From the definition of Fubini-Study metric \eqref{FSmetric}, the cost for this step is related to the quantum fidelity  between two extremely nearby pure states, \ie 
   \begin{equation}
   \begin{split}
   \delta \mC_{\mt{FS}} \(  \ket{ \Psi_{\mA\mA^c}(\s)}\)   &\equiv   \sqrt{2\(  1- |\langle \Psi_{\mA\mA^c}(\s)|\Psi_{\mA\mA^c}(\s+d\s)\rangle|^2 \) } \,,\\
    \delta \mP_{\mt{FS}}\( \hat{\rho}_\mA (\s)\) &= \sqrt{2\(  1- |\langle \tilde{\Psi}_{\mA\mA^c}(\s)|\tilde{\Psi}_{\mA\mA^c}(\s+d\s)\rangle|^2 \) } \,,\\
   \end{split}
   \end{equation}
   where the purification complexity $\mP_{\mt{FS}}$ from the Fubini-Study metric is associated with the optimal purification $\ket{\tilde{\Psi}_{\mA\mA^c}}$. on the other hand, we can also consider the same infinitesimal step and define the complexity of mixed states by considering the quantum Fisher information metric.  It is clear that Uhlmann's theorem ensures the inequality 
   \begin{equation}
   \begin{split}
   \delta \mathcal{D}_{\mt{IM}}\( \hat{\rho}_{\mA}(\s) \) &\equiv   \sqrt{ 2g^{\mt{IM}}_{\mu\nu}\, d{\lambda}^\mu\, d{\lambda}^\nu}= \sqrt{2 (1-F\( \hat{\rho}_{\mA}, \hat{\rho}_{\mA}(\s+d\s)) \)  }\,,\\
   & \le \delta \mP_{\mt{FS}}\( \hat{\rho}_\mA (\s)\)  \,,
   \end{split} 
   \end{equation}
   where we obtain the two near mixed states associated with pure states $\ket{\tilde{\Psi}_{\mA\mA^c}(\s+d\s)}$ and $\ket{\tilde{\Psi}_{\mA\mA^c}(\s)}$ by tracing out the ancillary system $\bmA^c$. Keeping doing this projection from optimal pure states $\ket{\tilde{\Psi}_{\mA\mA^c}(\s)}$ to the space of mixed states $\hr_{\mA}(\s)$ in subsystem $\mH_{\mA}$, we must be able to find a path in the space of mixed states with its length as the lowest bound of the purification complexity $\mP_{\mt{FS}}$ for arbitrary states $\hs_{\mA}, \hat{\rho}_{\mA}$. Recalling the fact that the complexity of mixed states from the quantum Fisher information metric is defined as the minimal geodesic length connecting a reference state and a target state, we finally arrive at the first conclusion for arbitrary mixed states, 
   	\begin{equation}\label{inequality_rho}
   	\mC_{\mt{IM}} \(\hs_{\mA}, \hat{\rho}_\mA \) \le \mathcal{D}_{\mt{IM}} \( \text{Projection of}\, \gamma_{\mt{FS}} \) \le  \mathcal{P}_{\mt{FS}}\(\hs_{\mA}, \hat{\rho}_\mA \) \equiv \min\limits_{\Phi}  \min\limits_{\Psi} \, \mathcal{C}_{\mt{FS}}\( \ket{\Psi_{\mA\mA^c}} \) \,,
   	\end{equation}
   which means that the purification complexity $\mathcal{P}_{\mt{FS}}$ is the upper bound of the complexity $\mC_{\mt{IM}}$ derived from quantum Fisher information metric. The above argument is illustrated by the projection from Hilbert space $\mathcal{H}_{\mA\mA^c}$ to $\mathcal{H}_{\mA}$ in the figure \ref{fig:map_inequality} . 
   It is stressed before that the quantum fidelity can always be saturated by choosing specific purifications. Then you may immediately face a puzzle: why we can find a lower value than the purification complexity $\mP_{\mt{FS}}$ even when  we have minimized the complexity from all possible purifications $\ket{\Psi_{\mA\mA^c}}$. Another natural question is that, by taking account of the inequality itself, when can we obtain the exact equality? All of these can be illustrated by stressing the difference between these two methods, which originates from the way of introducing the ancillae. The special point for the purification complexity is that we only introduce one specific optimal ancillary system $\bmA^c$ at the beginning and keep it in the full circuit. 
   For the circuit with complexity derived from the quantum Fisher information metric, it is possible that the quantum circuit may need different auxiliary systems after every step as shown in figure \ref{fig:circuit_fidelity}. From the viewpoint of optimal purification, this is because we have to introduce a special ancilla for every step to guarantee the fidelity between these purified states satisfying Uhlmann's fidelity, \ie \eqref{fidelity_def} which is a maximum for the purified states.

 However, Uhlmann's theorem also claims that the fidelity bound can be always saturated by taking some special purifications (see \eqref{optimal_vectors} for the explicit construction of these purified states in vector space). In other words, we can find a {\bf{continuous}} lift mapping from the geodesic in the space of quantum states $\hat{\rho}_\mA$ to a path in the Hilbert space $\mH_{\mA\mA^c}$ of purified states $\ket{\Psi_{\mA\mA^c}}$ \footnote{Especially, I would like to thank Juan Hernandez for illuminating discussions on that point.}. Due to the same infinitesimal complexity measure, it is obvious that the image after lift-map has the same distance as $\mC_{\mt{IM}}\(\hr_\mA\)$. 
  It is shown in the map from the left blue curve to the right blue curve in figure \ref{fig:map_inequality}\footnote{As discussed in figure \ref{fig:circuit_fidelity}, there are many ways to introduce the ancillae and simultaneously make Uhlmann's fidelity saturated (the lift-map is not injective). However, most of them after the lift-map only make discontinuous lines in $\mathcal{H}_{\mA\mA^c}$ but with the same length.}.  Again, we should notice that the purification complexity from the Fubini-Study metric is also defined as the minimal geodesic length. Comparing the geodesic distance with respect to the Fubini-Study metric and the distance of the image after the lift-map, we can also obtain another inequality  
  \begin{equation}
  \mC_{\mt{IM}} \(\hs_{\mA}, \hat{\rho}_\mA \)  =\mathcal{D}_{\mt{FS}} \( \text{Lift of}\, \gamma_{\mt{IM}} \) \ge   \mathcal{P}_{\mt{FS}}\(\hs_{\mA}, \hat{\rho}_\mA \)\,.
  \end{equation}
   Combining this new inequality with the first inequality from Uhlmann's theorem, we finally conclude that the complexity $\mC_{\mt{IM}}\( \hs_{\mA}, \hr_{\mA}\)$ derived from the quantum Fisher information metric is exactly the purification complexity measured by the Fubini-Study metric on purified states, \ie 
 	\begin{tcolorbox}
 	\begin{equation}\label{equality_rhoA}
 \mC_{\mt{IM}} \(\hs_{\mA},\hat{\rho}_\mA \)  =   \mathcal{P}_{\mt{FS}}\( \hs_{\mA},\hat{\rho}_\mA \) \equiv \min\limits_{\Phi}  \min\limits_{\Psi} \, \mathcal{C}_{\mt{FS}}\( \ket{\Phi_{\mA\mA^c}} ,\ket{\Psi_{\mA\mA^c}} \) \,,
 	\end{equation}
 	\end{tcolorbox}
where the target state and reference state are related to purified states in the extended system $\mH_{\mA\mA^c}$ by $\hr_{\mA}= \tr_{\mA^c}\( \ket{\Psi_{\mA\mA^c}} \bra{\Psi_{\mA\mA^c}}\)$ and $\hs_{\mA}= \tr_{\mA^c}\( \ket{\Phi_{\mA\mA^c}} \bra{\Phi_{\mA\mA^c}}\)$, respectively.
In the next section, we will take Gaussian mixed states as an explicit example to show that how the first equality holds after minimization and find the special purifications satisfying the bound from Uhlmann's fidelity along the whole geodesic $\hat{\rho}_{\mA}(\s)$. 

\subsubsection{The Non-increase of Purification Complexity}\label{sec:nonincreasing}
Instead of considering the optimal purified state $\ket{\Psi_{\mA\mA^c}}$ in the extended Hilbert space, we can start from generic mixed states $\hr_{\mA\mB}$ with $\tr_{\mB}\( \hr_{\mA\mB} \)= \hr_{\mA}$ in a bipartite Hilbert space $\mH_{\mA}\otimes\mH_{\mB}$ and assume the complexity from $\hs_{\mA\mB}$ to $\hr_{\mA\mB}$ is associated with the geodesic $\gamma\({\hr_{\mA\mB}}(\s)\)$. In order to show the consequence of partial trace on the complexity $\mC_{\mt{IM}}\(\hat{\sigma}_{\mA\mB}, \hat{\rho}_{\mA\mB}\)$, we can similarly trace out the system $\bmB$ along the geodesic  $\gamma\({\hr_{\mA\mB}}(\s)\)$, mapping the geodesic connecting $\sigma_{\mA\mB}$ and $\hr_{\mA\mB}$ to a special path in $\mH_\mA$. The non-decrease of fidelity under partial trace \eqref{fidelity_nondecreasing} gives rise to the monotone for circuit complexity by 
\begin{equation}
\mC_{\mt{IM}}\(\hat{\sigma}_{\mA\mB}, \hat{\rho}_{\mA\mB}\)  \ge \mathcal{D}_{\mt{IM}} \( \text{Projection of} \,\gamma\({\hr_{\mA\mB}}(\s)\) \) \ge   \mathcal{C}_{\mt{IM}}\( \hat{\sigma}_\mA,\hat{\rho}_\mA \)\,.
\end{equation}
where $\mathcal{D}_{\mt{IM}} $ denotes the length measured by the QFIM and we have used the fact that the projection of geodesic $\gamma\({\hr_{\mA\mB}}(\s)\)$ may not be a geodesic on the space of $\hr_{\mA}$ to obtain the second inequality. This non-increasing property of complexity $\mC_{\mt{IM}}$ looks obviously accord with our intuition because it is reasonable to expect the complexity for reduced states in a smaller Hilbert space is smaller. Furthermore, we can also consider the map of geodesic $\gamma \( \hr_{\mA}(\s)\)$ under any arbitrary quantum operation $\mathcal{E}$. From the non-decrease of quantum fidelity $F\( \hs_{\mA}, \hr_{\mA}\)$ under $\mathcal{E}$, one can also arrive at the most general non-increasing property of purification complexity $\mC_{\mt{IM}}$ by 
\begin{equation}
\mathcal{C}_{\mt{IM}}\( \hat{\sigma}_\mA,\hat{\rho}_\mA \) \ge \mathcal{D}_{\mt{IM}} \( \mathcal{E} \(({\hr_{\mA}}(\s)\) \) \ge \mathcal{C}_{\mt{IM}}\( \mathcal{E} \(\hat{\sigma}_\mA\),\mathcal{E} \(\hat{\rho}_\mA \)\)\,.
\end{equation}  
As a straightforward application, we can find the reversible quantum operation does not change the complexity since
\begin{equation}
\mC_{\mt{IM}}\(\hat{\sigma}_{\mA}, \hat{\rho}_{\mA}\) 
\ge \mC_{\mt{IM}}\(\mathcal{E}\( \hat{\sigma}_{\mA}\), \mathcal{E}\( \hat{\rho}_{\mA}\)\) \ge \mC_{\mt{IM}}\(\mathcal{E}^{-1}\comp\mathcal{E}\( \hat{\sigma}_{\mA}\), \mathcal{E}^{-1}\comp\mathcal{E}\( \hat{\rho}_{\mA}\)\)=\mathcal{C}_{\mt{IM}}\( \hat{\sigma}_\mA,\hat{\rho}_\mA \)\,.
\end{equation}
For example, any unitary operator is a reversible quantum operation and then we naturally have the unitary invariance of purification complexity
\begin{equation}
\mC_{\mt{IM}}\(\hat{\sigma}_{\mA}, \hat{\rho}_{\mA}\) 
= \mC_{\mt{IM}}\(U \hat{\sigma}_{\mA} U^\dagger, U \hat{\rho}_{\mA}U^\dagger\)\,,
\end{equation}
which can also be derived from the unitary invariance of the quantum fidelity and the fact that geodesic associated with complexity is chosen to be the one minimizing the distance.

Similar to the non-decreasing property of Uhlmann's fidelity in \eqref{fidelity_nondecreasing}, we summarize our observation as an universal conclusion that the purification complexity $\mC_{\mt{IM}}\(\hat{\sigma}_{\mA}, \hat{\rho}_{\mA}\) $ (for both pure states and mixed states) derived from the quantum Fisher information metric is {\bf no-increasing} under any trace-preserving quantum operations (quantum channel) acting on the reference state and target state simultaneously, \ie 

\begin{tcolorbox}
\begin{equation}\label{complexity_nonincrease}
\mC_{\mt{IM}}\(\hat{\sigma}_{\mA\mB}, \hat{\rho}_{\mA\mB}\)  \ge \mC_{\mt{IM}}\(\hat{\sigma}_{\mA}, \hat{\rho}_{\mA}\) = \mC_{\mt{IM}}\(U \hat{\sigma}_{\mA} U^\dagger, U \hat{\rho}_{\mA}U^\dagger\)
\ge \mC_{\mt{IM}}\(\mathcal{E}\( \hat{\sigma}_{\mA}\), \mathcal{E}\( \hat{\rho}_{\mA}\)\)\,.
\end{equation} 
\end{tcolorbox}  
Naively, a similar conclusion also holds for $\mC_{\mt{IM}}^{\kappa=2}$. Analogous to the triangle inequality of Von Neumann entropy or entanglement entropy
 \begin{equation}
 \big| S_{\mathrm{vN}}\(\hr_{\mA}\) -S_{\mathrm{vN}}\(\hr_{\mB}\)  \big| \le  S_{\mathrm{vN}}\(\hr_{\mA\mB}\) \le S_{\mathrm{vN}}\(\hr_{\mA}\) +S_{\mathrm{vN}}\(\hr_{\mB}\) \,, 
 \end{equation}
 the monotonicity of the purification complexity simply implies 
 \begin{equation}
 \big|  \mC_{\mt{IM}}\(\hat{\sigma}_{\mA}, \hat{\rho}_{\mA}\) -\mC_{\mt{IM}}\(\hat{\sigma}_{\mB}, \hat{\rho}_{\mB}\) \big|\le \mC_{\mt{IM}}\(\hat{\sigma}_{\mA\mB}, \hat{\rho}_{\mA\mB}\) \,.
 \end{equation}
 However, the subadditivity for purification complexity does not hold in general and will be discussed in the next subsection in details. 
 
Instead of applying quantum operations on both reference states and target states, we can also discuss the effect of quantum operations only on the target states or the reference states. In the space of quantum states in any Hilbert space $\mH_\mA$, there is an extremely simple state called the {\it maximally mixed state} defined by 
\begin{equation}
 \hs_{0,\mA}= \sum_{i}^{N_\mA}\frac{1}{N_{\mA}} \ket{\psi_i} \bra{\psi_i} = \frac{\mathbb{I}}{N_{\mA}} \,,
\end{equation}
where $N_{\mA}$ denotes the dimension of the Hilbert space $\mH_{\mA}$. It is easy to see the maximally mixed state has a fully degenerate spectrum ( \ie Schmidt coefficients take the same value) and its entropy 
\begin{equation}
S_{\mathrm{vN}}\( \hs_{0,\mA}\)= \tr_{\mA}\( \hs_0 \log \hs_0 \) =\log N_{\mA} \,,
\end{equation} 
 reaches the maximum entropy in a $N_{\mA}$-dimensional Hilbert space. As a result, this is a completely random state with zero information. For example, we can approach the maximally mixed state by taking the inverse temperature of a thermal state $\hat{\upsilon}_{\mathrm{th}}$ to zero, \ie 
 \begin{equation}
\lim_{\beta \to 0}\hat{\upsilon}_{\mathrm{th}}(\beta, \omega)
=  \lim_{\beta \to 0} \frac{1}{Z\(\beta \omega\)} \sum_{n=0}^{N_{\mA}} e^{-\beta \omega\,n} \ket{n}\!\bra{n} = \hs_{0,\mA}\,.
 \end{equation}
Taking the reference state (or target state) in a system $\bmA$ as the maximally mixed state and considering the unital quantum channels \footnote{Not all quantum operations are unital. A quantum operation is unital if it preserves the identity operator. In the operator-sum representation, the unital quantum channels satisfy $\sum_{k} \hat{M}_k \hat{M}^\dagger_k = \mathbb{I}$.}, the monotonicity of purification complexity reduces to 
\begin{equation}
\mC_{\mt{IM}}\( \hs_{0,\mA},  \hat{\rho}_{\mA}\) =\mC_{\mt{IM}}\( \hs_{0,\mA},  U\hat{\rho}_{\mA} U^\dagger\)\ge \mC_{\mt{IM}}\( \hs_{0,\mA}, \mathcal{E}\( \hat{\rho}_{\mA}\)\) \,,
\end{equation}
due to the invariance of maximally mixed states over any unital quantum channels.

From the above discussion, we have shown that the properties of the quantum fidelity are helpful to derive related properties for the purification complexity. Instead of using the non-decrease of fidelity, we can also adopt other properties of fidelity. As a result of the strong concavity of fidelity
\begin{equation}
\sum_i p_i F\(\hs_i,\hr_i \)\le F\(\sum_{i} p_i\hs_i, \sum_i p_i\hr_i \) \,, \qquad \text{with} \qquad \sum_i p_i= 1\,,
\end{equation}
we can find that the infinitesimal distance measures satisfy
\begin{equation}
\sum_i p_i \, ds^2_{\mt{IM}} \( \hr_{i}\( \s\),  \hr_{i}\( \s+d \s\) \)  \ge ds^2_{\mt{IM}} \( \sum_i p_i\hr_{i}\( \s\),  \sum_i p_i \hr_{i}\( \s+d \s\) \)  \,.
\end{equation}
Considering a sequence of reference states $\hs_i$ and targets states $\hr_i$ such that 
\begin{equation}
\hs_{\mt{R}}= \sum_i p_i \hs_i \,, \qquad \hr_{\mt{T}}= \sum_i p_i \hr_i \,, 
\end{equation}
and taking the integral along their respective geodesic $\gamma_{i}$ which connects $\hs_i$ to $\hr_{i}$, we can easily obtain another non-increasing behavior of the purification complexity
\begin{equation}
\sum_i p_i \mC^{\kappa=2}_{\mt{IM}}\( \hs_i, \hr_i \)  \ge  \mC^{\kappa=2}_{\mt{IM}}\( \hs_{\mt{R}}, \hr_{\mt{T}} \) \,.
\end{equation}
It means that the complexity $\mC^{\kappa=2}_{\mt{IM}}$ is jointly convex in reference states and target states. 
Taking the square root of the above inequality, we can find that the purification complexity $\mC_{\mt{IM}}$ satisfies 
	\begin{equation}
	\sum_i \sqrt{p_i} \mC_{\mt{IM}}\( \hs_i, \hr_i \)  \ge  \mC_{\mt{IM}}\( \hs_{\mt{R}}, \hr_{\mt{T}} \) \,,
	\end{equation}
which is reduced to 
\begin{equation}
\sum_i \sqrt{p_i} \mC_{\mt{IM}}\( \hs_{\mt{R}}, \hr_i \)  \ge  \mC_{\mt{IM}}\( \hs_{\mt{R}}, \hr_{\mt{T}} \) \,,\quad \text{with} \qquad  \hr_{\mt{T}}= \sum_i p_i \hr_i\,,
\end{equation}
after taking $\hs_{i}=\hs_{\mt{R}}$. 

Finally, we would like to remark that these properties of the purification complexity $\mC_{\mt{IM}}\( \hs, \hr\)$ with respective two quantum states are also shared by the quantum relative entropy $S\(\hr||\hs \)\equiv \tr\( \hr \(\log \hr - \log \hs\) \)$ in spirit. However, different from the complexity, quantum relative entropy is not symmetric and can not be considered as a distance measure between two quantum states. 

   \subsection{More on Purification Complexity of Mixed States}
   \subsubsection{Pure-State limit}\label{sec:purelimit}
    First of all, let's point out  the differences between pure-state complexity from the Fubini-Study metric and that from the quantum Fisher information metric.  Applying our previous conclusion on purification complexity to pure states, we can find 
   \begin{equation}
   \mC_{\mt{IM}}\( \ket{\Phi_{\mA}},\ket{\Psi_{\mA}} \) = \min\limits_{\Phi}\min\limits_{\Psi}\, \mathcal{C}_{\mt{FS}}\(\ket{\Phi_{\mA\mA^c}}, \ket{\Psi_{\mA\mA^c}} \) \le  \mathcal{C}_{\mt{FS}}\( \ket{\Phi_{\mA}} ,\ket{\Psi_{\mA}} \)\,.
   \end{equation}
   The above inequality implies that the pure-state limit of purification complexity may be different from the pure-state complexity derived from the Fubini-Study metric. You may feel surprised that why the pure-state limit of purification complexity is not reduced to the Fubini-Study complexity, in view of the fact that the quantum Fisher information metric for pure states is exactly equivalent to the Fubini-Study metric. However, it should fulfill the expectation because the geodesic in a higher dimensional manifold is not smaller than that on a reduced hypersurface.
   Furthermore, we can find two (equivalent) physical explanations for that discrepancy. Firstly, it is due to the fact that the geodesic for $\mathcal{C}_{\mt{FS}}\( \ket{\Phi_{\mA}} ,\ket{\Psi_{\mA}} \)$ is constrained on the space with only pure states in $\mH_{\mA}$. However, with the help of the ancillae, we are allowed to use all pure states in the extended Hilbert space $\mH_{\mA}\otimes \mH_{\mA^c}$. Since any product state $\ket{\Psi_{\mA}} \otimes \ket{\Psi_{\mA^c}}$ is also a "purification" of $\ket{\Psi_{\mA}}$, the two complexities for pure states are consistent if and only if the optimal purifications of 
   $\ket{\Phi_{\mA}},\ket{\Psi_{\mA}}$ are themselves. In other words, this equivalence only appears when the ancillae and entangled gates between $\mH_{\mA}$ and $\mH_{\mA^c}$ for pure states are useless. Without introducing the auxiliary system and purifications of pure states, we can focus on the subsystem $\mH_{\mA}$ and interpret the smaller complexity from $  \mC_{\mt{IM}}\( \ket{\Phi_{\mA}},\ket{\Psi_{\mA}} \) $ as the fact that we are allowed to evolve the pure reference state to pure target state by some mixed states. Again, we can find that the two complexities will be the same if the geodesic only goes along pure states, which equivalently means that we do not need entangled gates between the physical system and ancilla system from the view in the extended Hilbert space. In later examples, we will find that the Gaussian states happen to be that simple case because the circuit complexity of the factorized reference state and target state is just a direct sum of the complexity from every single mode as shown in \cite{qft1}.
   
   \subsubsection{Simplify the minimization}
  Although the pure-state limit of purification complexity does not always agree with the Fubini-Study complexity for pure states, we can further use this upper bound to simplify the process of minimization. Based on the monotonicity of purification complexity with respect to the partial trace, we can find the increasing sequence
  \begin{equation}
   \mC_{\mt{IM}} \(\hs_{\mA},\hat{\rho}_\mA \) \le \mathcal{C}_{\mt{IM}}\( \ket{\Phi_{\mA\mA^c}} ,\ket{\Psi_{\mA\mA^c}} \) \le \mathcal{C}_{\mt{FS}}\( \ket{\Phi_{\mA\mA^c}} ,\ket{\Psi_{\mA\mA^c}} \)\,.
  \end{equation}
   in discarding of the subsystem $\bmA^c$.
  Because we have shown that the minimization of $\mathcal{C}_{\mt{FS}}$ over all purifications exactly agrees with the purification complexity $ \mC_{\mt{IM}} \(\hs_{\mA},\hat{\rho}_\mA \)$, the minimum of $\mathcal{C}_{\mt{IM}}\( \ket{\Phi_{\mA\mA^c}} ,\ket{\Psi_{\mA\mA^c}} \)$ has to locate at the same value, \ie
  \begin{tcolorbox}
  	\begin{equation}\label{equality_rhoA2}
  	\begin{split}
  	\mC_{\mt{IM}} \(\hs_{\mA},\hat{\rho}_\mA \)  &=  \min\limits_{\Phi}  \min\limits_{\Psi} \, \mathcal{C}_{\mt{FS}}\( \ket{\Phi_{\mA\mA^c}} ,\ket{\Psi_{\mA\mA^c}} \) = \min\limits_{\Phi}  \min\limits_{\Psi} \, \mathcal{C}_{\mt{IM}}\( \ket{\Phi_{\mA\mA^c}} ,\ket{\Psi_{\mA\mA^c}} \)\\
  	&= \min\limits_{\Phi}  \, \mathcal{C}_{\mt{IM}}\( \ket{\Phi_{\mA\mA^c}} ,\ket{\Psi_{\mA\mA^c}} \)= \min\limits_{\Psi} \, \mathcal{C}_{\mt{IM}}\( \ket{\Phi_{\mA\mA^c}} ,\ket{\Psi_{\mA\mA^c}} \)\,.\\
  	\end{split}
  	\end{equation}
  	\end{tcolorbox}
  In order to simplify the double minimizations to one as shown in the above inequalities, we just note that the unitary invariance of purification complexity implies that we can relate the optimal purified states under the double minimizations to that with only one minimization. Taking the unitary operation $U_{\mA\mA^c}$ such that $U_{\mA\mA^c}\ket{\tilde{\Phi}_{\mA\mA^c}}=\ket{\Phi_{\mA\mA^c}}$, one can simplify the double minimizations by 
  \begin{equation}
  \begin{split}
    &\quad\min\limits_{\Phi}   \min\limits_{\Psi} \, \mathcal{C}_{\mt{IM}}\( \ket{\Phi_{\mA\mA^c}} ,\ket{\Psi_{\mA\mA^c}} \) 
    = \mathcal{C}_{\mt{IM}}\( \ket{\tilde{\Phi}_{\mA\mA^c}} ,\ket{\tilde{\Psi}_{\mA\mA^c}} \) \,,\\
    &=\mathcal{C}_{\mt{IM}}\( \ket{\Phi_{\mA\mA^c}} ,U_{\mA\mA^c}\ket{\tilde{\Psi}_{\mA\mA^c}} \) \,,\\
    &= \min\limits_{\Phi}  \, \mathcal{C}_{\mt{IM}}\( \ket{\Phi_{\mA\mA^c}} ,\ket{\Psi_{\mA\mA^c}} \)\,,  
  \end{split}
  \end{equation}
  which also works for one minimization over all purified target states $\ket{\Psi_{\mA\mA^c}}$. 
  
   \subsubsection{Mutual complexity of $\hr_{\mA\mB}$}
   Starting from the target state represented by a density operator $\hr_{\mA\mB}$ in the bipartite physical systems $\mH_{\bmA\bmB}$, we can also define the mixed-state complexity for two reduced density matrices 
   \begin{equation}
   \hat{\rho}_{\mA} = \tr_{\mB} \( \hr_{\mA\mB}\)\,, \quad  \hat{\rho}_{\mB} = \tr_{\mA} \(\hr_{\mA\mB} \)\,,
   \end{equation}
   in the subsystems $\bmA, \bmB$. 
   From the non-increase of purification complexity under the partial trace in \eqref{complexity_nonincrease}, it is direct to derive the inequality $\mC_{\mt{IM}} \(\hat{\rho}_\mA \)   \le \mC_{\mt{IM}} \( \hr_{\mA\mB}\)$ and also 
   \begin{equation}
    \mC_{\mt{IM}} \(\hat{\sigma}_\mA,\hat{\rho}_\mA \)  +  \mC_{\mt{IM}} \(\hat{\sigma}_\mB,\hat{\rho}_\mB \)   -2\mC_{\mt{IM}} \( \hs_{\mA\mB},\hr_{\mA\mB}\)  \le 0\,,
   \end{equation}
  where the reference states are given by $\hs_{\mA\mB}$ in the system $\bmA \bmB$, $\hs_{\mA}= \tr_{\mB}\( \hs_{\mA\mB} \)$ in a subsystem $\bmA$, and $\hs_{\mB}= \tr_{\mA}\( \hs_{\mA\mB}\)$ in a subsystem $\bmB$, respectively.
   On the other hand, we are interested in the non-trivial concept called mutual complexity \cite{Agon:2018zso,purification}. As a generalization of the mutual complexity for pure states $\ket{\Psi_{\mA\mB}}$, the authors of \cite{purification} propose to extend the mutual complexity to more generic quantum states with bipartition as
    \begin{equation}\label{IM_mutual_complexity}
 \Delta \mC_{\mt{IM}} =  \mC_{\mt{IM}} \(\hat{\sigma}_\mA,\hat{\rho}_\mA \)  +  \mC_{\mt{IM}} \(\hat{\sigma}_\mB,\hat{\rho}_\mB \)   -\mC_{\mt{IM}} \(\hs_{\mA\mB}, \hr_{\mA\mB}\)  \,,
 \end{equation}
  which quantifies the additional correlations between the subsystem $\bmA$ and $\bmB$. 
   Taking the complexity of states as that derived from the quantum Fisher information metric makes the above definition calculable. When $\Delta \mC> 0$ complexity is said to be subadditive, otherwise it is called to be superadditive when $\Delta \mC <0$.  As discussed before, we have another definition for pure-state complexity based on the Fubini-Study metric. Correspondingly, we can also define the mutual complexity for pure states $\ket{\Psi_{\mA\mB}}$ by 
   \begin{equation}
   \begin{split}
        \Delta \mC_{\mt{FS}} &=  \mP_{\mt{FS}} \(\hs_{\mA},\hat{\rho}_\mA \)  +  \mP_{\mt{FS}} \(\hs_{\mA},\hat{\rho}_\mB \)   -\mC_{\mt{FS}} \(\ket{\Phi_{\mA\mB}}, \ket{\Psi_{\mA\mB}}\)  \,,\\
        &= \mC_{\mt{IM}} \(\hs_{\mA},\hat{\rho}_\mA \)  +  \mC_{\mt{IM}} \(\hs_{\mA},\hat{\rho}_\mB \)   -\mC_{\mt{FS}} \(\ket{\Phi_{\mA\mB}}, \ket{\Psi_{\mA\mB}}\)  \,,\\
        &\le \Delta \mC_{\mt{IM}}  \,,
   \end{split}
   \end{equation}
   where we have used the fact $\mC_{\mt{IM}} \(\hs_{\mA}, \hat{\rho}_\mA \) =\mP_{\mt{FS}} \((\hs_{\mA}, \hat{\rho}_\mA \) $ for purification complexity of mixed states and also $\mC_{\mt{FS}} \( \Psi_{\mA\mB}\) \ge \mC_{\mt{IM}} \( \Psi_{\mA\mB}\) $ for pure states.
   
   Although we found that the mutual complexity \eqref{IM_mutual_complexity} is either always superadditive or always subadditive in general,  it is easy to get the subadditive mutual complexity in many simple cases due to the monotonicity of purification complexity $\mC_{\mt{IM}}$.
   If a quantum operation with $\mathcal{E}\( \hs_{\mA}\otimes \hs_{\mB}\)=\hs_{\mA\mB}, \mathcal{E}\( \hr_{\mA}\otimes \hr_{\mB}\)=\hr_{\mA\mB}$ exists, then one can easily confirm 
   \begin{equation}
      \Delta \mC_{\mt{IM}}\ge \mC_{\mt{IM}} \(\hs_{\mA}\otimes \hs_{\mB},  \hr_{\mA}\otimes \hr_{\mB}\) -\mC_{\mt{IM}} \(\hs_{\mA\mB}, \hr_{\mA\mB}\) \ge 0\,,
   \end{equation}
   by using the simple fact $\mC_{\mt{IM}} \(\hs_{\mA}\otimes \hs_{\mB},  \hr_{\mA}\otimes \hr_{\mB}\) \le \mC_{\mt{IM}} \(\hs_{\mA}, \hr_{\mA}\)  +\mC_{\mt{IM}} \(\hs_{\mB}, \hr_{\mB}\) $, and also applying the non-increase of purification complexity to derive the second inequality. For example, if the reference state and target state are both factorized (separable states) as $ \hs_{\mA\mB}=\hs_{\mA}\otimes \hs_{\mB}, \hr_{\mA\mB}= \hr_{\mA}\otimes \hr_{\mB}$, we have $\Delta \mC_{\mt{IM}} \ge 0$. 
   	
   It is intriguing to expect that the mutual complexity  $\Delta\mC_{\mt{IM}}\(\hr_{\mA\mB}\)$ is always subadditive. However, we can easily find a counterexample by relating the reference state to the target state in the way of $\sigma_{\mA\mB}=\hr_{\mA}\otimes \hr_{\mB}$. Then it is obvious that  
   	  \begin{equation}\label{counterexample}
   	\Delta \mC \( \hr_{\mA}\otimes \hr_{\mB}, \hr_{\mA\mB}\)  =\mC \(\hr_\mA,\hat{\rho}_\mA \)  +  \mC \(\hr_\mB,\hat{\rho}_\mB \)   -\mC\(\hr_{\mA}\otimes \hr_{\mB}, \hr_{\mA\mB}\) =-\mC\(\hr_{\mA}\otimes \hr_{\mB}, \hr_{\mA\mB}\)\le 0 \,,
   	\end{equation}
   because of the non-negativity of complexity. Finally, we should point out that the above example with $\Delta \mC \le 0$ exists for any potential definitions of complexity between two density operators. 
   
   \subsubsection{First Law of complexity for mixed states}
   Since we also define the complexity of mixed states as the geodesic distance, the idea about the first law of complexity \cite{firstlaw} also directly applies to the mixed states because they can be both considered as a similar classical mechanics problem. Therefore, perturbing the target state $\hat{\rho}_{\mA}$ by a small variation $\hat{\rho}_\mA+\delta \hat{\rho}_{\mA} \equiv \hat{\rho}_\mA(\lambda +\delta\lambda)$ with a fixed reference state, one can easily find that the variation of complexity for mixed states also satisfies {\it the first law of complexity} \cite{firstlaw,Bernamonti:2020bcf}
   \begin{equation}
   \begin{split}
   \delta \mC_{\mt{IM}}&= \mC_{\mt{IM}} \( \hat{\rho}_\mA+\delta \hat{\rho}_{\mA}\) -\mC_{\mt{IM}} \( \hat{\rho}_\mA\) \,,\\
   &=\(P_\mu \delta \lambda^\mu + \frac{1}{2} \delta P_\mu \delta \lambda^\mu+\cdots \) \bigg|_{s=1} \,,
   \end{split}  
   \end{equation}
   where the "momentum" $P_\mu$ is defined as
   \begin{equation}
   \begin{split}
   P_\mu&= \frac{\partial F_{\mt{IM}}}{\partial \dot{\lambda}^\mu}= \frac{2g^{\mt{IM}}_{\mu\nu}\, \dot{\lambda}^\nu}{ F_{\mt{IM}}}\,, \\
   \delta P_{\mu}&=\delta \lambda^{\nu} \frac{\partial^{2} F_{\mt{IM}}}{\partial \lambda^{\nu} \partial \dot{\lambda}^{\mu}}+\delta \dot{\lambda}^{\nu} \frac{\partial^{2} F_{\mt{IM}}}{\partial \dot{\lambda}^{\nu} \partial \dot{\lambda}^{\mu}} \,,\\
  \end{split}
   \end{equation}
   with respect to the complexity measure (cost function), \ie the proper distance with the QFIM, 
   $F_{\mt{IM}}= \sqrt{ 2g^{\mt{IM}}_{\mu\nu}\, \dot{\lambda}^\mu\, \dot{\lambda}^\nu}$.
   However, except for the similarity in form, we also want to point out an obvious difference between the first law of complexity for pure states and that for mixed states. The former only works for the perturbation from unitary transformations. As described in \cite{firstlaw,Bernamonti:2020bcf},  the variation of complexity is traced back to the change on unitary operator 
   \begin{equation}
    \ket{\Psi_\mt{T}} \longrightarrow \ket{\Psi_\mt{T}+\delta \Psi}\,,\quad \text{with} \quad  U_{\mt{TR}} \longrightarrow  U_{\mt{T}'\mt{R}}' =  U_{\mt{TR}} +\delta U \,.
   	\end{equation}
   	However, for the first law of complexity for mixed states, we can also interpret the change on target states as either unitary or non-unitary transformations. In a short sentence, the general quantum operation triggers a generic small variation of the mixed state by
   	\begin{equation}
   	\hat{\rho}_{\mA}   \longrightarrow  \hat{\rho}_{\mA} ' = 	\hat{\rho}_{\mA}  +\delta  	\hat{\rho}_{\mA}  = \sum_{i} \hat{M}_i \hat{\rho}_{\mA} \hat{M}_i^\dagger\,. 
   	\end{equation}
  
 %%%%%%%%%%%%%%%%%%%%%%%%%% 
 %%%%%%%%%%%%%%%%%%%%%%%%%%
  %%%%%%%%%%%%%%%%%%%%%%%%%% 
 %%%%%%%%%%%%%%%%%%%%%%%%%%
  %%%%%%%%%%%%%%%%%%%%%%%%%% 
 %%%%%%%%%%%%%%%%%%%%%%%%%%
\section{Application: Gaussian Mixed States}\label{sec:application}
 In the last section, we have shown that the complexity from the quantum Fisher information metric is the purification complexity in \eqref{equality_rhoA} and also \eqref{equality_rhoA2} with a simpler minimization. In this section, we would like to use the Gaussian mixed states as an explicit example to illustrate that the equivalence holds after the minimization on the Fubini-Study complexity over all purified states.
 \subsection{Geodesic and Complexity}
As the first application of the complexity  $\mC_{\mt{IM}}$ from the quantum Fisher information metric, we start from the one-mode Gaussian state (see \cite{purification} for more discussion about that simple mixed state)
\begin{equation}\label{one_mixed}
\hat{\rho}_1 = \hat S_1(r) \hat{\upsilon}_{\mathrm{th}} (\beta,\omega)\hat S_1^\dagger(r) \,.
\end{equation}
where the one-mode squeezing operator with a real parameter $r$ is defined as 
\begin{equation}\label{onemode_squeezed}
\begin{split}
\hat S_1(r) \equiv e^{-\frac{r}{2}\left({a_1^\dagger}{}^2-a_1^2\right)}
= e^{i \frac{r}2 \left(\hat{x}_1\hat{p}_1 + \hat{p}_1 \hat{x}_1 \right)}\,,
\end{split}
\end{equation}
and $\hat{\upsilon}_{\mathrm{th}}$ denotes the thermal state with inverse temperature $\beta$, \ie 
\begin{equation}\label{density_thermal}
\hat{\upsilon}_{\mathrm{th}}(\beta, \omega)
\equiv \frac{e^{-\beta \omega\, a^\dagger a}}{\tr (e^{-\beta \omega\, a^\dagger a})}=  \(1- e^{-\beta\omega}\) \sum_{n=0}^\infty e^{-\beta \omega\,n} \ket{n}\!\bra{n}\,.
\end{equation}
 For mixed Gaussian states, it is convenient to introduce new parameters 
\begin{equation}
\alpha= \frac{1}{2}\ln \frac{1 +e^{-\beta \omega/2}}{1-e^{-\beta \omega/2}}\,, \quad \bar{r}= r-\frac{1}{2}\ln \frac{\mu}{\omega} \,,
\end{equation}
where the $\mu$ is the characteristic frequency of the reference state $\ket{\psi_{\mt{R}}} \equiv \ket{0(\mu)}$ which is chosen to be a Gaussian pure state.
First of all, we need the quantum fidelity for squeezed thermal states \cite{fidelity} 
\begin{equation}\label{fidelity_Gaussian}
F(\hat{\rho}(\zeta_1, \beta_1), \hat{\rho}(\zeta_2, \beta_2)) =  \sqrt{\frac{2 \sinh \frac{\beta_1
			\omega}{2}\sinh \frac{\beta_2 \omega}{2}}{\sqrt{Y}-1}}\,,
\end{equation}
with complex squeezing parameters $\zeta_i=r_ie^{i\theta_i}$ and 
\begin{equation}
\begin{split}
Y =& \cos^2(\frac{\theta_1-\theta_2}{2}) \( \cosh^2 (r_1-r_2) \cosh^2 \(\frac{\beta_1+\beta_2}{2} \omega\)-\sinh^2 (r_1-r_2) \cosh^2 \( \omega \frac{\beta_1-\beta_2 }{2}\) \) +\\
& \sin^2(\frac{\theta_1-\theta_2}{2}) \( \cosh^2 (r_1 +r_2) \cosh^2\(\frac{\beta_1+\beta_2}{2} \omega\) -\sinh^2 (r_1+r_2)\cosh^2 \( \omega \frac{\beta_1-\beta_2}{2} \) \). 
\end{split}
\end{equation}
We can obtain the quantum Fisher information metric (Bures metric) by taking two nearby states 
\begin{equation}{\label{metric_3d}}
\begin{split}
ds^2 &=_2 1-F(\hat{\rho}, \hat{\rho}+\delta \rho)^2  \\
&=  \frac{\omega^2}{16 \sinh^2 \frac{\beta \omega }{2}} d\beta^2  + \frac{1}{8}\( 1 + \frac{1}{\cosh (\beta \omega)}\) \(  4dr^2 + \sinh^2(2r) d\theta^2 \)\\
&=d\alpha^2 +   \frac 14\( 1- \frac{2}{3+\cosh(4\alpha)} \)(4dr^2 + \sinh^2(2r) d\theta^2)\,,
\end{split}
\end{equation}
which  reduces to a hyperbolic geometry  $\mathbb{H}^2$ for pure states with $\beta = \infty$. Similar to the assumption in \cite{purification}, we can ignore the phase part associated with the angle $\theta$ and focus on the two dimensional metric 
\begin{equation}\label{infor_2D_metric}
ds^2_{\mt{IM}} = d\alpha^2 +   \( 1- \frac{2}{3+\cosh(4\alpha)} \) dr^2, 
\end{equation}
whose geodesic equations ($\s \in [0,1]$) reads
\begin{equation}\label{infor_geodesic_2D}
\begin{split}
\frac{\dot{r}(\s) (\cosh (4 \alpha(\s))+1)}{\cosh (4 \alpha(\s))+3} &= C_0 \,, \quad 
\frac{4 \dot{r}(\s)^2 \sinh (4 \alpha(\s))}{(\cosh (4 \alpha(\s))+3)^2}- \ddot{\alpha}(\s)=0.
\end{split}
\end{equation}
%%%%%%%%%%%%%%%%%

\begin{figure}[htbp]
	\centering\includegraphics[width=4.50in]{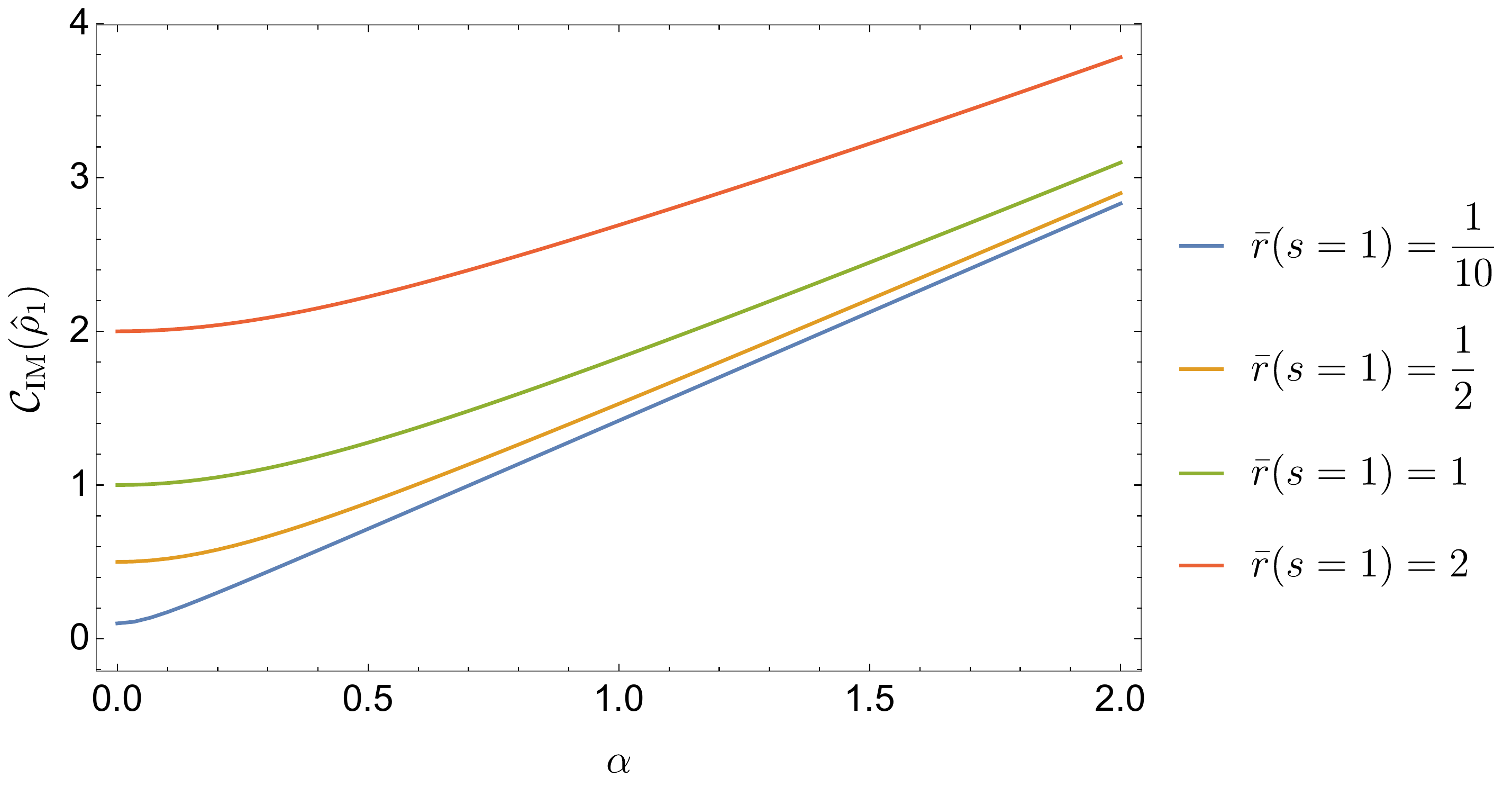}
	\caption{Circuit Complexity $\mC_{\mt{IM}}(\hat{\rho}_1)$ from the quantum Fisher information metric for one-mode Gaussian mixed state $\hat{\rho}_1\( r(\s=1),\alpha(\s=1)\)$ with different boundary value $\bar{r}(\s=1)$.}\label{fig:complexity_rho1}
\end{figure} 

%%%%%%%%%%%%%%%%%
%%%%%%%%%%%%%%%%%
Taking the special initial conditions $ \alpha(0)=0, \bar{r}(0)=0$ from our reference state $\ket{\psi_\mt{R}}= \ket{0(\mu)}$, we can find the analytic solutions of geodesic equations  
\begin{equation}\label{infor_solutions_2D}
\begin{split}
\alpha(\s)  &=  \frac{1}{2} \text{sech}^{-1}\left(\frac{\sqrt{C_1^2 \text{sech}^2\left(2 C_1 \s\right)}}{\sqrt{ C_1^2-C_0^2 \tanh ^2\left(2 C_1 \s\right)}}\right) , \\
\bar{r}(\s) &= \frac{1}{2} \left(2C_0 \s+\tanh ^{-1}\left(\frac{C_0 \tanh \left(2 C_1 \s\right)}{ C_1}\right)\right)\,.
\end{split}
\end{equation}
Imposing the boundary conditions, we can fix the constant $C_0, C_1$ as
\begin{equation}\label{boundary_equations}
\begin{split}
C_0 &= \pm   \sqrt{ C_1^2 \left( \coth ^2(2 C_1) -\text{csch}^2(2 C_1) \cosh^2 (2 \alpha(1))\right) }, \quad 
C_1= \frac{C_0 \tanh\( 2C_1\)}{\tanh \( 2\bar{r}(1)- 2C_0\)},  
\end{split}
\end{equation}
where the sign of the first equation depends on the sign of $\bar{r}(1)$ and the second transcendental equation cannot be solved analytically. However, one can still find that the length of geodesic and complexity from the quantum Fisher information metric satisfy
\begin{equation}
g_{\mu\nu}\, \dot{\lambda}^\mu\dot{\lambda}^\nu= \frac{\dot{r}(\s)^2 (\cosh (4 \alpha(\s))+1)}{\cosh (4 \alpha(\s))+3}+\dot{\alpha}(\s)^2 = C_0^2 +C_1^2, 
\end{equation}
and 
\begin{equation}\label{IM_complexity_rho1}
\mathcal{C}_\mt{IM}  \( \ket{0(\mu)},\hat{\rho}_1\)= \int_0^1\!\! ds\   \sqrt{ 2g_{\mu\nu}\, \dot{\lambda}^\mu\dot{\lambda}^\nu}= \sqrt{2C_0^2+2C_1^2 } \,,
\end{equation}
which is derived from the semi-analytical geodesic solution \eqref{infor_solutions_2D}. For later use, the geodesic solutions in \eqref{infor_solutions_2D} can be also rewritten in the form like
\begin{equation}\label{QFIM_solutions}
\begin{split}
\alpha (\s )&= \frac{1}{2} \cosh^{-1} \( \sqrt{\cosh^2(2C_1 \s)  - \frac{C_0^2}{C_1^2} \sinh^2 (2C_1  \s) }\) \,,\\
\bar{r}(\s)&=  C_0 \s + \frac{1}{4}\ln \(   \frac{\cosh (2C_1 \s) + \frac{C_0^2}{C_1^2} \sinh (2C_1\s)}{\cosh (2C_1 \s ) - \frac{C_0^2}{C_1^2}  \sinh (2C_1  \s) } \) \,. \\
\end{split}
\end{equation}
As a consistent check, we can consider the one-mode Gaussian pure state obtained by taking the following equivalent limits 
\begin{equation}
\beta \longrightarrow \infty \,, \quad T  \longrightarrow 0 \,, \quad  \alpha  \longrightarrow 0 \,.
\end{equation}
It is easy to find that the geodesic solution \eqref{infor_solutions_2D} reduces to $C_1^2 = C_0^2 $ and 
\begin{equation}\label{one_pure}
\begin{split}
\alpha(\s)&=0 \,, \bar{r}(\s)=2 C_0\s =\bar{r}\( \s=1\) \s\,,   \\
\mathcal{C}_\mt{IM}  \( \ket{0(\mu)},\ket{\psi_1}\)&=\sqrt{2C_0^2 +2C_1^2 }  = \bar{r}_1= \frac12 \left|\ln \frac{\omega_{1}}{\mu}\right| =\mathcal{C}_\mt{FS}  \(\ket{0(\mu)}, \ket{\psi_1}\)\,,
\end{split}
\end{equation}
which is the same as the results shown in \cite{qft1} and \cite{qft2}. Except for the pure-state limit, we can also easily obtain the numerical solutions for \eqref{boundary_equations} and the corresponding complexity by given  $\bar{r}\(\s=1 \), \alpha\(\s=1 \)$ for various target states. The numerical results are shown in the figure \ref{fig:complexity_rho1}.

From the geodesic solution \eqref{infor_solutions_2D}, we actually identify the evolution of mixed states in the optimal circuit from $\ket{0(\mu)}$ to $\hr_1$ as 
\begin{equation}
\hr_1(\s)=\hr_1\( r(\s),\alpha(\s)\)\,,
\end{equation}
without explicitly introducing the auxiliary system or performing any minimization process. 
In order to support our conclusion about the relation between purification complexity and complexity $\mC_{\mt{IM}}$ from quantum Fisher information metric, we would like to show that the analytical trajectory for $\bar{r}(s), \alpha(s)$ can be also subtracted from the optimal circuit for purified states, \eg the two-mode Gaussian pure states whose complexity has been discussed in \cite{qft1}. Comparing $\mC_{\mt{IM}}$ in \eqref{IM_complexity_rho1} with the purification complexity derived by minimizing the complexity of purified states, we will show that the two results are the same in the next subsection.

\begin{figure}[htbp]
	\centering\includegraphics[width=4.50in]{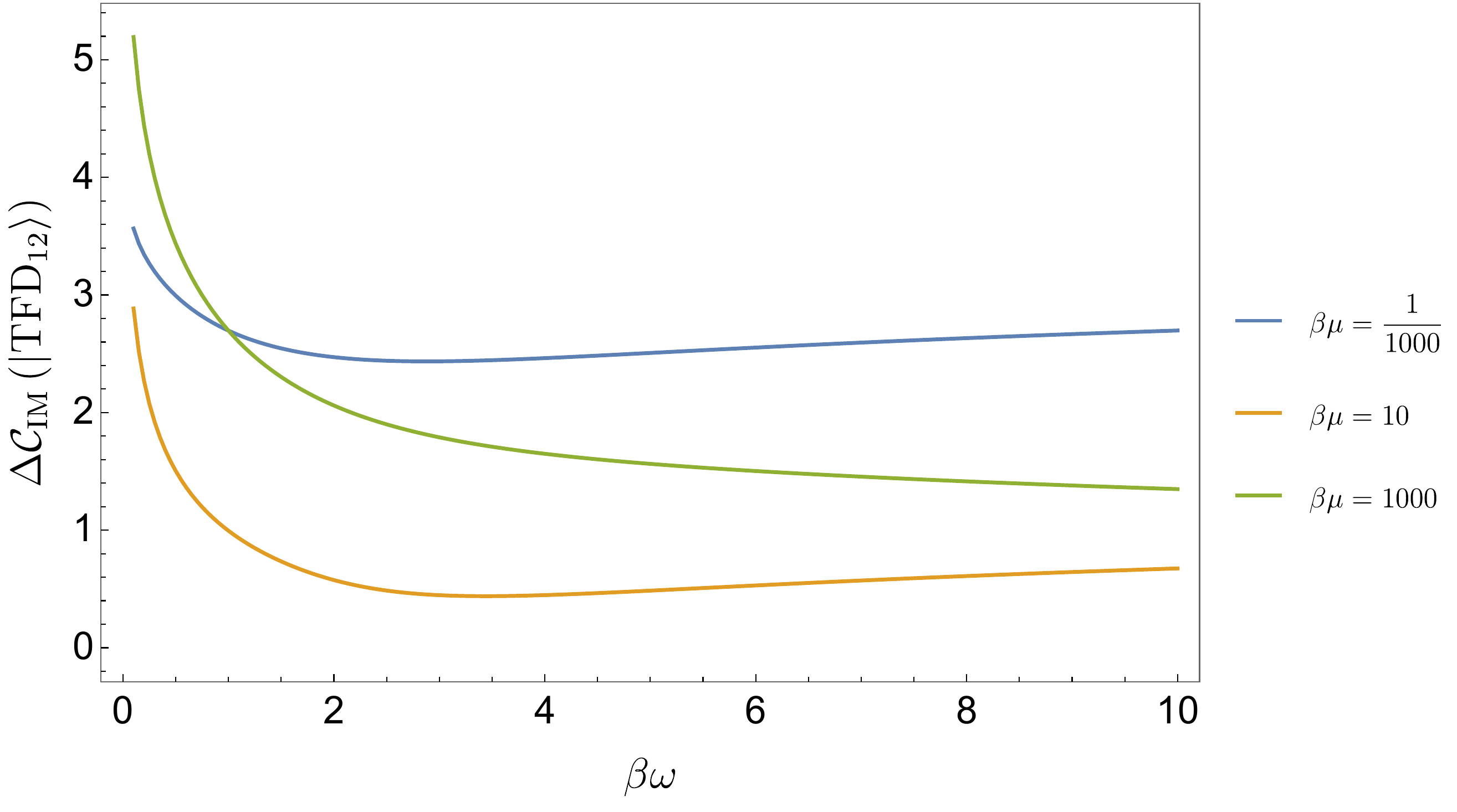}
	\caption{Mutual Complexity $\Delta \mC_{\mt{IM}}(\ket{\text{TFD}_{12}})=\Delta \mC_{\mt{FS}}(\ket{\text{TFD}_{12}})$ for TFD state is always subadditive.}\label{fig:mutualcomplexity_IM}
\end{figure}

Finally, we also note the mutual complexity of TFD state is sub-additive, \ie $\Delta \mC_{\mt{IM}} \ge 0$ as shown in figure \ref{fig:mutualcomplexity_IM}. From the viewpoint of purification complexity with $F_2$ cost function, the same result has been derived at section 7 in \cite{purification}.

\subsection{Optimal Purifications and Purification Complexity}\label{rem_qft1}
Before we move to the discussion about purified Gaussian states, we would like point out an important result from \cite{qft1} about pure Gaussian states, \ie the complexity of pure Gaussian state is factorized in the normal basis. As a result, we can find that the pure-state limit of $\mC_{\mt{IM}}$ equals to the complexity of any N-mode pure Gaussian state $\ket{\psi_\mt{N}}$ with Fubini-Study metric or $F_2$ cost function 
\begin{equation}\label{IM=FS}
\mC_{\mt{FS}} \( \ket{\psi_\mt{N}} \) =\mC_{2} \( \ket{\psi_\mt{N}} \) = \mC_{\mt{IM}} \( \ket{\psi_\mt{N}} \) \,,
\end{equation}
because the ancillae for pure Gaussian state cannot decrease the complexity.  This equivalence for the one-mode pure state has been shown in \eqref{one_pure}.

  \subsubsection{Reminiscence: Purified Gaussian States}
   In \cite{purification}, the purification complexity (with fixed reference state) related to different cost functions has been discussed in details by focusing on Gaussian mixed states and purified Gaussian states. Given the mixed state $\hat{\rho}_\mA$ for subsystem $\bmA$, we can obtain its purified state $\ket{\Psi}_{\mA\mA^c}$ by introducing ancillae, the auxiliary system $\bmA^c$. The purification complexity of $\hat{\rho}_\mA$ is defined as the minimal complexity of purified states, \ie 
	\begin{equation}
	\mP \left(\hat{\rho}_{\mA}\right) = \min\limits_{\mA^c}\,  \mC \left(\ket{\Psi}_{\mA\mA^c}\right) \,,
	\end{equation}	
	where the minimization is done over all possible purified states with $\tr_{\mA^c}\( \ket{\Psi}_{\mA\mA^c} \bra{\Psi}_{\mA\mA^c}\) = \hat{\rho}_{\mA}$. As we discussed in \eqref{equality_rhoA2}, this minimization on all purified target states is enough to get the purification complexity as we will explicitly show in this section.

Restricted on the Gaussian pure states, we can arrive at the one-parameter family of two-mode purified states (one mode is the ancilla)
\begin{equation}\label{Fock_psi12}
\ket{\Psi_{\mA\mA^c}} \rightarrow \ket{\psi}_{12} = \hat{S}_1(r)\,\hat{S}_2(s) \, \hat{S}_{12}(\alpha) \,\ket{0}_1\ket{0}_2\,,
\end{equation}
whose position-space wavefunction is described by 
\begin{equation}\label{pure1}
\begin{split}
\psi_{12}(x,y)  &= \left(\frac{(a-b)}{2b}\frac{k^2}{\pi^2}\right)^{1/4}\,  e^{-\frac{1}{2}\left[\left(a+b\right) x^2 + \frac{ k^2}{2b} y^2 + 2 k xy \right]} \\
&=\sqrt{\frac{ \omega}{\pi}}\, e^{\frac{r+s}{2}} \,
\exp\[-\frac{\omega }{2}\(\cosh 2\alpha\,  (e^{2r}x^2+  e^{2 s}y^2) - 2\,e^{r+ s}\sinh 2\alpha \,xy \)\]\,.
\end{split}
\end{equation}
We can also denote the pure Gaussian states \eqref{pure1} in the matrix representation as
\begin{equation}\label{matrix_purified}
\begin{split}
A^{ab} &=\omega  \left(
\begin{array}{cc}
\ e^{2 r} \cosh 2 \alpha & -e^{r+ s} \sinh 2 \alpha \\
-e^{r+s} \sinh 2 \alpha & \ e^{2 s} \cosh 2 \alpha \\
\end{array}
\right)
=\mu  \left(
\begin{array}{cc}
\ e^{2 \bar{r}} \cosh 2 \alpha & -e^{\bar{r}+ \bar{s}} \sinh 2 \alpha \\
-e^{\bar{r}+\bar{s}} \sinh 2 \alpha & \ e^{2 \bar{s}} \cosh 2 \alpha \\
\end{array}
\right)\,.
\end{split}
\end{equation}
Fixing the reference state as the unentangled state $\ket{0(\mu)} \otimes \ket{0(\mu)}$, the purification complexity of Gaussian mixed state $\hat{\rho}_1$ is derived as 
\begin{equation}
\mP\left(\ket{0(\mu)}, \hat{\rho}_1\right) = \min\limits_{s}\,  \mC \left(\ket{0(\mu), 0(\mu)}, \ket{\psi}_{12}\right) \,,
\label{walk2}
\end{equation}
where we can read the complexity of Gaussian pure states from \cite{qft1} and minimize its among the free parameter $s$.  For example, the authors in \cite{purification} found the purification complexity for the one-mode Gaussian mixed states with $F_1$-cost function reads
\begin{equation}\label{complexity_one_mode}
\mP_{1}\(\hat{\rho}_1\)  =
\left\{
\begin{array}{lr}
\frac 12  \ln \left(    \frac{{e^{-2 \bar{r} }\cosh 2 \alpha }-1}{1-{e^{2 \bar{r} } \cosh 2 \alpha }}\right), &~~~ 0\le  \alpha  \le -\bar{r} \,,\\
\\
2\alpha , &~~~ \alpha \ge |\bar{r} |\,,\\
\\
\frac 12 \ln \left(    \frac{{e^{2 \bar{r} } \cosh 2\alpha-1}}{{1-e^{-2 \bar{r} }\cosh 2\alpha}} \right) , &~~~ 0\le \alpha \le \bar{r} \,.
\end{array}
\right.
\end{equation}

\subsubsection{Optimal purification from minimization}
In order to compare with the complexity from the quantum Fisher information metric, it is natural to consider $F_2$-cost function or Fubini-Study metric. Before that, we simply review the optimal circuit found in \cite{qft1} for pure Gaussian states and apply the minimization to find the optimal trajectory for mixed states in the subsystem, \ie the one-mode state $\hr_1(\s)$.  From the unitary operations parametrized by a two-by-two matrix 
\begin{equation}\label{GL2_matrix}
U_{2} = e^y R(-x)\, S(\rho)\, R(z)=
e^y \begin{pmatrix}
\cos x & - \sin x\\
\sin x & \ \cos x
\end{pmatrix}
\begin{pmatrix}
e^{\rho} & 0 \\
0 & e^{-\rho} 
\end{pmatrix}
\begin{pmatrix}
\ \cos z & \sin z\\
-\sin z & \cos z
\end{pmatrix}\,,
\end{equation}
as the representation of the elements in $\text{GL}(2,\mathbb{R})$ group, the pure state  $\psi_{12}(x_+,x_-) = U_2\,\psi_\mt{R}(x_+,x_-)$ along the circuit (path) in this matrix representation is given by 
\begin{equation}\label{eq:A-target}
\begin{split}
A(\mathbf{y}(\s))&= U_2(\mathbf{y})\,A_\mt{R}\, U_2^T(\mathbf{y}) \\
&=\mu \left(
\begin{array}{ccc}
e^{2 y} (\cosh (2 \rho )+\cos (2x) \sinh (2 \rho )) & e^{2 y} \sin (2x) \sinh (2 \rho )  \\
e^{2 y} \sin (2x ) \sinh (2 \rho ) & e^{2 y} (\cosh (2 \rho )-\cos (2x) \sinh (2 \rho )) \\
\end{array}
\right)\,, 
\end{split}
\end{equation}
with boundary conditions fixed by the specific target state $A_{\mt{T}} \equiv 	A(\mathbf{y}(\s=1))$. It is found in \cite{qft1} that the optimal circuit is a straight line connecting the reference state and the target state in the norm basis, \ie 
\begin{equation}\label{geodesic-line}
\rho(\s) = \rho_1 \s\,,  \quad x(\s)= x_0= z\( \s \) \,, \quad  y(\s)= y_1 \s\,,
\end{equation}
with 
\begin{equation}
\label{eq:boundary-cond}
y_1 = \frac14 \log \( \frac{\omega_+\omega_-}{\mu^2}  \)\,, \quad \rho_1 = \frac14 \log \frac{ \omega_+}{\omega_-}  \,.
\end{equation}
According to the choice of cost functions, we can derive the complexity of two-mode Gaussian states as 
\begin{equation}\label{complexity_pure}
\begin{split}
\mC_{\mt{FS}}\( \ket{\psi_{12}}\) &=\mathcal{C}_{2}\( \ket{\psi_{12}}\) =
{\frac 12}\sqrt{ \(\ln \frac{\omega_+}{\mu}\)^2+\(\ln \frac{\omega_-}{\mu}\)^2}  \,,\\
\(\mC_{\mt{FS}}\( \ket{\psi_{12}}\) \)^2 &= \mathcal{C}_{\kappa=2}\( \ket{\psi_{12}}\) \,.  
\end{split}
\end{equation}
From the trajectory of two-mode Gaussian states, we can also explore the reduced trajectory of mixed Gaussian state $\hat{\rho}_1 (\s)$ by tracing out one of the two modes.  In our notations (or coordinates ) for mixed states $\hr_1$ defined in \eqref{one_mixed}, we can rewrite its purified states in matrix form as \eqref{matrix_purified} by
\begin{equation}
\begin{split}
\ket{\psi}_{12}  \quad \longrightarrow \quad A^{ab} (\s) 
&=\mu  \left(
\begin{array}{cc}
\ e^{2 \bar{r}} \cosh 2 \alpha & -e^{\bar{r}+ \bar{s}} \sinh 2 \alpha \\
-e^{\bar{r}+\bar{s}} \sinh 2 \alpha & \ e^{2 \bar{s}} \cosh 2 \alpha \\
\end{array}
\right)\,. 
\end{split}
\end{equation}
Comparing that with the coordinates system $\( \rho, y, x\)$ in \eqref{eq:A-target} from the representation of $\text{GL}(2,\mathbb{R})$ group, we can find the transformation 
\begin{equation}
\begin{split}
\bar{r} +\bar{s}&= 2y\,,  \qquad   \bar{r} - \bar{s} = \pm \frac{1}{2} \ln \(   \frac{\cosh 2\rho -\cos 2x \sinh 2\rho}{\cosh 2\rho + \cos 2x \sinh 2\rho } \) \,, \\
\alpha &= \frac{1}{2} \cosh^{-1} \( \sqrt{\cosh^22\rho -\cos^2 2x \sinh^2 2\rho}\) \,. 
\end{split} 
\end{equation}
Recalling the geodesic solution  \eqref{geodesic-line} for pure Gaussian states, we can derive the trajectory of mixed Gaussian states $\hat{\rho}_1 ( \bar{r},\alpha)$ as
\begin{equation}
\begin{split}
\alpha (\s )&= \frac{1}{2} \cosh^{-1} \( \sqrt{\cosh^2(2\rho_1 \s)  -\cos^2 2x_0 \sinh^2 (2\rho_1 \s) }\) \,,\\
\bar{r}(\s)&= y_1 \s + \frac{1}{4}\ln \(   \frac{\cosh (2\rho_1 \s) +\cos 2x_0 \sinh (2\rho_1 \s)}{\cosh (2\rho_1 \s ) - \cos 2x_0 \sinh (2\rho_1 \s) } \) \,, \\
\end{split}
\end{equation}
which exactly matches the geodesic path \eqref{QFIM_solutions} derived in 2D-manifold with the quantum Fisher information metric! But this is not the final answer for the purification complexity because we still need to find the optimal purification with a given target state $\hr_1\(\alpha\( \s=1\), \bar{r}\( \s=1\)\)$ and then it will determine the free parameter $\bar{s}$ for the optimal purification. Some analytical approximations have been discussed in \cite{purification}. Instead, we can also directly perform the numerical minimization.  With all given target states $\hr_1$ (as shown in figure \ref{fig:complexity_rho1} \footnote{
	We find the difference between that and the results from numerical minimization is at the order $10^{-15}$ which is just the machine precision. Decreasing machine precision also correspondingly decreases the difference.
}), we find that the minimization leads us to the same minimum for complexity as \eqref{IM_complexity_rho1}. As a summary, the minimization for the purification complexity of Gaussian mixed state $\hr_1$ simply shows the equivalence, \ie 
\begin{equation}\label{one_mode_equal}
\mP_{\mt{FS}} \(\ket{0(\mu)}, \hat{\rho}_1\)\equiv \min\limits_{\bar{s}} \,\mathcal{C}_{\mt{FS}}\( \ket{\psi_{12}}\)  =\min\limits_{\bar{s}} \,
{\frac 12}\sqrt{ \(\ln \frac{\omega_+}{\mu}\)^2+\(\ln \frac{\omega_-}{\mu}\)^2} = \mC_{\mt{IM}} \(\ket{0(\mu)},\hat{\rho}_1\)\,.
\end{equation}
This equivalence also means that the optimal circuit found from the geodesic associated with the quantum Fisher information metric is the same as that from the optimal circuit for two-mode pure Gaussian state by tracing out one extra ancillary mode. This simple example illustrates our main conclusion that the complexity (geodesic distance) associated with the quantum Fisher information metric is the purification complexity measured by the Fubini-Study metric. More generally, our proof in \eqref{equality_rhoA} also indicates the optimal purification for one-mode Gaussian state is actually the essential purification, \ie two-mode Gaussian state, confirming the expectation and assumption in \cite{purification}.
%%%%%%%%%%%%%%%%%%%%%%%%%
%%%%%%%%%%%%%%%%%%%%%%%%%
%%%%%%%%%%%%%%%%%%%%%%%%%

\subsection{Purified States with Uhlmann's fidelity}
In the last subsection, we have shown the equivalence between purification complexity $\mP_{\mt{FS}}$ and mixed-state complexity $\mC_{\mt{IM}}$ based on the quantum Fisher information metric. In this subsection, we explicitly construct the optimal purified states with the saturation of Uhlmann's fidelity. Furthermore, we also illustrate the quantum fidelity's bound, \ie Uhlmann's theorem for Gaussian state as stated in \eqref{Uhlmann_theorem} is satisfied by taking the two-mode pure Gaussian state as purification. Taking two arbitrary mixed Gaussian states $\hr_1\(r_1,\alpha_1\)$ and $\hr_1'\(r_2,\alpha_2\)$, their quantum fidelity is found to be  
\begin{equation}\label{fidelity_rho1}
F(\hat{\rho}_1, \hat{\rho}_1') =  \sqrt{\frac{2 \sinh \frac{\beta_1
			\omega}{2}\sinh \frac{\beta_2 \omega}{2}}{\sqrt{\( \cosh^2 (r_1-r_2) \cosh^2 \(\frac{\beta_1+\beta_2}{2} \omega\)-\sinh^2 (r_1-r_2) \cosh^2 \( \omega \frac{\beta_1-\beta_2 }{2}\) \)}-1}}\,,
\end{equation}
which should be equivalent to the fidelity between specific purified states according to Uhlmann's theorem.

First of all, we start from the simplest purification, \ie the two-mode Gaussian states \eqref{Fock_psi12}.
Noting that we can parametrize the wavefunction of purified Gaussian states as \eqref{pure1}
\begin{equation}\label{pure02}
\begin{split}
\psi_{12}(x,y)  &=\sqrt{\frac{ \omega}{\pi}}\, e^{\frac{r+s}{2}} \,
\exp\[-\frac{\omega }{2}\(\cosh 2\alpha\,  (e^{2r}x^2+  e^{2 s}y^2) - 2\,e^{r+ s}\sinh 2\alpha \,xy \)\]\,,
\end{split}
\end{equation}
it is easy to find the quantum fidelity between pure Gaussian states $ \psi_{12}(x,y;r_1,s_1,\alpha_1) $ and $ \psi_{12}'(x,y;r_2, s_2, \alpha_2) $ as
\begin{equation}
\begin{split}
&F(\lambda_1, \lambda_2)=\left| \langle{\psi} \ket{\psi'} \right |= \int_{-\infty}^{\infty} \,\int_{-\infty}^{\infty}  \psi_{12}(x,y;r_1,s_1,\alpha_1) \psi_{12}'(x,y;r_2, s_2, \alpha_2) \, dx\,dy\,,\\
&=\sqrt{\frac{2}{\cosh 2 \alpha _1 \cosh \left(2 \alpha _2\right) \cosh \left(r_1-r_2-s_1+s_2\right)+\cosh \left(r_1-r_2+s_1-s_2\right)-\sinh 2 \alpha _1 \sinh 2 \alpha _2}}\,,
\end{split}
\end{equation}
by a simple Gaussian integral. The maximal fidelity is decided by the saddle point with
\begin{equation}
\partial_{s_1} F(\lambda_1,\lambda_2)= 0  \qquad \text{and} \qquad  \partial_{s_2} F(\lambda_1,\lambda_2)= 0  \,.
\end{equation}
However, the above two derivative equations lead us to the same solution 
\begin{equation}
 \quad s_2 =s_1+ \frac{1}{2} \log \left(\frac{e^{2 r_1}+e^{2 r_2} \cosh 2 \alpha _1\cosh 2 \alpha _2}{e^{2 r_2}+e^{2 r_1} \cosh 2 \alpha _1 \cosh 2 \alpha _2}\right)\,.
\end{equation}
Generally, the maximum of $F(\lambda_1,\lambda_2)$ should be given by critical point with $ \partial_{s_1} F(\lambda_1,\lambda_2)= 0$ and $ \partial_{s_2} F(\lambda_1,\lambda_2)= 0 $ simultaneously. However, either condition is sufficient because of the unitary invariance of the fidelity.
Plugging the solutions of $s_1-s_2$ into the fidelity between pure states, we can find the maximum of fidelity as
\begin{equation}
\begin{split}
&\max\limits_{\ket{\psi_{12}},\ket{\psi_{12}'}} \, F(\lambda_1, \lambda_2)\\
&=\sqrt{ \frac{2}{\sqrt{ \cosh ^22 \alpha _1 \cosh ^22 \alpha _2+2 \cosh 2 \alpha _1 \cosh 2 \alpha _2 \cosh \left(2 r_1-2 r_2\right)+1}- \sinh 2 \alpha _1 \sinh 2\alpha _2}}\,,\\
\end{split}
\end{equation}
which equals Uhlmann's fidelity \eqref{fidelity_rho1} derived from two Gaussian mixed states $\hr\(r_1,\alpha_1\)$ and $\hr\(r_2,\alpha_2\)$. From this view of point, we can claim that the purification restricted on pure Gaussian states is enough to achieve the optimal purification for mixed Gaussian state $\hat{\rho}_1$, \ie satisfying the fidelity's bound in Uhlmann's theorem. That point illustrates why we can match the complexity and also the evolution path for mixed states $\hr_1(\s)$ with those derived from only two-mode Gaussian pure states as shown in the last subsection.

%%%%%%%%%%%%%%%%%%%%%%%%%%%%%%%%%%
%%%%%%%%%%%%%%%%%%%
%%%%%%%%%%%%%%%%%%%
%%%%%%%%%%%%%%%%%%%
\section{Comparison: Different Distances Measures for Mixed States}	
 Although we only pay attention to the quantum Fisher information metric or Bures metric in the last sections, there are also some other well-studied finite distances or metric structures for density matrices in the field of quantum information.  In previous studies on the geometry of quantum states, the distance measure is also an important concept, \eg \cite{bengtsson2017geometry,chruscinski2012geometric}. 
 Different from previous studies on the geometry of quantum states by defining finite distance structure between two quantum states with respective density operators $\hr$ and $\hs$, we prefer a local Riemannian geometry with positive definite metric since we can associate the geodesic with the optimal circuit connecting the reference state and target state. In this section, we simply introduce some other metrics on the space of density matrix and take them as the measure for complexity. We focus on comparing them with our proposal for purification complexity $\mC_{\mt{IM}}$. 
\subsection{Distances, Schatten Norms and Metrics}
 Starting from any hermitian operator $A$, we can find that the singular value decomposition is reduced to 
\begin{equation}
A= \sum_{i=1}^{r_{\mt{A}}}s_i \ket{x_i}\bra{x_i}\,,
\end{equation}
where $s_i$ are positive singular values of $A$ and $r_{\mt{A}}$ denotes its rank. A popular and generic norm structure is based on so-called Schatten p-norm (see \eg \cite{watrous2018theory}) defined by
	\begin{equation}
	\begin{split}
	||A||_p = \left[ \tr\( \( A^\ast A\)^{\frac{p}{2}} \) \right]^\frac{1}{p} =  \( \sum_k  s^p_k \)^{1/p}\,.\\
	\end{split}
	\end{equation}
	where the real and positive singular values $s_k$ are associated with the eigenvalues of $A^\dagger A$ in the spectral decomposition) by
	\begin{equation}
	s_k (A)=  \sqrt{\lambda_k (A^\dagger A)} =  \sqrt{\lambda_k (AA^\dagger)}\,, \qquad 1\le k \le \text{rank}(A) =r_{\mt{A}}\,. 
	\end{equation}
 For examples, one can obtain some well-known distance structures between any two density matrices $\hr, \hs$ by taking hermitian operator as $A=\hr-\hs$ with fixing the value of $p$, \eg 
	\begin{equation}
	\begin{split}
		||A||_1 &\longrightarrow   D_{\mt{Tr}} \( \hs_,\hat{\rho}\)= \frac 12 || \hat{\rho}- \hs||_1\,;\\
			||A||_2 &\longrightarrow  D_{\mt{HS}} \( \hs, \hat{\rho}\)= || \hat{\rho}- \hs||_2\,;\\
			 		||A||_{\infty}  &\longrightarrow        \text{Spectral norm}            \,.\\
		\end{split}
	\end{equation}
	Interestingly, Schatten norms present some nice properties :
	\begin{itemize}
		\item  The Schatten $p$-norm is non-increasing in $p$ : 
			\begin{equation}\label{non_increasing}
		|||A||_p  \ge ||A||_q\,, \qquad  1\le p\le q\le \infty\,.
		\end{equation}
		\item  The Schatten $p$-norm is submultiplicative 
		\begin{equation}
		||A\, B||_p  \le 	||A||_p \,||B||_p  \,.
		\end{equation}
		\item The non-zero operator $A$ with different norms $ (1\le p\le q\le \infty )$ satisfies 
		\begin{equation}
		||A||_p \le \(r_{\mt{A}}\)^{\frac{1}{p}-\frac{1}{q}}\,||A||_q \,, \longrightarrow  ||A||_1 \le \sqrt{r_{\mt{A}}}\,||A||_2\,, \quad  ||A||_2 \le \sqrt{r_{\mt{A}}}\,||A||_{\infty}\,.
		\end{equation}
		\item Schatten p-norm  is isometrically invariant 
		\begin{equation}
			||A||_p = ||BAC^\dagger||_p\,.
		\end{equation}
	\end{itemize}
For normalized density matrices, we can also define the normalized Schatten $p$-norms by 
\begin{equation}
\begin{split}
 \frac{1}{2^{1/p}}||\hr -\hs||_p = \frac{1}{2^{1/p}}  \left( \tr  | \hr -\hs|^p  \right)^\frac{1}{p}  \in [0,1] \,.
 \end{split}
\end{equation}
The analysis for different Schatten norms are very similar and we only take $p=1,2$ as examples in the following subsections.

 \subsubsection{Trace Distance, $p=1$}
 Beginning with the Schatten norm at $p=1$, we can define a finite distance between two arbitrary quantum states $\hs, \hat{\rho}$ by \footnote{As usual, one can define the positive square root by $|A|\equiv \sqrt{A^\dagger A}$ to simplify the notations.}
 \begin{equation}\label{TR_distance}
 D_{\mt{Tr}} \( \hs, \hat{\rho}\)= \frac 12 || \hat{\rho}- \hs ||_1=  \frac 12\tr \sqrt{  \(\hat{\rho}- \hs \)^2 } \,,
 \end{equation}
 which is known as the trace distance. It is easy to see that the trace distance between two pure states reduces to 
 \begin{equation}\label{TR_pure}
 \begin{split}
 D_{\mt{Tr}} \(  \ket{\psi}\bra{\psi},\ket{\phi}\bra{\phi}\) = \sqrt{1 -  F(\ket{\psi},\ket{\phi})^2}  \,.
 \end{split}
 \end{equation}
One can also find that the trace distance admits many similar properties \cite{NielsenChuang,wilde2013quantum} to our purification complexity $\mC_{\mt{IM}}$. 
However, noting we keep the convention in quantum information with a factor $\frac 12 $ for the normalized trace distance, one can easily show the trace distance is bounded by 
\begin{equation}
 0\le   D_{\mt{Tr}} \( \hs, \hat{\rho}\) \le 1 \,,
\end{equation}
which is obvious from the triangle inequality of Schatten norm. 
 Similar to the Uhlmann's fidelity, the trace distance is also monotone with respect to discarding of subsystems
 \begin{equation}
\begin{split}
D_{\mt{Tr}} \( \hat{\sigma}_\mA,\hat{\rho}_\mA\)  \le  D_{\mt{Tr}} \( \hat{\sigma}_{\mA\mA^c}, \hat{\rho}_{\mA\mA^c}\)\,.
\end{split}
\end{equation}
 The special relations between fidelity and trace distance are described by the following bound 
 \begin{equation}\label{trace_fidelity}
 \begin{split}
   1- F(\hat{\rho}, \hat{\sigma})  &\le      D_{\mt{Tr}} \( \hat{\rho}, \hat{\sigma}\) \le \sqrt{1 -  F(\hat{\rho}, \hat{\sigma})^2} \,,\\
   1-  D_{\mt{Tr}} \( \hat{\rho}, \hat{\sigma}\) &\le     F(\hat{\rho}, \hat{\sigma})   \le \sqrt{1 -   D_{\mt{Tr}} \( \hat{\rho}, \hat{\sigma}\)^2} \,,\\
   \end{split}
 \end{equation}
 where the second inequality is saturating if and only if we consider two pure states.
 These proofs are based on the properties of Schatten norm and can be found in the standard textbooks, \eg \cite{NielsenChuang,wilde2013quantum}.  Here we only sketch the proof of the second inequality 
 \begin{equation}
 \begin{split}
  D_{\mt{Tr}} \( \hat{\rho}, \hat{\sigma}\) \le  \frac 12 \big|\big| \ket{\Psi^{op}}\bra{\Psi^{op}}- \ket{\Phi^{op}}\bra{\Phi^{op}}\big|\big|_1 
  =  \sqrt{1 -  F(\hat{\rho}, \hat{\sigma})^2} 
   \le \frac 12 \big|\big| \ket{\Psi}\bra{\Psi}- \ket{\Phi}\bra{\Phi}\big|\big|_1  \,,\\
   \end{split}
 \end{equation}
 where pure states $\ket{\Psi},\ket{\Phi}$ are any purifications of respective mixed states $\hat{\rho},\hat{\sigma}$ and we only use the monotonicity of trace distance and Uhlmann's theorem for the existence of optimal pure states $\ket{\Psi^{op}}, \ket{\Phi^{op}}$ that saturate Uhlmann's fidelity $F\( \hr,\hs\)$. 

 From the perspective of complexity, we are more interested in the geodesic distance from the infinitesimal trace distance defined by 
 \begin{equation}\label{TR_metric}
 ds^2_{\mt{Tr}} \( \hat{\rho}, \hat{\rho}+\delta\hr\)= \frac 14  \( ||\delta \hr||_1 \)^2= \frac 14 \( \tr \sqrt{d\hat{\rho} d\hat{\rho} } \)^2 \equiv g^{\mt{Tr}}_{\mu\nu}\, \dot{\lambda}^\mu\dot{\lambda}^\nu d\s^2  \,.
 \end{equation}
 And it is obvious that only the lead order $\delta^{(1)}\hr$ will have contributions in the above metric form.
 Needless to say, we can also obtain the Fubini-Study metric from infinitesimal trace distance for two near pure states,  \ie  
 \begin{equation}\label{dsTR_pure}
 ds^2_{\mt{Tr}} \(\ket{\delta \psi}, \ket{\psi+\delta \psi}\) = ds^2_{\mt{FS}}  =_2 1- F^2(\ket{\psi},\ket{\psi+\delta\psi})  \,,
 \end{equation} 
 which provides a standard measure for us to fix the normalization factor when comparing different metrics. 
From the inequality \eqref{trace_fidelity}, we simply find the trace distance to be smaller than the quantum Fisher information metric 
 \begin{equation}\label{inequality_rhoA2}
  ds^2_{\mt{Tr}} \( \delta \hr\) \le  ds^2_{\mt{IM}} \( \delta \hr\)  \,.
 \end{equation}
 Noticing the normalization factor due to the distance for pure states, we similarly define the complexity of any quantum state $\hr_\mA$ in the Hilbert space $\mH_\mA$  by the trace metric  
 \begin{equation}
 	\mC_{\mt{Tr}} \(\hs_\mA, \hat{\rho}_\mA \)   \equiv  \int \sqrt{ 2ds^2_{\mt{Tr}}} =\int_0^1\!\! d\s\   \sqrt{ 2g^{\mt{Tr}}_{\mu\nu}\, \dot{\lambda}^\mu\dot{\lambda}^\nu} \,, 
 \end{equation}
 where the initial point and endpoint are determined by the reference state and target state. 
 Summarizing the inequalities we got before and also the equivalence for pure states \eqref{dsTR_pure}, we can have the following inequalities for complexity of a generic quantum state $\hr_\mA$
 	\begin{equation}\label{inequality_rhoA}
 	\mC_{\mt{Tr}} \(\hs_\mA, \hat{\rho}_\mA \)  \le 	\mC_{\mt{IM}} \(\hs_{\mA}, \hat{\rho}_\mA \)  =   \mathcal{P}_{\mt{FS}}\( \hs_\mA, \hat{\rho}_\mA \) 
 	\end{equation}
where the first equality is saturating if and only if the geodesic only goes through pure states.
 \subsubsection{Hilbert-Schmidt Distance, $p=2$}
 Taking $p=2$ for Schatten norm (\ie Frobenius norm), we arrive at the finite Hilbert-Schmidt distance between two arbitrary quantum states $\hat{\rho}, \hs$ 
 \begin{equation}\label{HS_distance}
  D_{\mt{HS}} \( \hs, \hat{\rho}\)= || \hat{\rho}- \hs||_2 = \sqrt{ \tr\(\hat{\rho}- \hs\)^2 } \,, 
 \end{equation}
 which is reduced to 
  \begin{equation}\label{HS_pure}
 D_{\mt{HS}} \( \ket{\psi}\bra{\psi},\ket{\phi}\bra{\phi}\)= \sqrt{2\(1- |\langle \psi | \phi \rangle|^2 \)} =\sqrt{2} D_{\mt{Tr}} \( \ket{\psi}\bra{\psi},\ket{\phi}\bra{\phi}\) \,,
 \end{equation}
  for two pure states. It is also straightforward to derive the infinitesimal metric 
 \begin{equation}\label{HS_metric}
  ds^2_{\mt{HS}} \( \hat{\rho}, \hat{\rho}+\delta\hr\)= \( ||\delta \hr||_2 \)^2=  \tr\( d\hat{\rho} \)^2=g^{\mt{HS}}_{\mu\nu}\, \dot{\lambda}^\mu\dot{\lambda}^\nu d\s^2 \,,
 \end{equation}
 by considering two near states. 
 Similarly, we can define the complexity of mixed states $\hr_\mA$ from Hilbert-Schmidt metric by
 \begin{equation}
 \mC_{\mt{HS}} \(\hs_\mA, \hat{\rho}_\mA \)   \equiv  \int \sqrt{ ds^2_{\mt{HS}}} =\int_0^1\!\! ds\   \sqrt{ g^{\mt{HS}}_{\mu\nu}\, \dot{\lambda}^\mu\dot{\lambda}^\nu} \,, 
 \end{equation} 
 where we do not need to add a factor again due to \eqref{HS_pure}. Here we can also compare the complexities from the trace metric and Hilbert-Schmidt metric. One may want to directly apply the non-increasing property of Schatten $p$-norm \eqref{non_increasing}.  However, we have stressed the complexity from various metrics should be normalized first to make them have the same results for pure states.  Then, we need to compare 
 \begin{equation}
 ds^2_{\mt{HS}} \( \hat{\rho}, \hat{\rho}+\delta\hr\)= \( ||\delta \hr||_2 \)^2 \,, \qquad  2ds^2_{\mt{Tr}} \( \hat{\rho}, \hat{\rho}+\delta\rho\)= \frac 12  \( ||\delta \hr||_1 \)^2 \,.
 \end{equation}
 Recalling the special property of Schatten $p$-norm, \ie $ ||A||_2 \le  ||A||_1 \le \sqrt{\text{rank}(A)}\,||A||_2$, 
 it is direct to show 
  \begin{equation}
 \frac{1}{\sqrt{2}}	\mC_{\mt{HS}} \(\hr_\mA, \hat{\rho}_\mA \)  \le	\mC_{\mt{Tr}} \(\hs_\mA, \hat{\rho}_A \) \le 	\mC_{\mt{HS}} \(\hs_\mA, \hat{\rho}_A \) \,,
 \end{equation}
 where the second inequality is true if we have $\text{rank}(\delta \hr_\mA) \ge 2$. 
Although we have the non-increasing property of Schatten norm, we have seen it is not easy to compare them properly after the normalization since it is hard to determine the rank of $\delta \hr_{\mA}$ along a general geodesic. Instead, we focus on analyzing the one-mode Gaussian density matrix as an explicit example.
 
 For a Gaussian state $\hr_{\mt{G}}$, its density matrix is equivalently described by the covariance matrix $\Sigma_{\mt{G}}$ (see appendix \ref{sec:app_QI} for more details). From the useful relation 
 \begin{equation}
\tr\( \hat{\rho}_{\mt{G}} \hat{\rho}_{\mt{G}}' \)=  \frac{1}{\sqrt{\det \frac 12 \(  \Sigma_{\mt{G}}+ \Sigma_{\mt{G}}' \) } } \,,
 \end{equation}
 we can find the Hilbert-Schmidt distance between two Gaussian states to be rewritten in form of the covariance matrix by  \cite{link2015geometry}
 \begin{equation}
  D_{\mt{HS}} \( \hat{\rho}_{\mt{G}}, \hat{\rho}_{\mt{G}}'\)= \sqrt{\frac{1}{\sqrt{ \det \Sigma_{\mt{G}}}} +\frac{1}{\sqrt{ \det \Sigma_{\mt{G}}'}}  - \frac{2}{\sqrt{\det \frac 12 \(  \Sigma_{\mt{G}}+ \Sigma_{\mt{G}}' \) } }   }  \,.
 \end{equation}
 Applying the formula $ \det \sigma = e^{\tr\( \ln \sigma\)}$ and its expansion 
 \begin{equation}
 \begin{split}
  \frac{1}{\sqrt{ \det \Sigma_{\mt{G}}'}}  &=  \frac{1}{\sqrt{ \det \Sigma_{\mt{G}} }} \exp \( -\frac 12 \tr  \ln  \( \mathbb{I} + \Sigma^{-1}_{\mt{G}}d\Sigma_{\mt{G}}  \) \) \\
  &= \frac{1}{\sqrt{ \det \Sigma_{\mt{G}} }}  \(  1 -\frac 12 \tr \( \Sigma^{-1}_{\mt{G}}d \Sigma_{\mt{G}}\) + \frac 14 \tr \( \Sigma^{-1}_{\mt{G}}d \Sigma_{\mt{G}}\)^2  +\frac 18 \[ \tr \( \Sigma^{-1}_{\mt{G}}d \Sigma_{\mt{G}}\)\]^2 \) \,,
  \end{split}
 \end{equation}
  one can obtain the infinitesimal metric 
 \begin{equation}
 \begin{split}
 ds^2_{\mt{HS}} \( \hat{\rho}_{\mt{G}}, \hat{\rho}_{\mt{G}}+\delta\hr_G\)&= \( ||\delta \hr_{\mt{G}} ||_2 \)^2=  \tr\( d\hat{\rho}_{\mt{G}} \)^2 \\
 &= \frac{1}{16 \sqrt{ \det \Sigma_{\mt{G}}}}  \(  2 \tr\( \( \Sigma_{\mt{G}}^{-1}d \Sigma_{\mt{G}} \)^2 \) +  \[ \tr\(  \Sigma_{\mt{G}}^{-1}d\Sigma_{\mt{G}} \)  \]^2 \) \,,
 \end{split}
 \end{equation}
 Taking the covariance matrix of the one-mode Gaussian state (see \eqref{CM1} and \eqref{CM_onemode2})
 \begin{equation}\label{CM_onemode}
 \Sigma_{\mt{G}} \( \hat{\rho}_1 \)
 = 
 \left(
 \begin{array}{cc}
 \frac{1}{\mu} e^{-2\bar{r}} \cosh 2 \alpha  & 0 \\
 0&  \mu e^{2\bar{r}} \cosh 2 \alpha\\
 \end{array}
 \right) \,,
 \end{equation}
 we obtain the Hilbert-Schmidt metric for Gaussian states 
 \begin{equation}
 \begin{split}
  ds^2_{\mt{HS}}
  &=\frac{1}{\cosh 2\alpha} \left(2 \tanh ^2(2 \alpha )d\alpha^2 +d\bar{r}^2\right) \,.
  \end{split}
 \end{equation}
 Comparing this with the quantum Fisher information metric for one-mode Gaussian state \eqref{infor_2D_metric}, one can easily show 
 \begin{equation}
  2ds^2_{\mt{IM} }(\delta \hr_1) -  ds^2_{\mt{HS}} (\delta \hr_1)  = 2\(1-\frac{\tanh ^22 \alpha}{\cosh 2\alpha}\) \,d\alpha^2 +\frac{4 \sinh ^2\alpha \(\cosh 2 \alpha+\cosh 4 \alpha+2\) }{\cosh 2\alpha(\cosh 4 \alpha+3)}\,d \bar{r}^2 \ge 0\,.
 \end{equation} 
 The difference on these two local measures implies we can always have the inequality for their geodesic distances with the same endpoints, \ie 
 \begin{equation}
 \mC_{\mt{HS}}\(\hat{\rho}_1', \hat{\rho}_1  \)  \le \mC_{\mt{IM}}\(\hat{\rho}_1', \hat{\rho}_1 \) \,,
 \end{equation}
 where the equality can be saturated if and only if $\hr_1', \hr_1$ are both pure states, \ie $\alpha_1'=0=\alpha_1$.

 \subsection{Bures Distance and Bures Metric}\label{sec:Bures}
 From the above comparisons after suitable normalizations, it is clear that no one can serve as the purification complexity of Fubini-Study complexity except for that from the quantum Fisher information metric. Although we introduce QFIM by considering Uhlmann's fidelity susceptibility, it can be also derived from a finite distance between two respective quantum states, \ie Bures distance, which is defined by \cite{bengtsson2017geometry,chruscinski2012geometric,bures1969extension}
 \begin{equation}
 \begin{split}
  D_{\mt{B}}\( \hr,\hs\)&= \sqrt{ \tr(\hr) +\tr (\hs) - 2F\(\hr,\hs \)} =   \sqrt{2\(1-F(\hr,\hs)\)} \,, \\
 \end{split}
 \end{equation}
 where we only consider normalized density matrices with $\tr\(\hr\)=1$ and the quantum fidelity $F\(\hr,\hs \)$ is given by \eqref{fidelity_def}. As before, it also reduces to the Fubini-Study distance for two pure states. From another definition of Bures distance \cite{chruscinski2012geometric}, \ie
 \begin{equation}
 \begin{split}
  D^2_{\mt{B}}\( \hr,\hs\) &=  \min\limits_{W_i}\, ||W_1-W_2||^2_{\mt{HS}} =  \min\limits_{W_i}\, \( \tr\( (W_1-W_2)^\dagger(W_1-W_2)\) \)\,, \\
  \end{split}
  \end{equation}
 where the minimization is taken over all Hilbert-Schmidt operator with $W^\dagger_1W_1 = \hr, W^\dagger_2W_2 = \hs$, it is clear that the Bures distance is the perfect analogue of Fubini-Study distance. Here we sketch the proof to show the above minimization results in the Uhlmann's fidelity, which also illustrates our motivation to choose the Uhlmann's fidelity. For arbitrary positive density matrix $\hr$, we can define a matrix $W$ such that  
 \begin{equation}
 W^\dagger W = \hr\,.
 \end{equation} 
  The matrix $W$ plays the role of the purification of $\hr$ and can be considered as a vector in Hilbert-Schmidt space. The freedom in purification is equivalent to the gauge symmetry $\hr = (UW)^\dagger(UW)$ with $U \in \text{U}(n)$.  A natural Euclidean distance between two vectors are defined by the root of 
  \begin{equation}
  ||W_1-W_2||^2_{\mt{HS}}= \tr(\hr) +\tr (\hs) - \( W_1^\dagger W_2 +W_1W_2^\dagger\)\,,
  \end{equation}
  with $W^\dagger_1W_1 = \hr\,,  W^\dagger_2W_2 = \hs$.
  The minimization for Bures metric between $\hr$ and $\hs$ is reduced to the maximization 
  \begin{equation}
   \max\limits_{W_i}\,\(\frac{1}{2}\tr\( W_1^\dagger W_2 +W_1W_2^\dagger \) \)=\max\limits_{W_i} \, \left|\tr\(W_1W_2^\dagger\)\right| = F\(\hr,\hs\) \,,
  \end{equation}
  which will be shown below to be the Uhlmann's fidelity.
 Applying the polar decomposition \footnote{Polar decomposition means that an arbitrary linear operator $W$ can be decomposed into product of unitary operator $U$ and positive operators such that $W= U\sqrt{W^\dagger W}= \sqrt{WW^\dagger}U$}, \ie $W_i= \sqrt{\hr_i} U_i$, one can get 
 \begin{equation}
 \tr\( W_1W_2^\dagger\) = \tr\( \sqrt{\hs}\sqrt{\hr} U_1U_2^\dagger \) \,. 
 \end{equation}
 Noting the existence of another polar decomposition $\sqrt{\hs}\sqrt{\hr}= \sqrt{\sqrt{\hs}\hr\sqrt{\hs}}U_{12}$\,, it is not hard to find that the special choice $U_{12}U_1U_2^\dagger= \mathbb{I}$  realizes the maximization with
 \begin{equation}
  F\(\hr,\hs\)= \tr\(\sqrt{\sqrt{\hs}\hat{\rho}\sqrt{\hs}}\)\,, 
 \end{equation}
 which is nothing but Uhlmann's fidelity \eqref{fidelity_def}. We also note the maximization condition also implies the two purifications are connected by the {\it geometric mean} (see \cite{erik} for more discussion about its application to the complexity of Gaussian states), \ie 
 \begin{equation}\label{optimal_vectors}
 \begin{split}
 \sqrt{\hs} W_1&=\sqrt{\sqrt{\hs}\hat{\rho}\sqrt{\hs}}\, U_2 \,,\\
  W_1 &=  \(\hs^{-\frac{1}{2}}\sqrt{\sqrt{\hs}\hat{\rho}\sqrt{\hs}}\,\hs^{-\frac{1}{2}} \) W_2\,.
 \end{split}
 \end{equation}
 where we have assumed the density matrices are positive definite to derive the second line. 
 Instead of the finite Bures distance between two density matrices, we prefer the geodesic distances on Riemannian geometry where geodesics can simulate the properties of optimal circuits. Then we focus on the infinitesimal metric from Bures distance, \ie Bures metric 
 \begin{equation}\label{def_Bures}
 ds^2_{\mt{B}} =_2  D^2_{\mt{B}}\( \hr,\hr+\delta \hr\) = 2\(1-\tr(\sqrt{\sqrt{\hat{\rho}}(\hr+\delta \hr)\sqrt{\hat{\rho}}}) \)\,,
 \end{equation}
which is equal to the quantum Fisher information metric or fidelity susceptibility of mixed states. Correspondingly, we can define the mixed-state complexity from the reference state $\hs_{\mt{R}}\( \lambda^\mu_0\)$ to the target state $\hat{\rho}_{\mt{T}}\( \lambda^\mu_1\)$ by 
\begin{equation}
\mathcal{C}_\mt{IM}\( \hs_{\mt{R}}, \hr_{\mt{T}} \) =\int_{\mt{R}}^{\mt{T}}\!\!    \sqrt{2\,ds^2_{\mt{IM}}}= \int_0^1\!\! d\s\,   \sqrt{ 2g_{\mu\nu}^{\mt{IM}}\, \dot{\lambda}^\mu\, \dot{\lambda}^\nu}\,,\qquad  \dot{\lambda}^\mu= \frac{d \lambda^\mu(\s)}{d\s }\,,
\end{equation}
which serves as the purification complexity as we have shown in previous sections.
Due to the appearance of two square roots of positive operators in the definition of quantum fidelity, the Bures metric is not easy to be written as a simple form of $\delta \hr$ like Schatten norms, While a more popular form for Bures metric or quantum Fisher information metric is taken as 
\begin{equation}\label{Bures_metric}
ds^2_{\mt{IM}}= \tr(G\hr G) = \frac 12 \tr\( Gd\hr\)\,,\\
\end{equation}
 where the hermitian operator $G$ known as {\it symmetric logarithmic derivative }is uniquely determined by the Laypunov equation, namely
 \begin{equation}
 d\hr = G\hr +\hr G\,. 
 \end{equation}
 In our coordinate system with $\hr(\lambda^\mu)$, the metric components read
 \begin{equation}
 g_{\mu\nu}^{\mt{IM}} = \frac{1}{2}\tr\big( \hr\left(  G_{\mu}G_{\nu}+G_{\nu}G_{\mu} \right)\big) \,, \qquad  d\hr = \partial_\mu \hr(\lambda) d \lambda^\mu\,, \quad G= G_\mu d\lambda^\mu\,,
 \end{equation}
 which is generally called {\it quantum Fisher information metric (matrix)} \cite{paris2009quantum}. 
 If we are restricted on pure states with $ \hr=\hr^2, \tr\hr =\tr \( \hr^2\)=1$, we can find 
 \begin{equation}
 \begin{split}
 d\hr &= \hr d\hr + d\hr \hr \,, \,,\\
  \tr\(  d\hr \)&=0=\tr\( \hr d\hr \)\,,\\
 \end{split}
 \end{equation}
 and obtain the symmetric logarithmic derivative as 
 \begin{equation}
 G= d\hr = \ket{d\psi} \bra{\psi} +  \ket{\psi} \bra{d\psi} \,.
 \end{equation}
 Correspondingly, the quantum Fisher information metric for pure states is simplified to be
 \begin{equation}
 \begin{split}
 ds^2_{\mt{IM}}&= \tr(G\hr G) = \tr\( \hr (d\hr)^2 \)=\frac 12 \tr\( d\hr d\hr\)  \equiv \frac 12 ds^2_{\mt{HS}}\,, \\
 &= \langle d\psi | d \psi \rangle - \langle \psi | d \psi \rangle \langle d\psi |  \psi \rangle \,.
 \end{split}
 \end{equation}
 which is nothing but the Fubini-Study metric as advertised in the introduction.  
  The "complexity" in calculations for quantum fidelity or Bures metric originates from the square root and also the non-commutation between $\hr $ and $\delta \hr$. If we focus on the special case where $\hr $, $\delta \hr$ commute, we can derive the explicit form for the QFIM by 
 \begin{equation}
 ds^2_{\mt{IM}}= \tr\( \(d\sqrt{\hr}\) ^2 \) = \frac{1}{4} \tr\( \hr^{-1} d\hr d\hr \)\,, 
 \end{equation}
 with 
 \begin{equation}
  \quad  [\delta \hr,\hr]=0\,, \qquad G= \frac{1}{2} \hr^{-1} d\hr\,.
 \end{equation}
 As expected, the QFIM in the above case actually reduces to the classical Fisher information matrix defined by
 \begin{equation}\label{classcial_Fisher}
 \begin{split}
 g_{\mu\nu} \( \lambda \) 
 &= \int dx \, P\(\lambda; x\) \frac{\partial^2 \ln P(\lambda; x)}{\partial \lambda^\mu  \partial \lambda^\nu} = \int dx \, P\(\lambda; x\) \frac{\partial \ln P(\lambda; x)}{\partial \lambda^\mu}\frac{\partial \ln P(\lambda; x)}{\partial \lambda^\nu} \,, \\
 \end{split}
 \end{equation}
 for any distribution $P(\lambda;x)$ on the parameter space $\lambda^\mu$.
 There are also some other simple forms for the quantum Fisher information metric like \cite{HUBNER1992239,paris2009quantum,Liu:2019xfr} 
 	\begin{equation}\label{Bures02}
 	\begin{split}
 	g_{\mu\nu}^{\mt{IM}} &= \frac{1}{2} \sum_{k,l}  \Re\( \frac{\bra{\psi_k} \partial_\mu \hr  \ket{\psi_l} \bra{\psi_l} \partial_\nu \hr  \ket{\psi_k}}{p_k +p_l} \) \,,\\
 	\end{split}
 	\end{equation}
 which is derived by rewriting the density matrix $\hr$ into the diagonal basis with $\hr= \sum_k p_k \ket{\psi_k}\bra{\psi_k}$. In the coordinate free form, we can obtain
 	\begin{equation}\label{Bures03}
 	\begin{split}
 	g_{\mu\nu}^{\mt{IM}} 
 	&= \frac{1}{2} \int^\infty_0 dt \tr\( e^{-\hr t} \partial_\mu \hr e^{-\hr t} \partial_\nu \hr\) \,,
 	\end{split}
 	\end{equation}
 which can be obtained by noticing the unique solution of Lyapunov equation, \ie 
 \begin{equation}
 d\hr = G\hr +\hr G\,, \qquad  G= \int^\infty_0 \(  e^{-t \hr} d\hr e^{-t \hr}   \)dt\,.
 \end{equation} 
 We present more details about these equivalent expressions in appendix \ref{sec:app_metric}.
 
 \subsection{Exercise: Single qubit}
   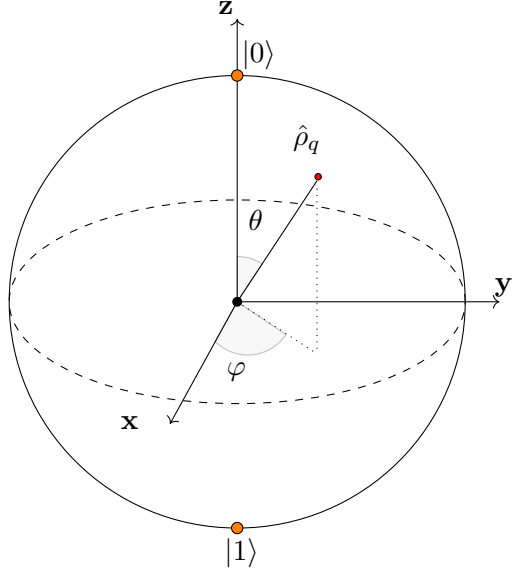
\begin{figure}[htbp]
  	\centering{ \begin{tikzpicture}[scale =1.5]
  		[line cap=round, line join=round, >=Triangle]
  		\clip(-2.19,-2.6) rectangle (2.66,2.8);
  		\draw [shift={(0,0)}, lightgray, fill, fill opacity=0.1] (0,0) -- (56.7:0.4) arc (56.7:90.:0.4) -- cycle;
  		\draw [shift={(0,0)}, lightgray, fill, fill opacity=0.1] (0,0) -- (-119:0.4) arc (-135.7:-33.2:0.4) -- cycle;
  		\draw(0,0) circle (2cm);
  		\draw [rotate around={0.:(0.,0.)},dash pattern=on 3pt off 3pt] (0,0) ellipse (2cm and 0.9cm);
  		\draw (0,0)-- (0.70,1.07);
  		\draw [->] (0,0) -- (0,2.5);
  		\draw [->] (0,0) -- (-0.59,-1.08);
  		\draw [->] (0,0) -- (2.3,0);
  		\draw [dotted] (0.7,1)-- (0.7,-0.46);
  		\draw [dotted] (0,0)-- (0.7,-0.46);
  		\draw (-0.18,-0.45) node[anchor=north west] {$\varphi$};
  		\draw (0.01,0.9) node[anchor=north west] {$\theta$};
  		\draw (-1.11,-0.92) node[anchor=north west] {$\mathbf {x}$};
  		\draw (2.17,0.3) node[anchor=north west] {$\mathbf {y}$};
  		\draw (-0.1,2.6) node{$\mathbf {z}$};
  		\draw (-0.2,-2) node[anchor=north west] {$\ket{1}$};
  		\draw (-0.05,+2.4) node[anchor=north west] {$\ket{0}$};
  		\draw (0.4,1.65) node[anchor=north west] {$\hr_q$};
  		\draw [fill=orange] (0,2.0) circle [radius=0.05];
  		\draw [fill=orange] (0,-2.0) circle [radius=0.05];
  		\scriptsize
  		\draw [fill] (0,0) circle (1.1pt);
  		\draw [fill=red] (0.71,1.105) circle (0.8pt);
  		\end{tikzpicture}}
  	\caption{Bloch ball consists all one qubit state $\hr_q$ with pure states as the Bloch sphere. The maximally mixed state locates at the center of sphere. North pole and south pole denotes the pure states $\ket{0}, \ket{1}$,  respectively. }\label{fig:qubit}
  \end{figure}

 In order to understand these different metrics, we discuss some results by taking single-qubit states (one fermionic mode) as a simple exercise. 
 The generic qubit state $\hr_{q}$ are parametrized by  a two-by-two matrix as \cite{chruscinski2012geometric}
 \begin{equation}
 \hr_q = \frac 12 \( \mathbb{I} + \vec{r}\cdot \vec{\sigma} \)=\frac{1}{2} \left(
 \begin{array}{cc}
 z+1 & x-i y\\
 x+i y & 1-z \\
 \end{array}
 \right)\,,
 \end{equation}
 with radial coordinate 
 \begin{equation}
 \vec{r}= \( x,y,z\)=\( r \sin \theta \cos \varphi, r \sin \theta \sin \varphi, r\cos \theta\), r=\sqrt{x^2+y^2+z^2} \le 1\,.
 \end{equation}
 In the above coordinate system, pure states are constrained by the condition $r=1$ and parametrized by 
 \begin{equation}
 \ket{\psi_q} = \cos \frac{\theta}{2} \ket{0} + e^{i \varphi}\sin \frac{\theta}{2} \ket{1}\,.
 \end{equation}
 All one qubit density matrices $\hr_q$ lie on or within the so-called Bloch ball as shown in figure \ref{fig:qubit}. The Bloch sphere is composed of all pure states with $r=1$.
 
  In order to derive the fidelity with respective two qubit states, we note a two-by-two matrix $M$ always satisfies 
 \begin{equation}
 M^2 -M\tr(M)  +\det M =0 \,, \qquad \(\tr(M)\)^2 = \tr\( M^2\) +2\det (M)\,,
 \end{equation}
 Taking $M= \sqrt{\sqrt{\hr_{q_1}}\hr_{q_2}\sqrt{\hr_{q_1}}}$ with arbitrary two qubit states $\hr_{q_1},\hr_{q_2}$,  it is easy to derive the explicit form for quantum fidelity 
 \begin{equation}
 	F(\hat{\rho}_{q_1}, \hr_{q_2})= \sqrt{ \tr(\hr_{q_1}\hr_{q_2}) +2\sqrt{\det \hr_{q_1}\det \hr_{q_2}}} \,.
 \end{equation}
 From the definition \eqref{def_Bures}, one can get the explicit forms 
 \begin{equation}
 F\( \hr_{q_1},\hr_{q_2}\) = \frac{1}{\sqrt{2}} \sqrt{ 1 + \vec{r}_1 \cdot \vec{r}_2 + \sqrt{(1-r_1^2)(1-r_2^2)}} \,,
 \end{equation}
 and the quantum Fisher information metric for single-qubit states as
 \begin{equation}\label{IM_qubit}
 \begin{split}
   ds^2_{\mt{IM}}\(\text{qubit} \)&=\frac 14 \(  (d\vec{r})^2 + \frac{ \(\vec{r} \cdot d \vec{r}\)^2}{1 -r^2}\)
   =\frac{1}{4} \( \frac{dr^2}{1-r^2} + r^2\( d\theta^2 +  \sin^2 \theta d\varphi^2  \)    \) \\
   &=\frac{1}{4} \( d \phi^2+ \sin^2 \phi  d\theta^2 +  \sin^2 \phi\sin^2 \theta d\varphi^2   \)   \,,
   \end{split}
 \end{equation}
 which is equal to a three-sphere with radius $\frac{1}{2}$ by redefining radial coordinate in Bloch ball as $r=\sin \phi$. 
 We can also rewrite this metric in the coordinate free form \cite{HUBNER1992239} 
 \begin{equation}
  ds^2_{\mt{IM}}\(\text{qubit} \)=\frac 14 \tr\( d\hr d\hr  + \(  d \sqrt{\det\hr}  \)^2\)   \,.
 \end{equation}
 It is interesting to note that the geometry for only pure states, \ie Fubini-Study metric for one qubit reduces to a two-dimensional sphere with radius $\frac{1}{2}$ as
 \begin{equation}
 \begin{split}
 ds^2_{\mt{FS}}
 =\frac{1}{4} \( d\theta^2 +  \sin^2 \theta d\varphi^2     \)\,. \\
 \end{split}
 \end{equation}
 See \cite{Brown:2019whu} for more discussion about the complexity geometry of pure states for a single qubit. 
 Here we would like to emphasize that the geodesic on the space of pure states with  Fubini-Study metric is the same as that in the full one-qubit space defined in \eqref{IM_qubit}. Because it is easy to find the extra equation of motion associated with the QFIM in $\eqref{IM_qubit}$ 
 \begin{equation}
  \ddot{r}(1-r^2) + 2r\dot{r}^2- r(1-r^2)^2\( \dot{\theta}^2 + \dot{\varphi}^2\sin^2 \theta  \)=0\,,
 \end{equation}
 always admits the trivial solution with $r(\s)=1$. Therefore, the geodesic connecting two pure states with respect to the quantum Fisher information metric actually moves on the Bloch sphere with only passing through pure states.  Correspondingly, we can also find the equivalence 
 \begin{equation}
 \mC_{\mt{IM}}\(  \ket{\psi_{q_1}},  \ket{\psi_{q_2}} \) =  \mC_{\mt{FS}}\(  \ket{\psi_{q_1}},  \ket{\psi_{q_2}} \) \,,
 \end{equation}
as what we also found for one-mode Gaussian state in \eqref{IM=FS}.
 
 In order to derive the Schatten norms between $\hr_{q_1}, \hr_{q_2}$, it is useful to notice the two singular values of matrix $(\hr_{q_1}-\hr_{q_2})$ are degenerate and read
 \begin{equation}
 s_1 =s_2= \frac{1}{2}\sqrt{(x_1-x_2)^2+(y_1-y_2)^2+(z_1-z_2)^2} = \frac{1}{2}|\vec{r}_1 -\vec{r}_2|\,. 
 \end{equation}
 Directly, we can find the finite distances from normalized Schatten norms between two qubit states same as
\begin{equation}
\frac{1}{2^{1/p}}||\hr_{q_1}-\hr_{q_2}||_p = \frac{1}{2}\( |\vec{r}_1 -\vec{r}_2|^p \)^\frac{1}{p} \,.
\end{equation}
 It obviously leads us to the same flat metric 
 \begin{equation}
 ds^2_p\( \text{qubit}\) = ds^2_{\mt{HS}} = 2 ds^2_{\mt{Tr}} = \frac{1}{4} \(  d\vec{r}\cdot d\vec{r}  \)=\frac{1}{4} \(dr^2 + r^2 d\theta^2 +  r^2\sin^2 \theta d\varphi^2     \) \\ \,.
 \end{equation}
 which is always smaller than that from the Bures metric as shown in \eqref{inequality_rhoA2}. This single qubit example is illuminating because the results can be generalized to more generic (finite) projective Hilbert space $\mathbb{CP}^{\mt{N}}$, see \cite{bengtsson2017geometry} for more details. 
%%%%%%%%%%%%%%%%%%%%%%%%%% 
%%%%%%%%%%%%%%%%%%%%%%%%%% 
%%%%%%%%%%%%%%%%%%%%%%%%%% 
%%%%%%%%%%%%%%%%%%%%%%%%%% 
%%%%%%%%%%%%%%%%%%%%%%%%%% 
%%%%%%%%%%%%%%%%%%%%%%%%%% 
 \section{Discussion}
We summarize the main results and discuss several interesting questions about purification complexity $\mC_{\mt{IM}}$ and holographic complexity as the future directions.  
   \begin{flushleft}
   \textbf{Summary of results}
   \end{flushleft}
   In this paper, we generalize the Fubini-Study method towards complexity to generic quantum states by using the quantum Fisher information metric $g_{\mu\nu}^{\mt{IM}}$. Due to Uhlmann's fidelity \eqref{Uhlmann_theorem}, we find that the complexity $\mC_{\mt{IM}}$ defined in \eqref{def_IFcomplexity} between arbitrary two quantum states exactly equals the purification complexity measured by Fubini-Study metric (or QFIM) on the extended Hilbert space for purified states, \ie \eqref{equality_rhoA2}
   \begin{equation}
   \begin{split}
   \mC_{\mt{IM}} \(\hs_{\mA},\hat{\rho}_\mA \)  
   &= \min\limits_{\Phi, \Psi} \, \mathcal{C}_{\mt{FS}}\( \ket{\Phi_{\mA\mA^c}} ,\ket{\Psi_{\mA\mA^c}} \)=  \min\limits_{\Phi,\Psi} \, \mathcal{C}_{\mt{IM}}\( \ket{\Phi_{\mA\mA^c}} ,\ket{\Psi_{\mA\mA^c}} \)\,.
   \end{split}
   \end{equation}
  Without explicitly introducing the auxiliary system and purifying the mixed states, our method avoids the challenging minimization over all purifications. This equivalence is illustrated by the example from Gaussian mixed states in \eqref{one_mode_equal}. 
  Furthermore, we also prove that this purification complexity $\mC_{\mt{IM}}$ is always non-increasing under any quantum operations such as partial trace in \eqref{complexity_nonincrease}. From this monotonicity, we also show the mutual complexity $\Delta\mC_{\mt{IM}}$ cannot be either subadditive or superadditive in general. 
  	\begin{flushleft}
   {\bf  Inequalities of purification complexity}
   \end{flushleft}
  	It is well-known that the entanglement entropy satisfies the subadditivity 
  	\begin{equation}
  	 S_{\mathrm{vN}}\(\hr_{\mA}\) +  S_{\mathrm{vN}}\(\hr_{\mB}\)  \ge  S_{\mathrm{vN}}\(\hr_{\mA\mB}\) \,,
  	\end{equation}
  	and also the strong subadditivity 
  	\begin{equation}
     \begin{split}
     S_{\mathrm{vN}}\(\hr_{\mA\mB}\) + S_{\mathrm{vN}}\(\hr_{\mB{\scriptscriptstyle\mathcal{C}}}\)  &\ge   S_{\mathrm{vN}}\(\hr_{\mA}\) + S_{\mathrm{vN}}\(\hr_{\scriptscriptstyle\mathcal{C}}\)  \,,\\
     S_{\mathrm{vN}}\(\hr_{\mA\mB}\) + S_{\mathrm{vN}}\(\hr_{\mB{\scriptscriptstyle\mathcal{C}}}\)  &\ge   S_{\mathrm{vN}}\(\hr_{\mA\mB{\scriptscriptstyle\mathcal{C}}}\) + S_{\mathrm{vN}}\(\hr_{\mB}\)  \,.\\
     \end{split}
  	\end{equation} 
  	 Taking the monotonicity of complexity under the partial trace (noting the similar monotonicity for von Neumann entropy  $S_{\mathrm{vN}}\(\hr_{\mA\mB}\) \ge S_{\mathrm{vN}}\(\hr_{\mB}\) $ is not true.), it is direct to show the counterpart of the first strong subadditivity for the purification complexity $\mC_{\mt{IM}}$ is also satisfied for generic density matrices, \ie 
  	\begin{equation}
  	 \mC_{\mt{IM}}\(\hs_{\mA\mB},\hr_{\mA\mB}\) + \mC_{\mt{IM}}\(\hs_{\mB{\scriptscriptstyle\mathcal{C}}},\hr_{\mB{\scriptscriptstyle\mathcal{C}}}\)  \ge   \mC_{\mt{IM}}\(\hs_{\mA},\hr_{\mA}\) + \mC_{\mt{IM}}\(\hs_{\scriptscriptstyle\mathcal{C}},\hr_{\scriptscriptstyle\mathcal{C}}\)  \,. \quad \( \cmark\)
  	\end{equation}
  	However, the second strong subadditivity for purification complexity 
  \begin{equation}
  \mC_{\mt{IM}}\(\hs_{\mA\mB},\hr_{\mA\mB}\) + \mC_{\mt{IM}}\(\hs_{\mB{\scriptscriptstyle\mathcal{C}}},\hr_{\mB{\scriptscriptstyle\mathcal{C}}}\)  \ge   \mC_{\mt{IM}}\(\hs_{\mA\mB{\scriptscriptstyle\mathcal{C}}},\hr_{\mA\mB{\scriptscriptstyle\mathcal{C}}}\) + \mC_{\mt{IM}}\(\hs_{\mB},\hr_{\mB}\) \,. \qquad \(\xmark\)
  \end{equation}
  is not obeyed by a general pair of density matrices $\hs_{\mA\mB{\scriptscriptstyle\mathcal{C}}},\hr_{\mA\mB{\scriptscriptstyle\mathcal{C}}}$ in a tripartite system. Because this strong subadditivity can be reduced to the subadditivity $\Delta \mC_{\mt{IM}} \ge 0$, \ie 
  	\begin{equation}
  	\mC_{\mt{IM}}\(\hs_{\mA},\hr_{\mA}\)+ \mC_{\mt{IM}}\(\hs_{\mB},\hr_{\mB}\) \ge \mC_{\mt{IM}}\(\hs_{\mA\mB},\hr_{\mA\mB}\) \,. \qquad \(\xmark\)
  	\end{equation}
  	 which does not always hold since we have found the existence of a counterexample as \eqref{counterexample}. However, we should also note that it is still possible to fix the additivity of the purification complexity by choosing a special reference state such as unentangled state (see figure \eqref{fig:mutualcomplexity_IM}) and maximally entangled state (see \eqref{mutual_max}). Especially, if we believe the holographic complexity may require a specific and trivial state as the universal reference state, it is still interesting to investigate whether "holographic complexity" is subadditive or superadditive. 
\begin{flushleft}
{ \bf Towards the complexity of generic quantum states in QFT}
\end{flushleft}
  As the first application of the purification complexity $\mC_{\mt{IM}}$, we only examine an extremely simple Gaussian state $\hr_1$ in section \ref{sec:application}. However, the proposal \eqref{def_IFcomplexity} is inspired by the Fubini-Study method for the complexity of pure states in QFT. Considering the purification complexity $\mC_{\mt{IM}}$ has gotten rid of the challenges in finding the optimal purification, it looks straightforward to apply the definition of purification complexity to a generic quantum state in QFT, \eg the reduced density operator for a subregion in the vacuum state of QFT. Although it is not easy to calculate the quantum fidelity \eqref{fidelity_def} or quantum Fisher information metric for arbitrary QFT states due to the appearance of the square root of the density operators, it is not so hard for a free quantum field theory. As a generalization of the Gaussian state $\hr_1$, one can consider the most general multimode Gaussian states $\hr_{G}$ defined by 
  \begin{equation}
  \hr_{\mt{G}} =  \frac{e^{-\beta \hat{H}_{\mt{G}}}}{\tr\(  e^{-\beta \hat{H}_{\mt{G}}} \)} \,,
  \end{equation} 
  where $\hat{H}_{\mt{G}}$ represents any quadratic Hamiltonian with $N$ bosonic (or fermionic) modes. Focusing on the free quantum field theory on a lattice, one can find its ground state, thermal states, and even reduced density operators for a subregion can be recast as Gaussian states $\hr_{\mt{G}}$ (see \eg \cite{qft1,purification,Chapman:2018hou}). With some efforts, the quantum fidelity and Bures metric for arbitrary two Gaussian (bosonic or fermionic) states have been derived in \eg  \cite{Gaussian_fidelity,Gaussian_fidelity02,carollo2018uhlmann,PhysRevA.71.032336}. With the knowledge of the QFIM for any Gaussian states, it is interesting to consider the purification complexity for thermal states and mixed Gaussian states in a free QFT as what has been discussed in \cite{purification} \footnote{Different from the purification complexity defined in \eqref{def_IFcomplexity}, the authors in \cite{erik} develop a direct way to calculate the complexity for arbitrary Gaussian states with taking the Fisher-Rao metric as the complexity measure.}. Rather than assuming a free theory, the author in \cite{Kirklin:2019ror} calculated the Uhlmann's fidelity for two holographic states based on a replica trick. It would be intriguing to further investigate the QFIM and purification complexity based on that result. 
 
\begin{flushleft}
	 { \bf Is quantum Fisher information metric holographic? } 
\end{flushleft}
  Our proposal for the purification complexity $\mC_{\mt{IM}}$ defined in \eqref{def_IFcomplexity} is based on the special properties of quantum Fisher information metric that is identified as Uhlmann's fidelity susceptibility. Recetly, different concepts associated with the geometry of quantum states also attract more attention in the field of AdS/CFT, see \eg \cite{Lashkari:2015hha,MIyaji:2015mia,Trivella:2016brw,Bak:2017rpp,Czech:2017zfq,Alishahiha:2017cuk,Banerjee:2017qti,Moosa:2018mik,Czech:2018kvg,Belin:2018fxe,Suzuki:2019xdq,Kirklin:2019ror,Erdmenger:2020vmo}.
  In \cite{MIyaji:2015mia} the authors consider the fidelity susceptibility (\ie the Fubini-Study metric) of ground states in CFT from a small perturbation by a primary operator and argue its gravity dual is the volume of maximal time slice in an AdS spacetime. As its generalization to mixed states, the authors in \cite{Banerjee:2017qti} think the holographic dual of the Fisher information for mixed states should be given by a regularized volume contained under the RT surface. Differently, the authors in \cite{Lashkari:2015hha} show the Fisher information metric for the vacuum density matrix in holographic CFT is dual to the canonical energy metric associated with the Rindler wedge in AdS. We should also note the Fisher information discussed in \cite{Lashkari:2015hha,Banerjee:2017qti} is derived from the second variation of relative entropy, as opposed to QFIM from Uhlmann's fidelity. But they are both reduced to the classical Fisher information when $\hr, \delta \hr$ commute as indicated in \eqref{classcial_Fisher}. Although these gravity duals have passed some quantitative tests in the vacuum states of CFT, it is also necessary to check different proposals by considering the realization of quantum information metric's universal properties (such as non-increase under any quantum operations) in the bulk geometry. On the other hand, it is also interesting to generalize these proposals from the vacuum state with $\lambda=0$ to more generic states with $g_{\mu\nu}^{\mt{IM}}\(\lambda \)$. In light of different proposals, a natural question arises: \\
  
  {\it For holographic states $\hr_{\mA}(\lambda)$ in the Hilbert space $\mH_\mA$, does quantum Fisher information $g_{\mu\nu}^{\mt{IM}} d\lambda^\mu d\lambda^\nu$ have a holographic dual in the bulk?} \\
  
  To be more specific, we have learned the holographic density operator for a subregion $\bmA$ is constrained by the modular Hamiltonian $\hK_{\mA}$ in the form like \cite{Jafferis:2015del}
   \begin{equation}
  -\log \hr_{\mA} \equiv \hK_{\mA} = \frac{\hat{A}_{\mathrm{ext}}\(\mathcal{E}_{\mA}\)}{4\Gn} + \hK_{\mathrm{bulk}} + \cdots +\mathcal{O}(\Gn) \,,
  \end{equation}
  where the $\hat{A}_{\mathrm{ext}}$ denotes the area operator associated with the extremal surface $\mE_{\mA}$ and $ \hK_{\mathrm{bulk}}$ is the bulk modular Hamiltonian of the bulk region enclosed by $\mE_{\mt{A}}$. Although the modular Hamiltonian generally is nonlocal and not easy to be derived except for several local cases \cite{Blanco:2013joa}, we can focus on a simple configuration for the holographic states $\ket{\Psi(\lambda)}$. Starting from the path integral for the ground state in  holographic conformal field theory on the boundary, we can turn on some sources in that path integral by inserting local (or global) Hermitian operators $\mathcal{O}^\mu$ and then obtain a natural class of excited states as 
  \begin{equation}
  \left\langle\varphi(\mathbf{x}) | \Psi(\lambda)\right\rangle=\int^{\phi\left(0, \mathbf{x}\right)=\varphi(\mathbf{x})} \mathcal{D} \phi \, e^{-\int_{-\infty}^{0} d \tau \int d^{d-1} \mathbf{x}\left(\mathcal{L}_{\mt{CFT}}[\phi]-\lambda^\mu \mO_{\mu}(x)\right)}
  \end{equation}
  whose gravity dual in an asymptotically AdS spacetime geometry corresponds to coherent states of classical bulk fields dual to $\mO^\mu$ \cite{Botta-Cantcheff:2015sav,Marolf:2017kvq,Arias:2020qpg}. Then the holographic mixed states $\hr_{\mA}(\lambda)$ we are interested in can be constructed by the path integral for a reduced density matrix with a cut along subregion $\bmA$. In that configuration, the calculations for the quantum Fisher information metric is actually related to the correlation functions of operator $\mO^\mu$ on these excited states.  
 \begin{flushleft}
 	  {\bf  Is purification complexity holographic?}
 \end{flushleft} 
  Besides these holographic proposals to quantum Fisher information, most recent research on complexity is motivated by holographic conjectures such as complexity=volume (CV) \cite{Susskind:2014rva,Stanford:2014jda} and also complexity=action (CA) \cite{Brown:2015bva}.  As extensions of these conjectures for pure states to mixed states, the gravitational dual of the mixed-state complexity associated with reduced density operators for subregions on the boundary of asymptotically AdS spaces are proposed to be subregion volume=complexity (sub-CV) \cite{Alishahiha:2015rta,Carmi:2016wjl} and subregion action=complexity (sub-CA) \cite{Carmi:2016wjl}. For more studies on that direction, see \eg \cite{Ben-Ami:2016qex,Agon:2018zso,Caceres:2018blh,Alishahiha:2018lfv,Braccia:2019xxi,purification,Abt:2018ywl} and references therein. Based on these proposals for subregion complexity, \cite{Caceres:2018blh} has studied the additivity properties and examined whether they are holographic purification complexity, \ie the minimum holographic complexity (CA or CV) among all holographic purifications \footnote{In \cite{Agon:2018zso, Caceres:2018blh}, the definition of purification complexity is restricted on all purifications of $\hr_{\mA}$ with no separable factors which are also purifications. In our definition, we do not need to impose this constrain because of the non-increasing of $\mC_{\mt{IM}}$.}. Instead of starting from holographic conjectures for complexity, we would like to ask another question standing on the boundary: \\
  
  {\it In a given Hilbert space $\mH_\mA$ of holographic states $\hr_{\mA}(\lambda)$, is purification complexity $\mC_{\mt{IM}}\( \hs_{\mA},\hr_{\mA}\)$ from the quantum Fisher information metric holographic?}\\
  
  We hope to get back to these questions in the near future.

	\acknowledgments
	It is a pleasure to thank Giuseppe Di Giulio, Juan Hernandez and Erik Tonni for initial collaboration and many useful discussions on this project. I also thank Bartlomiej Czech and Ziwen Liu for conversations. Especially, I would like to thank Shao-Jiang Wang for carefully reading the manuscript and my supervisor, Rob Myers for many helpful comments and suggestions on the draft.  Research at Perimeter Institute is supported in part by the Government of Canada through the Department of Innovation, Science and Economic Development Canada and by the Province of Ontario through the Ministry of Economic Development, Job Creation and Trade.

	\begin{appendix}

	\appendix

%%%%%%%%%%%%%%%%%%%%%%%%%%%%%%%%%%%%

\section{Background}\label{sec:app_QI}
 In this appendix, we provide a minimal introduction to some notations and terminologies in quantum information, which are used in the main content. 
\subsection{Gaussian State and Covariance Matrix}
 It is known that the any Gaussian states $\hr_{\mt{G}}$ can be equivalently described by its covariance matrix $\Sigma_{\mt{G}}$. See \cite{RevModPhys.84.621,ferraro2005gaussian,serafini2017quantum} for more details about Gaussian states. 
 Considering any Gaussian state with $N$ modes,  we can find $N$ pairs of the standard self-adjoint canonical operators $\hat{x}_i, \hat{p}_i$ with the canonical commutation relations 
\begin{equation}
 [\hat{x}_i, \, \hat{p}_j] = i\delta_{ij} \hbar\,, \qquad \hbar=1\,, \quad \hat{a}_{i} =\frac{\hat{x}_i+i\hat{p}_i}{\sqrt{2}}\,,
\end{equation}
whose vector form is defined to be
\begin{equation}
 [\hat{\R}, \, \hat{\R}^\intercal] =i \Omega \equiv i  \bigoplus_{k=1}^n \Omega_k\,,\quad \Omega_k=\left(
 \begin{array}{cc}
 0 & 1 \\
 -1 & 0 \\
 \end{array}
 \right) \,, \quad \hat{\R}= \(\hat{x}_1, \hat{p}_1, \hat{x}_2,\hat{p}_2,\cdots \hat{x}_n,\hat{p}_n \)^\intercal\,, 
\end{equation}
where $\Omega$ is the symplectic form satisfying $\Omega^\intercal\Omega =-\Omega^2= \mathbb{I}_{2n}$. 
%One can also re-order the basis to $ \mathbf{S}= \(\hat{x}_1, \hat{x}_2,\cdots \hat{x}_n,\hat{p}_1,\hat{p}_2\cdots \hat{p}_n \)^T$ with the new canonical commutation relations 
%\begin{equation}
%[\hat{\mathbf{S}}, \, \hat{\mathbf{S}}^T] = i  J = i \left(
%\begin{array}{cc}
%0 & 1 \\
%-1 & 0 \\
%\end{array}
%\right)  \otimes \mathbb{I}_n \,,\qquad J=\left(
%\begin{array}{cc}
%0_n & \mathbb{I}_{n} \\
%-\mathbb{I}_{n} & 0_n \\
%\end{array}
%\right) \,.
%\end{equation}
The {\it covariance matrix} (CM) $\Sigma_{\mt{G}}$ of any Gaussian state $\hat{\rho}_{\mt{G}}$ is defines as 
\begin{equation}\label{CM}
\begin{split}
\(\Sigma_{\mt{G}}\)_{ij} &\equiv  \tr\( \hr_{\mt{G}}  \left\{\(\hat{\R} -\langle\hat{\R} \rangle\),\(\hat{\R} -\langle\hat{\R} \rangle \)^\intercal  \right\}\)=\left\langle \hat{\R}_i  \hat{\R}_j + \hat{\R}_j \hat{\R}_i\right\rangle -2\langle  \hat{\R}_i \rangle \langle  \hat{\R}_j \rangle\,.\\
\end{split}
\end{equation}
For example, the covariance matrix $\Sigma_1$ for a single mode reads 
\begin{equation}
\Sigma_1 =  2 \left(
\begin{array}{cc}
\left\langle \hat{x}^2\right\rangle -\langle \hat{x}\rangle ^2 & \langle \{ \hat{x},\hat{p}\}\rangle-\langle \hat{p}\rangle  \langle \hat{x}\rangle  \\
\langle \{ \hat{x},\hat{p}\}\rangle-\langle \hat{p}\rangle  \langle \hat{x}\rangle   & \left\langle \hat{p}^2\right\rangle -\langle \hat{p}\rangle ^2    \\
\end{array}
\right)\,,
\end{equation}
The physical Gaussian state $\hat{\rho}_{\mt{G}}$ with covariance matrix $\Sigma_{\mt{G}}$ should also satisfy the uncertainty principle 
\begin{equation}
 \Sigma_{\mt{G}} + i \Omega \ge 0\,,
\end{equation}
which is invariant under the symplectic transformations. For a single-mode quantum state, the physical constrains are reduced to 
\begin{equation}
\det \Sigma_1 \ge 1 \,, \qquad \Sigma_1 \ge 0\,.
\end{equation}
As a consequence of Williamson theorem, the covariance matrix $\Sigma_{\mt{G}}$ of the most general Gaussian state $\hat{\rho}_{\mt{G}}$ can be decomposed as
\begin{equation}
\Sigma_{\mt{G}} \( \hr_{\mt{G}}\)= S \bigoplus_{k=1}^n 
\left( \begin{array}{cc}
 \nu_k  & 0 \\
 0& \nu_k \\
 \end{array}
 \right) S^\intercal\,, \qquad  S \in \text{Sp}(2n, \mathbb{R})\,, \qquad \nu_k \ge 1 \,.
\end{equation}
where  $\nu_k$ are the symplectic eigenvalues of CM. Correspondingly, we can also obtain the decomposition of generic Gaussian state \cite{RevModPhys.84.621,ferraro2005gaussian,serafini2017quantum} 
\begin{equation}
\begin{split}
\hat{\rho}_{\mt{G}} &= \hat{D}^\dagger \hat{S}^{\dagger} \( \bigotimes^n_k \hat{\upsilon}_{\mathrm{th}}(\beta_k, \omega_k)  \)\hat{D}\hat{S} \,, \\
\end{split}
\end{equation}
where $\hat{D}, \hat{S}$ denote the displacement operator and squeezing operator, respectively, and $\hat{\upsilon}_{\mathrm{th}}$ is the thermal density matrix defined in \eqref{density_thermal} with the inverse temperature $\beta_k \omega_k$ associated with symplectic values of the covariance matrix  by $\nu_k= \coth \( \frac{\beta_k\omega_k}{2}\) =  \cosh 2\alpha_k$. 
From the covariance matrix, we can easily distinguish pure Gaussian states and mixed Gaussian states by considering its determinant, \ie 
\begin{equation}
\det \( \Sigma \) = 
\begin{cases}
+1\,, \text{pure},\\
>1\,, \text{mixed}\,.\\
\end{cases}
\text{with} \qquad    \tr\( \hat{\rho}_{\mt{G}}^2 \)= \prod^n_{k=1} \frac{1}{\nu_k} = \frac{1}{\sqrt{\det \Sigma_{\mt{G}}}} \,.
\end{equation}
For later use, one can also find
\begin{equation}
\tr\( \hat{\rho}_{\mt{G}} \hat{\rho}_{\mt{G}}' \)=  \frac{1}{\sqrt{\det \frac 12 \(  \Sigma_{\mt{G}}+ \Sigma_{\mt{G}}' \) } } \,.
\end{equation}
In this paper, we also discuss the partial trace with discarding some modes in the full system. In the representation of Gaussian states with the vector $\hat{\R}$ and covariance matrix, it is easy to see the action for tracing out a subsystem.   
Diving the N-mode system $\hr_{\mA\mB}$ into two parts with a $n$-mode system $\bmA$ and a $m$-mode system $\bmB$,  we can decompose the covariance matrix in the way like
\begin{equation}\label{AB_CM}
\Sigma_{\mA\mB}=\left(
\begin{array}{cc}
\Sigma_{nn} & \Sigma_{nm} \\
\Sigma_{mn} & \Sigma_{mm} \\
\end{array}
\right) \,, \qquad 
 \hat{\R}_{\mA\mB}=\left(
\begin{array}{cc}
\hat{\R}_{n}\\
\hat{\R}_{m} \\
\end{array}
\right) \,,
\end{equation}
where the $\Sigma_{nm}$ denotes a $2n$-by-$2m$ matrix. Then the reduced density matrix $\hr_{\mA} =\tr_{\mB}\( \hr_{\mA\mB}\)$ for the subsystem $\bmA$ is easily obtained by 
\begin{equation}
\Sigma_{\mA} =\Sigma_{nn} \,, \qquad \text{with} \quad  \hat{\R}_{\mA} = \hat{\R}_{n}\,.
\end{equation}

%%%%%%%%%%%%%%%%%%%%%
%%%%%%%%%%%%%%%%%%%%%
%%%%%%%%%%%%%%%%%%%%%
\subsection{Density Matrix and Covariance Matrix for $\hr_1$}
As the simplest Gaussian state, we consider the one-mode Gaussian state $\hr_1$ in section \ref{sec:application} and define its density operator as 
\begin{equation}
\hat{\rho}_1 = \hat S_1(r) \hat{\upsilon}_{\mathrm{th}} (\beta,\omega)\hat S_1^\dagger(r) \,,
\end{equation}  
whose density matrix function can be rewritten in general as a Gaussian function, \ie  
\begin{equation}\label{Gaussian_1mode}
\rho_1(x,x') = \sqrt{\frac{a -b}{\pi}}  \exp \( -\frac12 \(ax^2 + a x'^2\) + bxx' \) \,.
\end{equation}
The two representations are connected by the transformation (see the section 2 in \cite{purification})
\begin{equation}\label{transformation}
a= e^{2r} \omega \coth \beta \omega \,, \qquad  b = \frac{e^{2r} \omega }{\sinh \beta \omega }\,.
\end{equation}
It is also convenient to derive the fidelity of Gaussian states by considering the density matrix function. For example, it is obvious to find the covariance matrix $\Sigma_1$ of $\hr_1$ as 
\begin{equation}\label{CM1}
\Sigma_1
=\left(
\begin{array}{cc}
\frac{1}{a-b} & 0 \\
0& a+b \\
\end{array}
\right)\,,
\end{equation}
which equals the form \eqref{CM_onemode} by the transformation \eqref{transformation}.
Explicitly, we need to define the multiplication between two density matrices and its square root from 
\begin{equation}
\bra{x'} \hat{\rho}_1\hat{\rho}_1' \ket{x} = \int^\infty_{-\infty} \rho_1(x',y) \rho_1'(y,x) dy,
\end{equation}
and 
\begin{equation}
\bra{x'} \hat{\rho}_1 \ket{x} =\bra{x} \sqrt{\hat{\rho}_1} \sqrt{\hat{\rho}_1} \ket{x} = \int^\infty_{-\infty } \sqrt{\rho_1}(x',y) \sqrt{\rho_1}(y,x) dy. 
\end{equation}
From the simple Gaussian integral, one can derive the square root of a Gaussian density matrix \eqref{Gaussian_1mode} as
\begin{equation}
\sqrt{\rho_1}(x,x' ) = \exp\(-\frac{a+b}{2} \(x^2 +x'^2\) + \sqrt{2b(a+b)}   xx'  + C\)\,.
\end{equation}
where $C$ is the normalization factor. Hence, the quantum fidelity can be obtained from its definition \eqref{fidelity_def} by the Gaussian integral as \cite{gaussian} 
\begin{equation}
\begin{split}
F(\hr_1, \hr_1')  &= \sqrt{\frac{2}{\sqrt{\det{(\Sigma_1+\Sigma_1')} +(\det{\Sigma_1}-1)(\det{\Sigma_1'}-1)}-\sqrt{(\det{\Sigma_1}-1)(\det{\Sigma_1'}-1)}} } \\
\end{split}
\end{equation}
For pure states with $\det \Sigma_1 =1$, the quantum fidelity is simplified as 
\begin{equation}
\begin{split}
F(\ket{\psi_1}, \ket{\psi_1'})  = \sqrt{  \frac{2}{\sqrt{\det{(\Sigma_1+\Sigma_1')} }}} \,.\\
\end{split}
\end{equation} 
For the most general one-mode Gaussian states $\hr_1$ defined in \eqref{Gaussian_1mode} or \eqref{one_mixed}, the quantum fidelity between two Gaussian states is expressed as 
\begin{equation}\label{simple-fidelity}
F(a_1,b_1;a_2,b_2)= \sqrt{\frac{2}{\sqrt{\frac{\left(a_1+a_2+b_1-b_2\right) \left(a_1+a_2-b_1+b_2\right)}{\left(a_1-b_1\right) \left(a_2-b_2\right)}}-2 \sqrt{\frac{b_1 b_2}{\left(a_1-b_1\right) \left(a_2-b_2\right)}}}}
\end{equation}
and the quantum Fisher information metric from Uhlmann's fidelity is the following 
\begin{equation}
ds^2 = \frac{1}{8} \( \frac{1}{(a-b)^2} da^2 - \frac{2}{(a-b)^2} da db +\(\frac{1}{(a-b)^2}  +\frac{1}{ab} \)  db^2  \),
\end{equation}
which is the same as the result \eqref{metric_3d} used in the main text by taking the transformation \eqref{transformation}. 
%%%%%%%%%%%%%%%%%%%%%%%%%%%
%%%%%%%%%%%%%%%%%%%%%%%%%%%
%%%%%%%%%%%%%%%%%%%%%%%%%%%

\subsection{Quantum Operation (Quantum Channel)}\label{sec:app_operation}
 It is obvious that we can use unitary operations to realize the transformations from a pure state to another one. For a generic quantum state $\hr_{\mA}$ in a principle system $\bmA$, we need to introduce a more general transformation beyond unitraies as 
 \begin{equation}\label{quantum_operations}
  \hr_{\mA}' =\mE\( \hr_{\mA} \) \,,
 \end{equation}
 where the map $\mE$ is called a quantum operation. In the literatures of quantum computation, a quantum operation is also called a \text{\it quantum channel} \footnote{In some literatures, the term "quantum operation" specifically refers to completely positive (CP) and non-trace-increasing maps on the space of density matrices. Instead the term "quantum channel" refers to CPTP. In this paper, we only consider CPTP and it is referred to as "quantum operation".}. 
 With one more terminology used in the main content, we only focus on the quantum operation defined as the {\it completely positive trace-preserving map} (CPTP map) 
 \begin{equation}
 \mathcal{E} : \hr_{\mA} \longrightarrow \mathcal{E}\( \hr_{\mA}\)\,,
 \end{equation}
 with $\tr\(\hr_\mA \)=\tr\(\mE\(\hr_\mA\) \)$.  As it is known, \eg \cite{NielsenChuang,wilde2013quantum}, the \text{\it quantum operation} formalism \eqref{quantum_operations} can be represented in different but equivalent ways. 
 For example, we can consider the quantum operation $\mathcal{E}\( \hr_{\mA}\)$ on density operators $\hr_{\mA}$ as the unitary transformation with ancillae (or environment) in the extended Hilbert space $\mathcal{H}_{\mA}\otimes \mathcal{H}_{\mA^c} $, \ie 
\begin{equation}
\mathcal{E}\( \hr_{\mA}\)= \tr_{\mA^c} \( U_{\mA\mA^c}\(  \hat{\rho}_{\mA}\otimes \hr_{\mA^c}\)U^\dagger_{\mA\mA^c}  \)\,,
\end{equation}
where the density operator $\hr_{\mA^c}$ denotes an initial state for the auxiliary system and $\tr_{\mA^c}$ traces out the ancilla part. Taking the initial state as any pure state $\ket{\psi_0}$ in its orthogonal basis, it is easy to find that the reduced density operator after tracing out  $\mH_{\mA^c}$ reads 
\begin{equation}\label{operator_sum}
\begin{split}
\mathcal{E}\( \hr_{\mA}\)&= \tr_{\mA^c} \( U_{\mA\mA^c}\(  \hat{\rho}_{\mA}\otimes \ket{\psi_0}\bra{\psi_0}\)U_{\mA\mA^c}^\dagger  \)\equiv  \sum_{k} \hat{M}_k \hr_\mA \hat{M}_k^\dagger \,,\\
\end{split}
\end{equation}
with $\hat{M}_k\equiv \bra{\psi_k} U_{\mA\mA^c}\ket{\psi_0}$ defined as the operation elements for this quantum operation $\mE$. This representation \eqref{operator_sum} is known as the {\it operator-sum representation} describing the dynamics of the principal system $\bmA$ without having to explicitly consider any properties of the auxiliary system $\bmA^c$. More importantly, this special representation benefits us from avoiding purifying the system $\bmA$ and making our interpretation to the purification complexity $\mC_{\mt{IM}}$ not require any explicit purifications. 
Furthermore, we can also consider the measurements on the principle system by taking the outcome as $\hr_{k}$ with probability $p(k)$ after measurement. Obviously, the redefinitions 
\begin{equation}
\hr_{k} =\frac{\hat{M}_k \hr_\mA \hat{M}_k^\dagger }{\tr \( \hat{M}_k \hr_\mA \hat{M}_k^\dagger  \)} \,, \qquad  p(k)= {\tr \( \hat{M}_k \hr_\mA \hat{M}_k^\dagger  \)}\,,
\end{equation}
relate the quantum operation $\mE$ to measurements without reporting outcomes by rewriting the quantum operation as 
\begin{equation}\label{measurement}
\begin{split}
\mathcal{E}\( \hr_{\mA}\)&= \sum_k p(k) \hr_k \,.\\
\end{split}
\end{equation}
In order to describe the transformation from a physical and normalized state to another one, the trace-preserving quantum operations are restricted by the normalization condition 
\begin{equation}
 \sum_{k} \hat{M}^\dagger_k \hat{M}_k = \mathbb{I} \,.
\end{equation} 

In the following, we use Gaussian states as an example to illustrate the quantum operations acting on quantum states can be understood as the unitary operations acting on purified states in the extended Hilbert space. For any N-mode Gaussian state $\hr_{\mt{G}}$, the CPTP map $\mE\(\hr_{\mt{G}}\)$ (also called bosonic Gaussian channel) is completely characterized by two real $2N$-by-$2N$ matrices $\mathbf{T},\mathbf{N}$ acting on its vector and covariance matrix in the following way \cite{serafini2017quantum}
\begin{equation}\label{Gaussian_CP}
\begin{split} 
&\hat{\R}_{\mt{G}}  \longrightarrow   \mathbf{T} \hat{\R}_{\mt{G}} \,,\\
&\Sigma_{\mt{G}} \longrightarrow   \mathbf{T} \,\Sigma_{\mt{G}}\,  \mathbf{T}^{\intercal} + \mathbf{N} \,,
\end{split}
\end{equation}
where the real matrices $\mathbf{T}, \mathbf{N}$ are constrained by the complete positivity condition 
\begin{equation}
\mathbf{N} + i \Omega  \ge i \mathbf{T} \Omega \mathbf{T}^{\intercal} \,.
\end{equation}
 As we have shown in \eqref{operator_sum}, the action of a CP map on any Gaussian state $\hr_{\mA}$ with $n$ modes can be obtained by tracing out the ancillae (with $m$ modes) after the unitary operation on the global system $\bmA\bmA^c$ where the evolution in the full system is parametrized by the $2(n+m)$-by-$2(n+m)$ symplectic matrix $S_{\mA\mA^c}$ acting on the extended covariance matrix, \ie 
 $S_{\mA\mA^c} \( \Sigma_{\mA} \otimes \Sigma_{\mA^c}\) S^\dagger_{\mA\mA^c}$. 
 Considering the bipartition of the extended system $\bmA\bmA^c$, we can divide the full symplectic matrix into four sub-matrices 
 \begin{equation}
 S_{\mA\mA^c} = \left(
 \begin{array}{cc}
 S_{\mA}& S_{\mA\mA^c}\\
 S_{\mA^c\mA}& S_{\mA^c} \\
 \end{array}
 \right) \,,
 \end{equation}  
 corresponding to the bipartite covariance matrix shown in \eqref{AB_CM}.
 Applying the symplectic condition for $S_{\mA\mA^c}$ on the purified Gaussian states $\hr_{\mA\mA^c}$, we can find the following constrains 
 \begin{equation}
 S_{\mA\mA^c} \, \Omega \, S_{\mA\mA^c}^{\intercal} = \left(
 \begin{array}{cc}
 S_{\mA} \Omega_n S_{\mA}^{\intercal}+S_{\mA\mA^c} \Omega_m S_{\mA\mA^c}^{\intercal}& S_{\mA} \Omega_n S_{\mA^c\mA}^{\intercal}+S_{\mA\mA^c} \Omega_m S_{\mA^c}^{\intercal}\\
S_{\mA^c\mA} \Omega_n S_{\mA}^{\intercal}+S_{\mA^c} \Omega_m S_{\mA\mA^c}^{\intercal}& S_{\mA^c\mA} \Omega_n S_{\mA^c\mA}^{\intercal}+S_{\mA^c} \Omega_m S_{\mA^c}^{\intercal}\\
 \end{array}
 \right) 
 =
\left(
\begin{array}{cc}
\Omega_{n}& 0\\
0& \Omega_{m}
\end{array}
\right) \,.
 \end{equation}
 From the above equation, it is easy to find that the $S_{\mA}$ has to be symplectic if $S_{\mA\mA^c}=0$. More generally, after tracing out the auxiliary system $\bmA^c$ with $m$ modes, \ie 
 \begin{equation}
 \mE\( \Sigma_\mA \) = \tr_{\mA^c} \( S_{\mA\mA^c} \( \Sigma_{\mA} \otimes \Sigma_{\mA^c}\) S^\dagger_{\mA\mA^c}  \)\,,
 \end{equation}
  we can find that the generic Gaussian CP map \eqref{Gaussian_CP} acting on the reduced density matrix $\hr_{\mA}$ is obtained by 
	\begin{equation}
	\mathbf{T} = S_{\mA}\,, \quad \mathbf{N} = S_{\mA\mA^c} \Sigma_{\mA^c} S_{\mA\mA^c}^{\intercal} \,.
	\end{equation}
  which illustrates the connections between the quantum operations (Gaussian channels) and unitary operations with ancillae. Specifically, we can find that the $\mathbf{T}$-part provides the full information of the operations acting only on the principle system $\bmA$ while the crossing $\mathbf{N}$-part encodes the information of entangled gates. From the above identifications, it is also obvious that the unitary operation in the full system $\bmA\bmA^c$ is not unique because the quantum operations on $\hr_{\mA}$ are only sensitive to the sub-matrix $S_{\mA}$ and $S_{\mA\mA^c}$, reflecting the freedom in purifications. 
  
   In order to clarify these concepts, we take $\hr_{\mA}$ as the one-mode Gaussian state $\hr_1$ (defined in \eqref{one_mixed} and \eqref{Gaussian_1mode}) as an example and consider the two-mode Gaussian states $\ket{\psi_{12}}$ (see \eqref{Fock_psi12} and \eqref{pure1}) as purified states in $\bmA\bmA^c$ system. The symplectic matrices for one-mode squeezing operator $\hat{S}_1(r)$ and two-mode squeezing operator $\hat{S}_{12}(\alpha)$ are expressed as  
\begin{equation}
S_{1}(r) :=\left(\begin{array}{cccc} 
e^{-r} & {0} & 0& {0} \\ 
{0} &e^{r}& {0} & 0\\ 
0 & {0} &1 & {0} \\ 
{0} & 0& {0} &1
\end{array}\right)\,, 
\quad S_{11^c}(\alpha) :=\left(\begin{array}{cccc}{\cosh \alpha} & {0} & {\sinh  \alpha} & {0} \\ {0} & {\cosh  \alpha} & {0} & {-\sinh  \alpha} \\ {\sinh  \alpha} & {0} & {\cosh  \alpha} & {0} \\ {0} & {-\sinh  \alpha} & {0} & {\cosh  \alpha}\end{array}\right)\,,
%S_{\mA}= \left(\begin{array}{cccc}{\cosh \alpha} & {0} \\ {0} & {\cosh  \alpha} \end{array}\right) \,,
\end{equation}
where it is obvious that the one-mode squeezing operator $\hat S_1(r)$ \eqref{onemode_squeezed} only acts on a single mode. 
More explicitly, we choose the reference state as $\ket{0(\mu)}\otimes\ket{0(\mu)}$. The combined operation $S_{1}(\bar{r}) S_{11^c}\(\alpha\)$ on the two-mode initial state leads to 
\begin{equation}
S_{1}S_{11^c} \( \Sigma_{1} \otimes \Sigma_{1^c}\)  S_{11^c}^{\intercal}S^\intercal_{1} =
\left(
\begin{array}{cccc}
\frac{e^{-2\bar{r}}\cosh (2 \alpha )}{\mu } & 0 & \frac{e^{-\bar{r}}\sinh (2 \alpha )}{\mu } & 0 \\
0 & \mu  e^{2\bar{r}}\cosh (2 \alpha ) & 0 & - \mu e^{\bar{r}} \sinh (2\alpha )  \\
\frac{e^{-\bar{r}}\sinh (2 \alpha )}{\mu } & 0 & \frac{\cosh (2 \alpha )}{\mu } & 0 \\
0 & -\mu e^{\bar{r}} \sinh (2\alpha ) & 0 & \mu  \cosh (2 \alpha ) \\
\end{array}
\right)\,.
\end{equation}
After tracing out the second mode, \ie the ancilla, we obtain the Gaussian CP map $\mE$ acting on the principle mode  as  
\begin{equation}
\begin{split}
\mathcal{E}\( \Sigma_{1} \)&=\mathbf{T} \,\Sigma_{1}\,  \mathbf{T}^{\intercal} + \mathbf{N} =\left(
\begin{array}{cc}
\frac{e^{-2\bar{r}}}{\mu}  \cosh 2 \alpha  & 0 \\
0&  \mu e^{2\bar{r}}\cosh 2 \alpha\\
\end{array}
\right) \,,\\
\end{split}
\end{equation}
which is noting but the covariance matrix \eqref{CM_onemode} of the one-mode mixed states $\hr_1$. As a result, we can also identify the corresponding Gaussian CP map by 
\begin{equation}\label{CM_onemode2}
\mathbf{T}= \left(
\begin{array}{cc}
e^{-\bar{r}}\cosh\alpha & 0 \\
0 & e^{\bar{r}}\cosh \alpha \\
\end{array}
\right)\,, \quad 
\mathbf{N}= \left(
\begin{array}{cc}
\frac{e^{-2\bar{r}}\sinh ^2\alpha }{\mu } & 0 \\
0 & \mu e^{2\bar{r}}  \sinh ^2 \alpha \\
\end{array}
\right)\,.
\end{equation}
%%%%%%%%%%%%%%%%%%%%%%%%%%
%%%%%%%%%%%%%%%%%%%%%%%%%%
%%%%%%%%%%%%%%%%%%%%%%%%%%
\section{Quantum Fisher Information Metric and Bures Metric}\label{sec:app_metric}

In the literatures of quantum information or quantum estimation (\eg \cite{paris2009quantum,Liu:2019xfr}), the quantum Fisher information metric (QFIM) is defined in various ways and also different from Bures metric. In this section, we show they are equivalent up to a irrelevant constant factor and also list some equivalent expressions for the QFIM. 

In order to show some explicit forms of Bures metric defined by \eqref{def_Bures}
\begin{equation}
ds^2_{\mt{B}} =_2  D^2_{\mt{B}}\( \hr,\hr+\delta \hr\) = 2\(1-\tr(\sqrt{\sqrt{\hat{\rho}}(\hr+\delta \hr)\sqrt{\hat{\rho}}}) \)\,,
\end{equation}
we can start from the series expansion 
\begin{equation}
\sqrt{\sqrt{\hat{\rho}}(\hr+\delta \hr)\sqrt{\hat{\rho}}}  \approx \hr + \hat{X} + \hat{Y} + \mathcal{O}(\delta \hr^3)\,,
\end{equation}
where we keep the first two orders, \ie $\hat{X} \sim \delta \hr $, $\hat{Y} \sim \delta \hr\delta \hr $. And we note that there are simple constrains $\tr d\hr =0=\tr \hat{X}$ due to the normalization condition $\tr \hr =1=\tr \( \hr +d\hr\)$.
Taking the square of the above series expansion, one can simply find 
\begin{equation}
\begin{split}
\sqrt{\hat{\rho}}\,d \hr\,\sqrt{\hat{\rho}} &= \hr \hat{X} + \hat{X} \hr\,,\\
\hat{X}^2 + \hr \hat{Y} + \hat{Y} \hr &=0 \,.\\
\end{split}
\end{equation}
Choosing the basis of the density matrix $\hr $ by $\hr = p_i \ket{\psi_i}\bra{\psi_i}$, we can obtain the basis-dependent results 
\begin{equation}
\begin{split}
\bra{\psi_i}\hat{X} \ket{\psi_j} &= \frac{\sqrt{p_i}\sqrt{p_j}}{p_i+p_j}  \bra{\psi_i}d\hat{\rho} \ket{\psi_j}  \,,\\
\bra{\psi_i}\hat{Y} \ket{\psi_j} &= -\frac{\bra{\psi_i}\hat{X}\hat{X}  \ket{\psi_j} }{p_i+p_j} \,, \\
\end{split}
\end{equation}
which simply result in 
\begin{equation}
\begin{split}
\tr \, \hat{Y}   &= - \sum_{i,j}\, \frac{1}{2p_i} \bra{\psi_i}\hat{X} \ket{\psi_j}\bra{\psi_j}\hat{X} \ket{\psi_i}=-\frac{1}{4}\sum_{i,j}\,  \frac{\left| \bra{\psi_i}d\hat{\rho} \ket{\psi_j} \right|^2}{p_i +p_j}\,.
\end{split}
\end{equation}
Correspondingly, the Bures metric defined in \eqref{def_Bures} reads \cite{HUBNER1992239}
\begin{equation}
\begin{split}
ds^2_{\mt{B}} &= 2\(1- \tr(\hr + \hat{X} +\hat{Y} )\) = -2 \tr \,\hat{Y} \\
&= \frac{1}{2}\sum_{i,j}\,  \frac{\left| \bra{\psi_i}d\hat{\rho} \ket{\psi_j} \right|^2}{p_i +p_j}\,,
\end{split}
\end{equation} 
which is the form shown in \eqref{Bures02}. Furthermore, we can also expand the variation $d \hr=\partial_\mu \hr \, d\lambda^\mu$ in a specific basis. Noting the basis for $\hr(\lambda) = p_i (\lambda) \ket{\psi_i (\lambda)} \bra{\psi_i (\lambda)}$ also depend on the parameters $\lambda^\mu$ like the spectrum of $\hr$, \ie $p_i (\lambda^\mu)$, one can find 
\begin{equation}
 \partial_\mu \hr (\lambda) = \sum_{j} \( \partial_\mu p_j \ket{\psi_i} \bra{\partial_\mu\psi_i }    + p_j\ket{\psi_i} \bra{\psi_i}  + p_j \ket{\psi_i} \bra{\partial_\mu\psi_i }   \) \,.
\end{equation}
Then the Bures metric is rewritten as \cite{Liu:2019xfr}
\begin{equation}
 g_{\mu\nu} = \sum_i \frac{\partial_\mu p_i \partial_\nu p_i }{p_i} + \sum_{i\ne j} \frac{(p_i -p_j)^2}{p_i + p_j} \( \langle\psi_i \ket{\partial_\mu\psi_j } \langle \partial_\nu\psi_j \ket{\psi_i }   + \langle\psi_i \ket{\partial_\nu\psi_j } \langle \partial_\mu\psi_j \ket{\psi_i }    \)\,,
\end{equation}
where the first term is the same as the {\bf classical} Fisher information metric defined in \eqref{classcial_Fisher} and the second terms count the quantum contributions.  Following the popular conventions in quantum information or quantum estimation (see \eg \cite{paris2009quantum,Liu:2019xfr}), Bures metric is related to the {\bf quantum} Fisher information metric ($H_{\mu\nu}$) in the way of 
\begin{equation}
 g_{\mu\nu} =\frac{1}{4}  H_{\mu\nu}\,.
\end{equation}  
In the main content, we don't distinguish the Bures metric and QFIM by simply taking $ds^2_{\mt{IM}}=ds^2_{\mt{B}}$ because we need to normalize various metrics before taking them as the complexity measure in \eqref{def_IFcomplexity}.
On the other hand, the above expressions explicitly depend on the choice of basis for density matrix and have assumed that the dimension of Hilbert space is finite. 
In order to find a basis-independent expression without assuming the dimension of Hilbert space, we can introduce the symmetric logarithmic derivative and redefine the first order variation $\hat{X}$ by
\begin{equation}
\hat{G}=\hr^{-\frac{1}{2}} \hat{X} \hr^{-\frac{1}{2}} \,. 
\end{equation} 
 Obviously, one can get the following constrain equations
 \begin{equation}
 \begin{split}
  d\hr &= G\hr +\hr G\\
  0 &= \hr^{-1} \hat{X}^2 + \hat{Y} + \hat{\rho}^{-1} \hat{Y} \hr  \longrightarrow  -2 \tr \hat{Y} = \tr\( \hr^{-1} \hat{X}^2 \) \,.
 \end{split}
 \end{equation}
 Finally, we can rewrite the Bures metric into the new form 
 \begin{equation}
 \begin{split}
 ds^2_{\mt{B}} &=-2 \tr \hat{Y} = \tr\(\hr^{-\frac{1}{2}} \hat{X} \hat{X}\hr^{-\frac{1}{2}}\)= \tr \( \hr^{-\frac{1}{2}} \hat{X}\hr^{-\frac{1}{2}} \hr \hr^{-\frac{1}{2}}\hat{X}\hr^{-\frac{1}{2}} \) \\
 &= \tr(G\hr G) = \frac 12 \tr\( Gd\hr\)\,.
 \end{split}
 \end{equation}
 With the help of the unique solution of Lyapunov equation, \ie 
 \begin{equation}
 d\hr = G\hr +\hr G\,, \qquad  G= \int^\infty_0 \(  e^{-t \hr} d\hr e^{-t \hr}   \)dt\,,
 \end{equation} 
 we can calculate the Bures metric by the integral
 \begin{equation}
ds^2_{\mt{B}}= \frac{1}{2} \int^\infty_0 \, \tr\( e^{-\hr t} d\hr e^{-\hr t} d\hr\) dt \,,
 \end{equation}
 which is obviously basis-independent.

  %%%%%%%%%%%%%%%%%%%%%%%%%
 %%%%%%%%%%%%%%%%%%%%%%%%%
  %%%%%%%%%%%%%%%%%%%%%%%%%
 %%%%%%%%%%%%%%%%%%%%%%%%%
 %%%%%%%%%%%%%%%%%%%%%%%%%
  %%%%%%%%%%%%%%%%%%%%%%%%%
   %%%%%%%%%%%%%%%%%%%%%%%%%
    %%%%%%%%%%%%%%%%%%%%%%%%%

\end{appendix}

\bibliographystyle{JHEP}
\bibliography{biography_mixed}

\providecommand{\href}[2]{#2}\begingroup\raggedright\begin{thebibliography}{10}

\bibitem{Ryu:2006bv}
S.~Ryu and T.~Takayanagi, \emph{{Holographic derivation of entanglement entropy
  from AdS/CFT}},
  \href{http://dx.doi.org/10.1103/PhysRevLett.96.181602}{\emph{Phys. Rev.
  Lett.} {\bfseries 96} (2006) 181602},
  [\href{https://arxiv.org/abs/hep-th/0603001}{{\ttfamily hep-th/0603001}}].

\bibitem{Rangamani:2016dms}
M.~Rangamani and T.~Takayanagi, \emph{{Holographic Entanglement Entropy}},
  vol.~931.
\newblock Springer, 2017,
  \href{http://dx.doi.org/10.1007/978-3-319-52573-0}{10.1007/978-3-319-52573-0}.

\bibitem{Blanco:2013joa}
D.~D. Blanco, H.~Casini, L.-Y. Hung and R.~C. Myers, \emph{{Relative Entropy
  and Holography}},
  \href{http://dx.doi.org/10.1007/JHEP08(2013)060}{\emph{JHEP} {\bfseries 08}
  (2013) 060}, [\href{https://arxiv.org/abs/1305.3182}{{\ttfamily 1305.3182}}].

\bibitem{Faulkner:2013ica}
T.~Faulkner, M.~Guica, T.~Hartman, R.~C. Myers and M.~Van~Raamsdonk,
  \emph{{Gravitation from Entanglement in Holographic CFTs}},
  \href{http://dx.doi.org/10.1007/JHEP03(2014)051}{\emph{JHEP} {\bfseries 03}
  (2014) 051}, [\href{https://arxiv.org/abs/1312.7856}{{\ttfamily 1312.7856}}].

\bibitem{Almheiri:2014lwa}
A.~Almheiri, X.~Dong and D.~Harlow, \emph{{Bulk Locality and Quantum Error
  Correction in AdS/CFT}},
  \href{http://dx.doi.org/10.1007/JHEP04(2015)163}{\emph{JHEP} {\bfseries 04}
  (2015) 163}, [\href{https://arxiv.org/abs/1411.7041}{{\ttfamily 1411.7041}}].

\bibitem{Swingle:2009bg}
B.~Swingle, \emph{{Entanglement Renormalization and Holography}},
  \href{http://dx.doi.org/10.1103/PhysRevD.86.065007}{\emph{Phys. Rev. D}
  {\bfseries 86} (2012) 065007},
  [\href{https://arxiv.org/abs/0905.1317}{{\ttfamily 0905.1317}}].

\bibitem{VanRaamsdonk:2010pw}
M.~Van~Raamsdonk, \emph{{Building up spacetime with quantum entanglement}},
  \href{http://dx.doi.org/10.1142/S0218271810018529}{\emph{Gen. Rel. Grav.}
  {\bfseries 42} (2010) 2323--2329},
  [\href{https://arxiv.org/abs/1005.3035}{{\ttfamily 1005.3035}}].

\bibitem{Harlow:2018fse}
D.~Harlow, \emph{{TASI Lectures on the Emergence of Bulk Physics in AdS/CFT}},
  \href{http://dx.doi.org/10.22323/1.305.0002}{\emph{PoS} {\bfseries TASI2017}
  (2018) 002}, [\href{https://arxiv.org/abs/1802.01040}{{\ttfamily
  1802.01040}}].

\bibitem{Susskind:2018pmk}
L.~Susskind, \emph{{Three Lectures on Complexity and Black Holes}},  10, 2018.
\newblock \href{https://arxiv.org/abs/1810.11563}{{\ttfamily 1810.11563}}.

\bibitem{qft1}
R.~Jefferson and R.~C. Myers, \emph{{Circuit complexity in quantum field
  theory}}, \href{http://dx.doi.org/10.1007/JHEP10(2017)107}{\emph{JHEP}
  {\bfseries 10} (2017) 107},
  [\href{https://arxiv.org/abs/1707.08570}{{\ttfamily 1707.08570}}].

\bibitem{Susskind:2014rva}
L.~Susskind, \emph{{Computational Complexity and Black Hole Horizons}},
  \href{http://dx.doi.org/10.1002/prop.201500092}{\emph{Fortsch. Phys.}
  {\bfseries 64} (2016) 24--43},
  [\href{https://arxiv.org/abs/1403.5695}{{\ttfamily 1403.5695}}].

\bibitem{Stanford:2014jda}
D.~Stanford and L.~Susskind, \emph{{Complexity and Shock Wave Geometries}},
  \href{http://dx.doi.org/10.1103/PhysRevD.90.126007}{\emph{Phys. Rev. D}
  {\bfseries 90} (2014) 126007},
  [\href{https://arxiv.org/abs/1406.2678}{{\ttfamily 1406.2678}}].

\bibitem{Brown:2015bva}
A.~R. Brown, D.~A. Roberts, L.~Susskind, B.~Swingle and Y.~Zhao,
  \emph{{Holographic Complexity Equals Bulk Action?}},
  \href{http://dx.doi.org/10.1103/PhysRevLett.116.191301}{\emph{Phys. Rev.
  Lett.} {\bfseries 116} (2016) 191301},
  [\href{https://arxiv.org/abs/1509.07876}{{\ttfamily 1509.07876}}].

\bibitem{Nielsen}
M.~R. Dowling and M.~A. Nielsen, \emph{The geometry of quantum computation},
  {\emph{Quantum Info. Comput.} {\bfseries 8} (Nov., 2008) 861–899}.

\bibitem{Nielsen02}
M.~A. Nielsen, \emph{A geometric approach to quantum circuit lower bounds},
  {\emph{Quantum Info. Comput.} {\bfseries 6} (May, 2006) 213–262}.

\bibitem{qft2}
S.~Chapman, M.~P. Heller, H.~Marrochio and F.~Pastawski, \emph{{Toward a
  Definition of Complexity for Quantum Field Theory States}},
  \href{http://dx.doi.org/10.1103/PhysRevLett.120.121602}{\emph{Phys. Rev.
  Lett.} {\bfseries 120} (2018) 121602},
  [\href{https://arxiv.org/abs/1707.08582}{{\ttfamily 1707.08582}}].

\bibitem{Caputa:2017urj}
P.~Caputa, N.~Kundu, M.~Miyaji, T.~Takayanagi and K.~Watanabe, \emph{{Anti-de
  Sitter Space from Optimization of Path Integrals in Conformal Field
  Theories}},
  \href{http://dx.doi.org/10.1103/PhysRevLett.119.071602}{\emph{Phys. Rev.
  Lett.} {\bfseries 119} (2017) 071602},
  [\href{https://arxiv.org/abs/1703.00456}{{\ttfamily 1703.00456}}].

\bibitem{Czech:2017ryf}
B.~Czech, \emph{{Einstein Equations from Varying Complexity}},
  \href{http://dx.doi.org/10.1103/PhysRevLett.120.031601}{\emph{Phys. Rev.
  Lett.} {\bfseries 120} (2018) 031601},
  [\href{https://arxiv.org/abs/1706.00965}{{\ttfamily 1706.00965}}].

\bibitem{coherent}
M.~Guo, J.~Hernandez, R.~C. Myers and S.-M. Ruan, \emph{{Circuit Complexity for
  Coherent States}},
  \href{http://dx.doi.org/10.1007/JHEP10(2018)011}{\emph{JHEP} {\bfseries 10}
  (2018) 011}, [\href{https://arxiv.org/abs/1807.07677}{{\ttfamily
  1807.07677}}].

\bibitem{Hackl:2018ptj}
L.~Hackl and R.~C. Myers, \emph{{Circuit complexity for free fermions}},
  \href{http://dx.doi.org/10.1007/JHEP07(2018)139}{\emph{JHEP} {\bfseries 07}
  (2018) 139}, [\href{https://arxiv.org/abs/1803.10638}{{\ttfamily
  1803.10638}}].

\bibitem{Khan:2018rzm}
R.~Khan, C.~Krishnan and S.~Sharma, \emph{{Circuit Complexity in Fermionic
  Field Theory}},
  \href{http://dx.doi.org/10.1103/PhysRevD.98.126001}{\emph{Phys. Rev. D}
  {\bfseries 98} (2018) 126001},
  [\href{https://arxiv.org/abs/1801.07620}{{\ttfamily 1801.07620}}].

\bibitem{Bhattacharyya:2018bbv}
A.~Bhattacharyya, A.~Shekar and A.~Sinha, \emph{{Circuit complexity in
  interacting QFTs and RG flows}},
  \href{http://dx.doi.org/10.1007/JHEP10(2018)140}{\emph{JHEP} {\bfseries 10}
  (2018) 140}, [\href{https://arxiv.org/abs/1808.03105}{{\ttfamily
  1808.03105}}].

\bibitem{Chapman:2018hou}
S.~Chapman, J.~Eisert, L.~Hackl, M.~P. Heller, R.~Jefferson, H.~Marrochio
  et~al., \emph{{Complexity and entanglement for thermofield double states}},
  \href{http://dx.doi.org/10.21468/SciPostPhys.6.3.034}{\emph{SciPost Phys.}
  {\bfseries 6} (2019) 034},
  [\href{https://arxiv.org/abs/1810.05151}{{\ttfamily 1810.05151}}].

\bibitem{Ali:2018fcz}
T.~Ali, A.~Bhattacharyya, S.~Shajidul~Haque, E.~H. Kim and N.~Moynihan,
  \emph{{Time Evolution of Complexity: A Critique of Three Methods}},
  \href{http://dx.doi.org/10.1007/JHEP04(2019)087}{\emph{JHEP} {\bfseries 04}
  (2019) 087}, [\href{https://arxiv.org/abs/1810.02734}{{\ttfamily
  1810.02734}}].

\bibitem{Caputa:2017yrh}
P.~Caputa, N.~Kundu, M.~Miyaji, T.~Takayanagi and K.~Watanabe, \emph{{Liouville
  Action as Path-Integral Complexity: From Continuous Tensor Networks to
  AdS/CFT}}, \href{http://dx.doi.org/10.1007/JHEP11(2017)097}{\emph{JHEP}
  {\bfseries 11} (2017) 097},
  [\href{https://arxiv.org/abs/1706.07056}{{\ttfamily 1706.07056}}].

\bibitem{Bhattacharyya:2018wym}
A.~Bhattacharyya, P.~Caputa, S.~R. Das, N.~Kundu, M.~Miyaji and T.~Takayanagi,
  \emph{{Path-Integral Complexity for Perturbed CFTs}},
  \href{http://dx.doi.org/10.1007/JHEP07(2018)086}{\emph{JHEP} {\bfseries 07}
  (2018) 086}, [\href{https://arxiv.org/abs/1804.01999}{{\ttfamily
  1804.01999}}].

\bibitem{Camargo:2018eof}
H.~A. Camargo, P.~Caputa, D.~Das, M.~P. Heller and R.~Jefferson,
  \emph{{Complexity as a novel probe of quantum quenches: universal scalings
  and purifications}},
  \href{http://dx.doi.org/10.1103/PhysRevLett.122.081601}{\emph{Phys. Rev.
  Lett.} {\bfseries 122} (2019) 081601},
  [\href{https://arxiv.org/abs/1807.07075}{{\ttfamily 1807.07075}}].

\bibitem{Camargo:2019isp}
H.~A. Camargo, M.~P. Heller, R.~Jefferson and J.~Knaute, \emph{{Path integral
  optimization as circuit complexity}},
  \href{http://dx.doi.org/10.1103/PhysRevLett.123.011601}{\emph{Phys. Rev.
  Lett.} {\bfseries 123} (2019) 011601},
  [\href{https://arxiv.org/abs/1904.02713}{{\ttfamily 1904.02713}}].

\bibitem{Caputa:2018xuf}
P.~Caputa, M.~Miyaji, T.~Takayanagi and K.~Umemoto, \emph{{Holographic
  Entanglement of Purification from Conformal Field Theories}},
  \href{http://dx.doi.org/10.1103/PhysRevLett.122.111601}{\emph{Phys. Rev.
  Lett.} {\bfseries 122} (2019) 111601},
  [\href{https://arxiv.org/abs/1812.05268}{{\ttfamily 1812.05268}}].

\bibitem{Erdmenger:2020sup}
J.~Erdmenger, M.~Gerbershagen and A.-L. Weigel, \emph{{Complexity measures from
  geometric actions on Virasoro and Kac-Moody orbits}},
  \href{https://arxiv.org/abs/2004.03619}{{\ttfamily 2004.03619}}.

\bibitem{Flory:2020eot}
M.~Flory and M.~P. Heller, \emph{{Complexity and Conformal Field Theory}},
  \href{https://arxiv.org/abs/2005.02415}{{\ttfamily 2005.02415}}.

\bibitem{Agon:2018zso}
C.~A. Agón, M.~Headrick and B.~Swingle, \emph{{Subsystem Complexity and
  Holography}}, \href{http://dx.doi.org/10.1007/JHEP02(2019)145}{\emph{JHEP}
  {\bfseries 02} (2019) 145},
  [\href{https://arxiv.org/abs/1804.01561}{{\ttfamily 1804.01561}}].

\bibitem{purification}
E.~Caceres, S.~Chapman, J.~D. Couch, J.~P. Hernandez, R.~C. Myers and S.-M.
  Ruan, \emph{{Complexity of Mixed States in QFT and Holography}},
  \href{http://dx.doi.org/10.1007/JHEP03(2020)012}{\emph{JHEP} {\bfseries 03}
  (2020) 012}, [\href{https://arxiv.org/abs/1909.10557}{{\ttfamily
  1909.10557}}].

\bibitem{bengtsson2017geometry}
I.~Bengtsson and K.~{\.Z}yczkowski, \emph{Geometry of quantum states: an
  introduction to quantum entanglement}.
\newblock Cambridge university press, 2017.

\bibitem{chruscinski2012geometric}
D.~Chruscinski and A.~Jamiolkowski, \emph{Geometric phases in classical and
  quantum mechanics}, vol.~36.
\newblock Springer Science \& Business Media, 2012.

\bibitem{NielsenChuang}
M.~A. Nielsen and I.~L. Chuang, \emph{{Quantum Computation and Quantum
  Information}}.
\newblock Cambridge University Press, Cambridge, 2000.

\bibitem{uhlmann1976transition}
A.~Uhlmann, \emph{The transition probability in the state space of c-algebra},
  {\emph{Reports on Mathematical Physics} {\bfseries 9} (1976) 273--279}.

\bibitem{fidelity_review}
S.-J. GU, \emph{Fidelity approach to quantum phase transitions},
  \href{http://dx.doi.org/10.1142/S0217979210056335}{\emph{International
  Journal of Modern Physics B} {\bfseries 24} (2010) 4371--4458},
  [\href{https://arxiv.org/abs/0811.3127}{{\ttfamily 0811.3127}}].

\bibitem{watrous2018theory}
J.~Watrous, \emph{The theory of quantum information}.
\newblock Cambridge University Press, 2018.

\bibitem{wilde2013quantum}
M.~M. Wilde, \emph{Quantum information theory}.
\newblock Cambridge University Press, 2013.

\bibitem{jozsa1994fidelity}
R.~Jozsa, \emph{Fidelity for mixed quantum states}, {\emph{Journal of modern
  optics} {\bfseries 41} (1994) 2315--2323}.

\bibitem{PhysRevLett.76.2818}
H.~Barnum, C.~M. Caves, C.~A. Fuchs, R.~Jozsa and B.~Schumacher,
  \emph{Noncommuting mixed states cannot be broadcast},
  \href{http://dx.doi.org/10.1103/PhysRevLett.76.2818}{\emph{Phys. Rev. Lett.}
  {\bfseries 76} (Apr, 1996) 2818--2821}.

\bibitem{Nielsen:1996uh}
M.~A. Nielsen, \emph{{The Entanglement fidelity and quantum error correction}},
   \href{https://arxiv.org/abs/quant-ph/9606012}{{\ttfamily quant-ph/9606012}}.

\bibitem{bures1969extension}
D.~Bures, \emph{An extension of kakutani's theorem on infinite product measures
  to the tensor product of semifinite w*-algebras}, {\emph{Transactions of the
  American Mathematical Society} {\bfseries 135} (1969) 199--212}.

\bibitem{Liu:2019xfr}
J.~Liu, H.~Yuan, X.-M. Lu and X.~Wang, \emph{{Quantum Fisher information matrix
  and multiparameter estimation}},
  \href{http://dx.doi.org/10.1088/1751-8121/ab5d4d}{\emph{J. Phys. A}
  {\bfseries 53} (2020) 023001},
  [\href{https://arxiv.org/abs/1907.08037}{{\ttfamily 1907.08037}}].

\bibitem{firstlaw}
A.~Bernamonti, F.~Galli, J.~Hernandez, R.~C. Myers, S.-M. Ruan and J.~Simon,
  \emph{{First Law of Holographic Complexity}},
  \href{http://dx.doi.org/10.1103/PhysRevLett.123.081601}{\emph{Phys. Rev.
  Lett.} {\bfseries 123} (2019) 081601},
  [\href{https://arxiv.org/abs/1903.04511}{{\ttfamily 1903.04511}}].

\bibitem{Bernamonti:2020bcf}
A.~Bernamonti, F.~Galli, J.~Hernandez, R.~C. Myers, S.-M. Ruan and J.~Simón,
  \emph{{Aspects of The First Law of Complexity}},
  \href{https://arxiv.org/abs/2002.05779}{{\ttfamily 2002.05779}}.

\bibitem{fidelity}
J.~Twamley, \emph{Bures and statistical distance for squeezed thermal states},
  {\emph{Journal of Physics A: Mathematical and General} {\bfseries 29} (1996)
  3723}.

\bibitem{link2015geometry}
V.~Link and W.~T. Strunz, \emph{Geometry of gaussian quantum states},
  {\emph{Journal of Physics A: Mathematical and Theoretical} {\bfseries 48}
  (2015) 275301}.

\bibitem{erik}
G.~Di~Giulio and E.~Tonni, \emph{{to appear soon}}, .

\bibitem{paris2009quantum}
M.~G. Paris, \emph{Quantum estimation for quantum technology},
  {\emph{International Journal of Quantum Information} {\bfseries 7} (2009)
  125--137}.

\bibitem{HUBNER1992239}
M.~Hübner, \emph{Explicit computation of the bures distance for density
  matrices},
  \href{http://dx.doi.org/https://doi.org/10.1016/0375-9601(92)91004-B}{\emph{Physics
  Letters A} {\bfseries 163} (1992) 239 -- 242}.

\bibitem{Brown:2019whu}
A.~R. Brown and L.~Susskind, \emph{{Complexity geometry of a single qubit}},
  \href{http://dx.doi.org/10.1103/PhysRevD.100.046020}{\emph{Phys. Rev. D}
  {\bfseries 100} (2019) 046020},
  [\href{https://arxiv.org/abs/1903.12621}{{\ttfamily 1903.12621}}].

\bibitem{Gaussian_fidelity}
L.~Banchi, P.~Giorda and P.~Zanardi, \emph{Quantum information-geometry of
  dissipative quantum phase transitions},
  \href{http://dx.doi.org/10.1103/PhysRevE.89.022102}{\emph{Phys. Rev. E}
  {\bfseries 89} (Feb, 2014) 022102}.

\bibitem{Gaussian_fidelity02}
L.~Banchi, S.~L. Braunstein and S.~Pirandola, \emph{Quantum fidelity for
  arbitrary gaussian states},
  \href{http://dx.doi.org/10.1103/PhysRevLett.115.260501}{\emph{Phys. Rev.
  Lett.} {\bfseries 115} (Dec, 2015) 260501}.

\bibitem{carollo2018uhlmann}
A.~Carollo, B.~Spagnolo and D.~Valenti, \emph{Uhlmann curvature in dissipative
  phase transitions}, {\emph{Scientific reports} {\bfseries 8} (2018) 9852}.

\bibitem{PhysRevA.71.032336}
H.~Nha and H.~J. Carmichael, \emph{Distinguishing two single-mode gaussian
  states by homodyne detection: An information-theoretic approach},
  \href{http://dx.doi.org/10.1103/PhysRevA.71.032336}{\emph{Phys. Rev. A}
  {\bfseries 71} (Mar, 2005) 032336}.

\bibitem{Kirklin:2019ror}
J.~Kirklin, \emph{{The Holographic Dual of the Entanglement Wedge Symplectic
  Form}}, \href{http://dx.doi.org/10.1007/JHEP01(2020)071}{\emph{JHEP}
  {\bfseries 01} (2020) 071},
  [\href{https://arxiv.org/abs/1910.00457}{{\ttfamily 1910.00457}}].

\bibitem{Lashkari:2015hha}
N.~Lashkari and M.~Van~Raamsdonk, \emph{{Canonical Energy is Quantum Fisher
  Information}}, \href{http://dx.doi.org/10.1007/JHEP04(2016)153}{\emph{JHEP}
  {\bfseries 04} (2016) 153},
  [\href{https://arxiv.org/abs/1508.00897}{{\ttfamily 1508.00897}}].

\bibitem{MIyaji:2015mia}
M.~Miyaji, T.~Numasawa, N.~Shiba, T.~Takayanagi and K.~Watanabe,
  \emph{{Distance between Quantum States and Gauge-Gravity Duality}},
  \href{http://dx.doi.org/10.1103/PhysRevLett.115.261602}{\emph{Phys. Rev.
  Lett.} {\bfseries 115} (2015) 261602},
  [\href{https://arxiv.org/abs/1507.07555}{{\ttfamily 1507.07555}}].

\bibitem{Trivella:2016brw}
A.~Trivella, \emph{{Holographic Computations of the Quantum Information
  Metric}}, \href{http://dx.doi.org/10.1088/1361-6382/aa69a6}{\emph{Class.
  Quant. Grav.} {\bfseries 34} (2017) 105003},
  [\href{https://arxiv.org/abs/1607.06519}{{\ttfamily 1607.06519}}].

\bibitem{Bak:2017rpp}
D.~Bak and A.~Trivella, \emph{{Quantum Information Metric on $\mathbb{R} \times
  S^{d-1}$}}, \href{http://dx.doi.org/10.1007/JHEP09(2017)086}{\emph{JHEP}
  {\bfseries 09} (2017) 086},
  [\href{https://arxiv.org/abs/1707.05366}{{\ttfamily 1707.05366}}].

\bibitem{Czech:2017zfq}
B.~Czech, L.~Lamprou, S.~Mccandlish and J.~Sully, \emph{{Modular Berry
  Connection for Entangled Subregions in AdS/CFT}},
  \href{http://dx.doi.org/10.1103/PhysRevLett.120.091601}{\emph{Phys. Rev.
  Lett.} {\bfseries 120} (2018) 091601},
  [\href{https://arxiv.org/abs/1712.07123}{{\ttfamily 1712.07123}}].

\bibitem{Alishahiha:2017cuk}
M.~Alishahiha and A.~Faraji~Astaneh, \emph{{Holographic Fidelity
  Susceptibility}},
  \href{http://dx.doi.org/10.1103/PhysRevD.96.086004}{\emph{Phys. Rev. D}
  {\bfseries 96} (2017) 086004},
  [\href{https://arxiv.org/abs/1705.01834}{{\ttfamily 1705.01834}}].

\bibitem{Banerjee:2017qti}
S.~Banerjee, J.~Erdmenger and D.~Sarkar, \emph{{Connecting Fisher information
  to bulk entanglement in holography}},
  \href{http://dx.doi.org/10.1007/JHEP08(2018)001}{\emph{JHEP} {\bfseries 08}
  (2018) 001}, [\href{https://arxiv.org/abs/1701.02319}{{\ttfamily
  1701.02319}}].

\bibitem{Moosa:2018mik}
M.~Moosa and I.~Shehzad, \emph{{Is volume the holographic dual of fidelity
  susceptibility?}},  \href{https://arxiv.org/abs/1809.10169}{{\ttfamily
  1809.10169}}.

\bibitem{Czech:2018kvg}
B.~Czech, L.~Lamprou and L.~Susskind, \emph{{Entanglement Holonomies}},
  \href{https://arxiv.org/abs/1807.04276}{{\ttfamily 1807.04276}}.

\bibitem{Belin:2018fxe}
A.~Belin, A.~Lewkowycz and G.~Sárosi, \emph{{The boundary dual of the bulk
  symplectic form}},
  \href{http://dx.doi.org/10.1016/j.physletb.2018.10.071}{\emph{Phys. Lett. B}
  {\bfseries 789} (2019) 71--75},
  [\href{https://arxiv.org/abs/1806.10144}{{\ttfamily 1806.10144}}].

\bibitem{Suzuki:2019xdq}
Y.~Suzuki, T.~Takayanagi and K.~Umemoto, \emph{{Entanglement Wedges from the
  Information Metric in Conformal Field Theories}},
  \href{http://dx.doi.org/10.1103/PhysRevLett.123.221601}{\emph{Phys. Rev.
  Lett.} {\bfseries 123} (2019) 221601},
  [\href{https://arxiv.org/abs/1908.09939}{{\ttfamily 1908.09939}}].

\bibitem{Erdmenger:2020vmo}
J.~Erdmenger, K.~T. Grosvenor and R.~Jefferson, \emph{{Information geometry in
  quantum field theory: lessons from simple examples}},
  \href{https://arxiv.org/abs/2001.02683}{{\ttfamily 2001.02683}}.

\bibitem{Jafferis:2015del}
D.~L. Jafferis, A.~Lewkowycz, J.~Maldacena and S.~J. Suh, \emph{{Relative
  entropy equals bulk relative entropy}},
  \href{http://dx.doi.org/10.1007/JHEP06(2016)004}{\emph{JHEP} {\bfseries 06}
  (2016) 004}, [\href{https://arxiv.org/abs/1512.06431}{{\ttfamily
  1512.06431}}].

\bibitem{Botta-Cantcheff:2015sav}
M.~Botta-Cantcheff, P.~Martínez and G.~A. Silva, \emph{{On excited states in
  real-time AdS/CFT}},
  \href{http://dx.doi.org/10.1007/JHEP02(2016)171}{\emph{JHEP} {\bfseries 02}
  (2016) 171}, [\href{https://arxiv.org/abs/1512.07850}{{\ttfamily
  1512.07850}}].

\bibitem{Marolf:2017kvq}
D.~Marolf, O.~Parrikar, C.~Rabideau, A.~Izadi~Rad and M.~Van~Raamsdonk,
  \emph{{From Euclidean Sources to Lorentzian Spacetimes in Holographic
  Conformal Field Theories}},
  \href{http://dx.doi.org/10.1007/JHEP06(2018)077}{\emph{JHEP} {\bfseries 06}
  (2018) 077}, [\href{https://arxiv.org/abs/1709.10101}{{\ttfamily
  1709.10101}}].

\bibitem{Arias:2020qpg}
R.~Arias, M.~Botta-Cantcheff, P.~J. Martinez and J.~F. Zarate, \emph{{Modular
  Hamiltonian for (holographic) excited states}},
  \href{https://arxiv.org/abs/2002.04637}{{\ttfamily 2002.04637}}.

\bibitem{Alishahiha:2015rta}
M.~Alishahiha, \emph{{Holographic Complexity}},
  \href{http://dx.doi.org/10.1103/PhysRevD.92.126009}{\emph{Phys. Rev. D}
  {\bfseries 92} (2015) 126009},
  [\href{https://arxiv.org/abs/1509.06614}{{\ttfamily 1509.06614}}].

\bibitem{Carmi:2016wjl}
D.~Carmi, R.~C. Myers and P.~Rath, \emph{{Comments on Holographic Complexity}},
  \href{http://dx.doi.org/10.1007/JHEP03(2017)118}{\emph{JHEP} {\bfseries 03}
  (2017) 118}, [\href{https://arxiv.org/abs/1612.00433}{{\ttfamily
  1612.00433}}].

\bibitem{Ben-Ami:2016qex}
O.~Ben-Ami and D.~Carmi, \emph{{On Volumes of Subregions in Holography and
  Complexity}}, \href{http://dx.doi.org/10.1007/JHEP11(2016)129}{\emph{JHEP}
  {\bfseries 11} (2016) 129},
  [\href{https://arxiv.org/abs/1609.02514}{{\ttfamily 1609.02514}}].

\bibitem{Caceres:2018blh}
E.~Cáceres, J.~Couch, S.~Eccles and W.~Fischler, \emph{{Holographic
  Purification Complexity}},
  \href{http://dx.doi.org/10.1103/PhysRevD.99.086016}{\emph{Phys. Rev. D}
  {\bfseries 99} (2019) 086016},
  [\href{https://arxiv.org/abs/1811.10650}{{\ttfamily 1811.10650}}].

\bibitem{Alishahiha:2018lfv}
M.~Alishahiha, K.~Babaei~Velni and M.~R. Mohammadi~Mozaffar, \emph{{Black hole
  subregion action and complexity}},
  \href{http://dx.doi.org/10.1103/PhysRevD.99.126016}{\emph{Phys. Rev. D}
  {\bfseries 99} (2019) 126016},
  [\href{https://arxiv.org/abs/1809.06031}{{\ttfamily 1809.06031}}].

\bibitem{Braccia:2019xxi}
P.~Braccia, A.~L. Cotrone and E.~Tonni, \emph{{Complexity in the presence of a
  boundary}}, \href{http://dx.doi.org/10.1007/JHEP02(2020)051}{\emph{JHEP}
  {\bfseries 02} (2020) 051},
  [\href{https://arxiv.org/abs/1910.03489}{{\ttfamily 1910.03489}}].

\bibitem{Abt:2018ywl}
R.~Abt, J.~Erdmenger, M.~Gerbershagen, C.~M. Melby-Thompson and C.~Northe,
  \emph{{Holographic Subregion Complexity from Kinematic Space}},
  \href{http://dx.doi.org/10.1007/JHEP01(2019)012}{\emph{JHEP} {\bfseries 01}
  (2019) 012}, [\href{https://arxiv.org/abs/1805.10298}{{\ttfamily
  1805.10298}}].

\bibitem{RevModPhys.84.621}
C.~Weedbrook, S.~Pirandola, R.~Garc\'{\i}a-Patr\'on, N.~J. Cerf, T.~C. Ralph,
  J.~H. Shapiro et~al., \emph{Gaussian quantum information},
  \href{http://dx.doi.org/10.1103/RevModPhys.84.621}{\emph{Rev. Mod. Phys.}
  {\bfseries 84} (May, 2012) 621--669}.

\bibitem{ferraro2005gaussian}
A.~Ferraro, S.~Olivares and M.~G. Paris, \emph{Gaussian states in continuous
  variable quantum information},
  \href{https://arxiv.org/abs/quant-ph/0503237}{{\ttfamily quant-ph/0503237}}.

\bibitem{serafini2017quantum}
A.~Serafini, \emph{Quantum Continuous Variables: A Primer of Theoretical
  Methods}.
\newblock CRC Press, 2017.

\bibitem{gaussian}
H.~Scutaru, \emph{Fidelity for displaced squeezed thermal states and the
  oscillator semigroup}, {\emph{Journal of Physics A: Mathematical and General}
  {\bfseries 31} (1998) 3659}.

\end{thebibliography}\endgroup

\end{document}